\documentclass[12pt]{article}
\usepackage{epsfig,cite}
\usepackage {amsmath}
\usepackage{amssymb}
\include{epsf}
\newcount\timecount
\newcount\hours \newcount\minutes  \newcount\temp \newcount\pmhours
\hours = \time
\divide\hours by 60
\temp = \hours
\multiply\temp by 60
\minutes = \time
\advance\minutes by -\temp
\def\hour{\the\hours}
\def\minute{\ifnum\minutes<10 0\the\minutes
            \else\the\minutes\fi}
\def\clock{
\ifnum\hours=0 12:\minute\ AM
\else\ifnum\hours<12 \hour:\minute\ AM
      \else\ifnum\hours=12 12:\minute\ PM
            \else\ifnum\hours>12
                 \pmhours=\hours
                 \advance\pmhours by -12
                 \the\pmhours:\minute\ PM
                 \fi
            \fi
      \fi
\fi
}

\def\monthname{\relax\ifcase\month 0/\or January\or February\or
   March\or April\or May\or June\or July\or August\or September\or
   October\or November\or December\else\number\month/\fi}

\def\bold#1{\setbox0=\hbox{$#1$}%
     \kern-.025em\copy0\kern-\wd0
     \kern.05em\copy0\kern-\wd0
     \kern-.025em\raise.0433em\box0 }

\newcommand{\ETslash}{/ \hspace{-.7em} E_T}

\def\beq{\begin{equation}}
\def\eeq{\end{equation}}

\def\ga{\mathrel{\raise.3ex\hbox{$>$\kern-.75em\lower1ex\hbox{$\sim$}}}}
\def\la{\mathrel{\raise.3ex\hbox{$<$\kern-.75em\lower1ex\hbox{$\sim$}}}}
\def\gev{{\rm \, Ge\kern-0.125em V}}
\def\tev{{\rm \, Te\kern-0.125em V}}
\def\gyr{{\rm \, G\kern-0.125em yr}}

\def\tb{\tan \beta}

\def\gappeq{\mathrel{\rlap {\raise.5ex\hbox{$>$}}
{\lower.5ex\hbox{$\sim$}}}}
\def\lappeq{\mathrel{\rlap{\raise.5ex\hbox{$<$}}
{\lower.5ex\hbox{$\sim$}}}}
\def\Toprel#1\over#2{\mathrel{\mathop{#2}\limits^{#1}}}

\def\m12{m_{1\!/2}}
\def\bea{\begin{eqnarray}}
\def\eea{\end{eqnarray}}

\begin{document}
\begin{titlepage}
\pagestyle{empty}
\baselineskip=21pt
\rightline{KCL-PH-TH/2014-17, LCTS/2014-16, CERN-PH-TH/2014-067}
\rightline{UMN--TH--3333/14, FTPI--MINN--14/11}
\vskip 0.75in
\begin{center}
{\large{\bf The Extent of the Stop Coannihilation Strip}}
\end{center}
\begin{center}
\vskip 0.3in
{\bf John~Ellis}$^{1,2}$,
{\bf Keith~A.~Olive}$^{3,4}$
and {\bf Jiaming~Zheng}$^3$
\vskip 0.3in
{\small {\it
$^1${Theoretical Particle Physics and Cosmology Group, Department of
  Physics, \\ King's College London, London WC2R 2LS, United Kingdom}\\
$^2${Theory Division, CERN, CH-1211 Geneva 23, Switzerland}\\
$^3${School of Physics and Astronomy, University of Minnesota, Minneapolis, MN 55455, USA}\\
$^4${William I. Fine Theoretical Physics Institute, School of Physics and Astronomy,\\
University of Minnesota, Minneapolis, MN 55455, USA}\\
}}

\vskip 0.75in
{\bf Abstract}
\end{center}
\baselineskip=18pt \noindent
%%%%%%%%%%%%%%%%%%%%%%%%%%%%%%%%%%%%%%%%%%%%%%%%%%%%%%%%%%%%%%%%%%%%%

Many supersymmetric models such as the CMSSM feature a strip in parameter space where the lightest neutralino $\chi$
is identified as the lightest supersymmetric particle (LSP), the lighter stop squark 
${\tilde t_1}$ is the next-to-lightest supersymmetric particle (NLSP), and the relic $\chi$ 
cold dark matter density is brought into the range allowed by astrophysics and cosmology 
by coannihilation with the lighter stop squark ${\tilde t_1}$ NLSP. We calculate the
stop coannihilation strip in the CMSSM, incorporating Sommerfeld enhancement effects,
and explore the relevant phenomenological constraints and phenomenological signatures. 
In particular, we show that the ${\tilde t_1}$ may weigh several TeV, 
and its lifetime may be in the nanosecond range, features that are more general than the
specific CMSSM scenarios that we study in this paper.

%%%%%%%%%%%%%%%%%%%%%%%%%%%%%%%%%%%%%%%%%%%%%%%%%%%%%%%%%%%%%%%%%%%%%
\vfill
%\leftline{CERN-PH-TH/2005-173}
\leftline{April 2014}
\end{titlepage}
\baselineskip=18pt
%%%%%%%%%%%%%%%%%%%%%%%%%%%%%%%%%%%%%%%%%%%%%%%%%%%%%%%%%%%%%%%%%%%%%

\section{Introduction}

The non-appearance of supersymmetry during Run~1 of the LHC has given
many theorists pause for thought. However, they should be encouraged by
the fact that the Higgs boson has been discovered~\cite{lhch} within the mass range
predicted by simple supersymmetric models~\cite{mh,mh2loop,mhiggsAEC,susycompare},
and that its principal production
and decay modes have occured at rates similar to those predicted for the Higgs
boson of the Standard Model, also as predicted by simple supersymmetric models.
The search for supersymmetry will continue during Run~2 of the LHC at higher 
energies and luminosities, which will have greatly-extended physics reach
compared to Run~1. It is important that this renewed experimental effort be
matched by a thorough theoretical exploration of the different possible
phenomenological signatures.

Many supersymmetric models, such as the constrained minimal supersymmetric extension of the Standard Model 
(CMSSM)~\cite{funnel,cmssm},
incorporate $R$-parity conservation, in which case the lightest supersymmetric particle (LSP) is stable
and could provide astrophysical dark matter~\cite{cmssmwmap,eo6,elos,ehow+}. 
We assume here that the LSP is the
lightest neutralino $\chi$~\cite{ehnos}. There are several regions of the CMSSM parameter space
where the relic $\chi$ density may fall within the range allowed by astrophysical and
cosmological observations. Among the possibilities that have been most studied are
the strip where stau-$\chi$ coannihilation is important \cite{stau-co}, the funnel where there is rapid
$\chi \chi$ annihilation via direct-channel heavy Higgs poles~\cite{funnel,efgosi}, and the focus-point
region where the $\chi$ acquires a significant Higgsino component~\cite{fp}. The purpose of
this paper is to pay closer attention to another possibility, namely the strip in the CMSSM
parameter space where stop-$\chi$ coannihilation is important~\cite{stop,eos,deos,Harz}.

Generally speaking, the allowed parameter space of the CMSSM for any fixed values of 
$\tan \beta$ and $A_0/m_0$ may be viewed as a wedge in the $(m_{1/2}, m_0)$ plane. Low values
of $m_0/m_{1/2}$ are excluded because there the LSP is the lighter stau slepton, which is charged
and hence not a suitable dark matter candidate. The stau coannihilation strip runs along
the boundary of this forbidden region~\cite{stau-co}. High values of $m_0/m_{1/2}$ are also
generically excluded, though for varying reasons. At low $A_0/m_0$, the reason is that no
consistent electroweak vacuum can be found at large $m_0/m_{1/2}$, and close to the
boundary of this forbidden region the Higgs superpotential mixing parameter $\mu$
becomes small, the Higgsino component of the $\chi$ gets enhanced, and one encounters
the focus-point strip~\cite{fp}. However, when $A_0/m_0$ is larger, the issue at large $m_0/m_{1/2}$
is that the LSP becomes the lighter stop squark ${\tilde t_1}$, which is also not a suitable 
dark matter candidate. Close to this boundary of the CMSSM wedge, the ${\tilde t_1}$ is the
next-to-lightest supersymmetric particle (NLSP), and the relic $\chi$ density may be brought
into the cosmological range by ${\tilde t_1} \chi$ coannihilation~\cite{stop,eos,deos}. 
The length of the ${\tilde t_1} \chi$
coannihilation strip is increased by Sommerfeld enhancements in some ${\tilde t_1} {\tilde t_1}^\star$
annihilation channels~\cite{as,somm,dgs}, which we include in our analysis.

In this paper we study the extent to which portions of this ${\tilde t_1} \chi$ strip may be
compatible with experimental and phenomenological constraints as well as the cosmological
dark matter density, paying particular attention to the constraint imposed by the LHC
measurement of the mass of the Higgs boson. Other things being equal, the measurement
$m_H = 125.9 \pm 0.4$~GeV tends to favour larger values of $A_0$ such as those featuring
a ${\tilde t_1} \chi$ coannihilation strip, reinforcing our interest in this region of the CMSSM
parameter space \cite{eo6,elos,ehow+,post-mh}. 
We use {\tt FeynHiggs~2.10.0} to calculate the lightest supersymmetric
Higgs mass and to estimate uncertainties in this calculation \cite{fh2100}.
We find that the stop coannihilation strip may extend up to $m_{1/2} \simeq 13000$~GeV,
corresponding to $m_\chi = m_{\tilde t_1} \simeq 6500$~GeV, that the end-point of the stop
coannihilation strip may be compatible with the LHC measurement of $m_h$ for $\tan \beta = 40$ or 
large $A_0/m_0 = 5.0$
within the {\tt FeynHiggs~2.10.0} uncertainty,
and that the stop lifetime may extend into the nanosecond range.

The layout of this paper is as follows. In Section~2 we review relevant general features of the
CMSSM, setting the ${\tilde t_1} \chi$ coannihilation strip in context and
describing our treatment of Sommerfeld enhancement effects. Then, in Section~3 we
study the possible extent of this strip and the allowed range of the ${\tilde t_1}$ mass. Although our specific numerical
studies are the framework of the CMSSM, we emphasize that our general conclusions have broader
validity. In Section~4 we discuss ${\tilde t_1}$ decay signatures, which are also not specific to
the CMSSM, and in Section~5 we summarize our conclusions.

\section{Anatomy of the Stop Coannihilation Strip}

We work in the framework of the CP-conserving CMSSM, in which the soft supersymmetry-breaking parameters
$m_{1/2}, m_0$ and $A_0$ are assumed to be real and universal at the GUT scale. We treat $\tan \beta$ as
another free parameter and use the renormalization-group equations (RGEs) and the electroweak
vacuum conditions to determine the Higgs superpotential mixing parameter $\mu$ and the corresponding
soft supersymmetry-breaking parameter $B$ (or, equivalently, the pseudoscalar Higgs mass $M_A$).
We concentrate in the following on the choices $\mu > 0$ and $A_0 > 0$.

\subsection{Sommerfeld Effect}

We evaluate the dark matter density in the regions of the stop coannihilation strips
including the Sommerfeld effect, which may enhance the annihilation rates at low
velocities, and is particularly
relevant for strongly-interacting particles such as the stop squark. 
As we discuss in more detail below, the general effect of including the Sommerfeld factors
is to increase substantially the length of the stop coannihilation strip.

In general, the
Sommerfeld effect modifies s-wave cross-sections by factors \cite{as}
\begin{equation}
F(s) \; \equiv \frac{- \pi s}{1 - e^{\pi s}} : \; s \; \equiv \; \frac{\alpha}{\beta} \, ,
\label{Sommerfeld}
\end{equation}
where $\beta$ is the annihilating particle velocity and
$\alpha$ is the coefficient of a Coulomb-like potential whose sign is chosen so that $\alpha < 0$
corresponds to attraction. In the case of annihilating particles with strong interactions, 
the Coulomb-like potential may be written as \cite{coul}
\begin{equation}
V \; = \; \frac{\alpha_3}{2 r} \left[ C_f - C_i - C_i^\prime \right] \, ,
\label{Casimirs}
\end{equation}
where $\alpha_3$ is the strong coupling strength at the appropriate scale,
$C_i$ and $C_i^\prime$ are the quadratic Casimir coefficients of the annihilating coloured
particles, and $C_f$ is the quadratic Casimir coefficient of a specific final-state colour representation.
In our case, we always have $C_i = C_i^\prime = C_3 = 4/3$. In ${\tilde t_1} - {\tilde t_1}^\star$ annihilations the
possible s-channel states are singlets with $C_1 = 0$ and octets with $C_8 = 3$, whereas in
${\tilde t_1} - {\tilde t_1}$ annihilations Bose symmetry implies that
the only possible final colour state is a sextet with $C_6 = 10/3$.
The factors in the square parentheses $[ ... ]$ for the singlet, octet and sextet final states are
therefore $-8/3, + 1/3$ and $+ 2/3$, respectively, corresponding to $\alpha = - 4 \alpha_3/3, \alpha_3/6$
and $\alpha_3/3$, respectively. Only the singlet final state exhibits a Sommerfeld
enhancement: s-wave annihilations in the other two colour states actually exhibit suppressions.

We implement the Sommerfeld effects in the {\tt SSARD} code~\cite{SSARD} for calculating the relic
dark matter density, which is based on a non-relativistic expansion for annihilation cross-sections:
\begin{equation}
\langle \sigma v \rangle \; = \; a + b x + \dots \, ,
\label{ab}
\end{equation}
where $\langle ... \rangle$ denotes an average over the thermal distributions of the annihilating
particles, the coefficient $a$ represents the contribution of the s-wave cross-section, $x \equiv T/m$, and the dots
represent terms of higher order in $x$. When $\alpha < 0$ in (\ref{Sommerfeld}), 
as in the singlet final state discussed above, the leading term in (\ref{ab}) acquires a singularity
\begin{equation}
a \to a \frac{\sqrt{2 \pi}}{x} + \dots \, ,
\label{singularity}
\end{equation}
where the dots again represent terms of higher order in $x$.

Along the stop coannihilation strip, the dominant ${\tilde t_1} - {\tilde t_1}^\star$ s-wave annihilation
cross-sections are typically those into colour-singlet pairs of Higgs bosons ($\sim 60 - 70$\% in the CMSSM before
incorporating the Sommerfeld effect) and into gluon pairs ($\sim 20 - 30 $\%),
which are a mixture of 2/7 colour-singlet and 5/7 colour-octet final states, followed by the colour-octet
$Z$ + gluon final state ($\sim 5$\% in the CMSSM). We have implemented the Sommerfeld effects for these
${\tilde t_1} - {\tilde t_1}^\star$ final states, and also for ${\tilde t_1} - {\tilde t_1} \to t + t$ annihilations,
whose s-wave annihilation cross-section $\sim 5$\% of the total ${\tilde t_1} - {\tilde t_1}^\star$ 
s-wave annihilation cross-section before including the Sommerfeld effect. 

We emphasize that the Sommerfeld factors in different channels
depend only on the final states, and are independent
of the specific CMSSM scenario that we study. We also emphasize that many other supersymmetric
models feature the same suite of final states in stop-neutralino coannihilation. Moreover, some of the
couplings to these final states are universal, e.g., ${\tilde t_1} - {\tilde t_1}^\star$ annihilations
to gluon pairs mediated by crossed-channel ${\tilde t_1}$ exchange and direct-channel gluon exchange.
The similarities imply that results resembling ours
would hold in many related supersymmetric models~\footnote{We take the opportunity to recall that
radiative corrections to stop coannihilation processes have been calculated in~\cite{Harz}. Their
effects are, in general, smaller than other uncertainties in our calculations and are not included in our analysis.}

\subsection{The End-Point of the Stop Coannihilation Strip}

As we shall also see, there are differences in the lengths of of the stop
coannihilation strips for different values of the model parameters. Looking at
the dominant ${\tilde t_1} - {\tilde t_1}^\star$ annihilation mechanisms, it is clear that
the matrix elements for annihilations to some final states are universal, e.g., to gluon pairs.
However, the dominant ${\tilde t_1} - {\tilde t_1}^\star$ annihilations to pairs of Higgs bosons
are model-dependent. The dominant contributions to ${\tilde t_1} - {\tilde t_1}^\star \to h + h$ annihilation,
 in the notation of the Appendix in~\cite{eos}, are
 $\textrm{I}\times\textrm{I},\textrm{II}\times\textrm{II},\textrm{I}\times\textrm{II},\textrm{I}\times\textrm{III}$
 and $\textrm{II}\times\textrm{III}$ with $i=2$, corresponding to $t-$ and $u$-channel exchanges
 of the heavier stop ${\tilde t_2}$, the exchange of the lighter stop exchange being suppressed by $\sin\theta_t$,
 where $\theta_t$ is the ${\tilde t_1} - {\tilde t_2}$ mixing angle.
The ${\tilde t_1} - {\tilde t_2}^\star - h$ coupling takes the form
\begin{equation}
C_{\tilde{t}_1-\tilde{t}_2-h} \; \sim \; \frac{\mu\sin\alpha-A_t\cos\alpha}{2m_W\sin\beta}\cos2\theta_t  \, ,
\label{t1t2h}
\end{equation}
which depends on $A_t$, $\sin \beta$, the Higgs mixing angle $\alpha$ and $\mu$, as well as $\theta_t$, and the annihilation
cross-section also depends on $m_{\tilde{t}_2}$. The ${\tilde t_1} - {\tilde t_1}^\star \to h + h$ annihilation rate is 
therefore model-dependent,
depending primarily on the combination $C_{\tilde{t}_1-\tilde{t}_2-h}/m_{\tilde{t}_2}$, which causes
$m_\chi$ at the tip of the stop coannihilation strip to vary as we see later.

\section{Representative Parameter Planes in the CMSSM}

\subsection{$(m_{1/2}, m_0)$ Planes}

\begin{figure}
\vspace{-4cm}
\begin{center}
\begin{tabular}{c c}
\includegraphics[height=8cm]{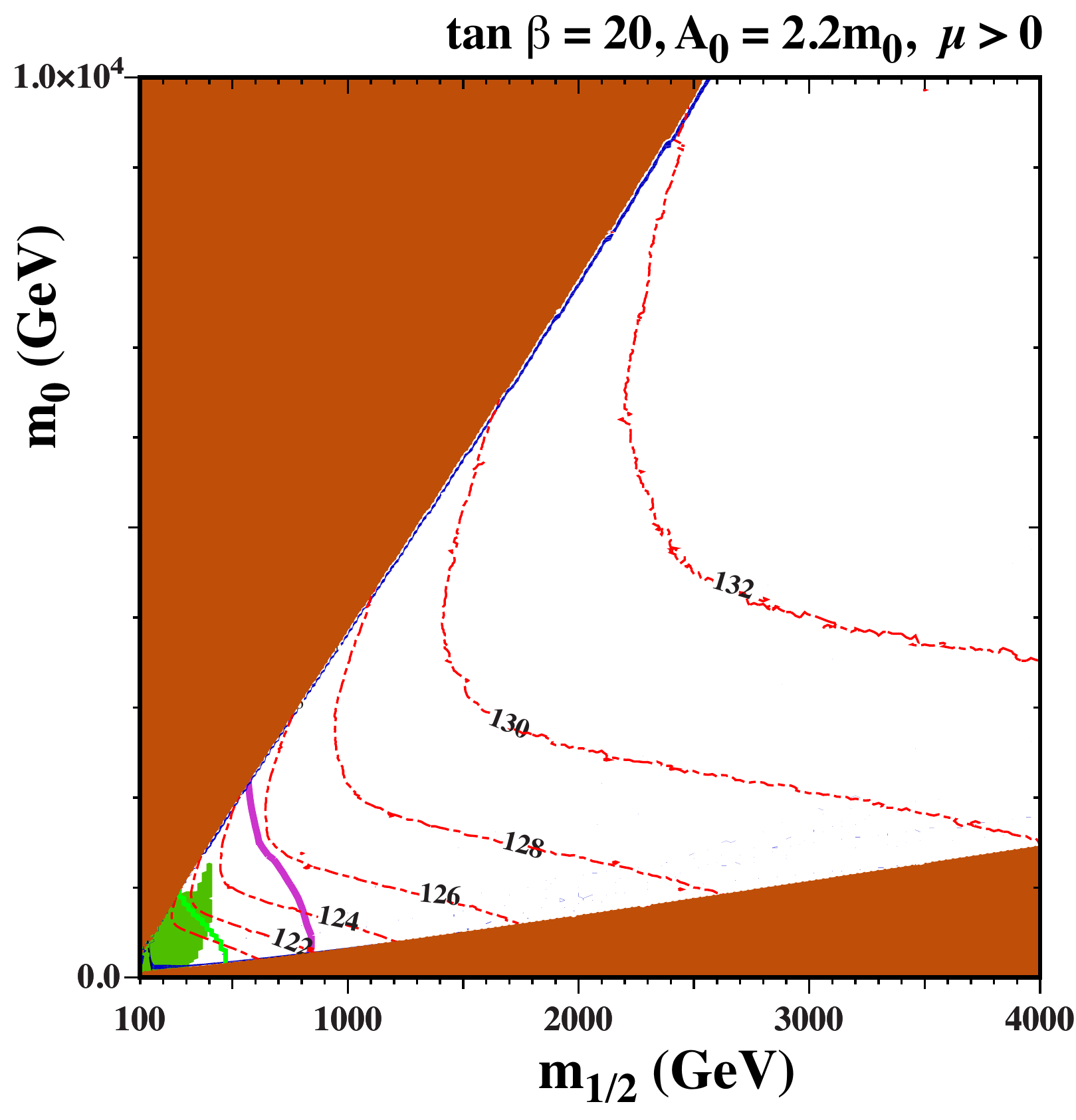} & 
%\hspace{-2cm}
\includegraphics[height=8cm]{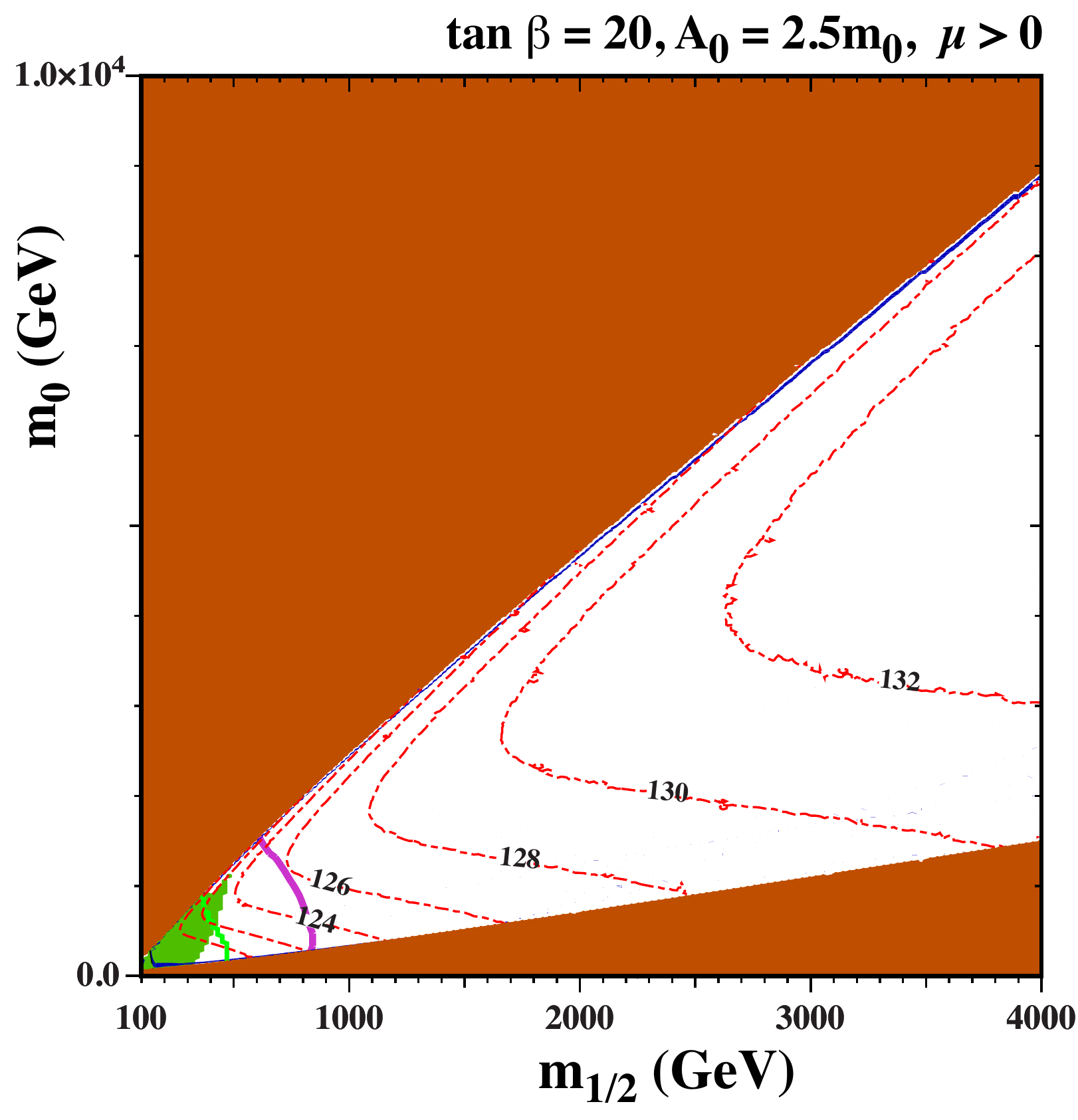} \\
\end{tabular}
\end{center}   
%\vspace{-5cm}
\begin{center}
\begin{tabular}{c c}
\includegraphics[height=8cm]{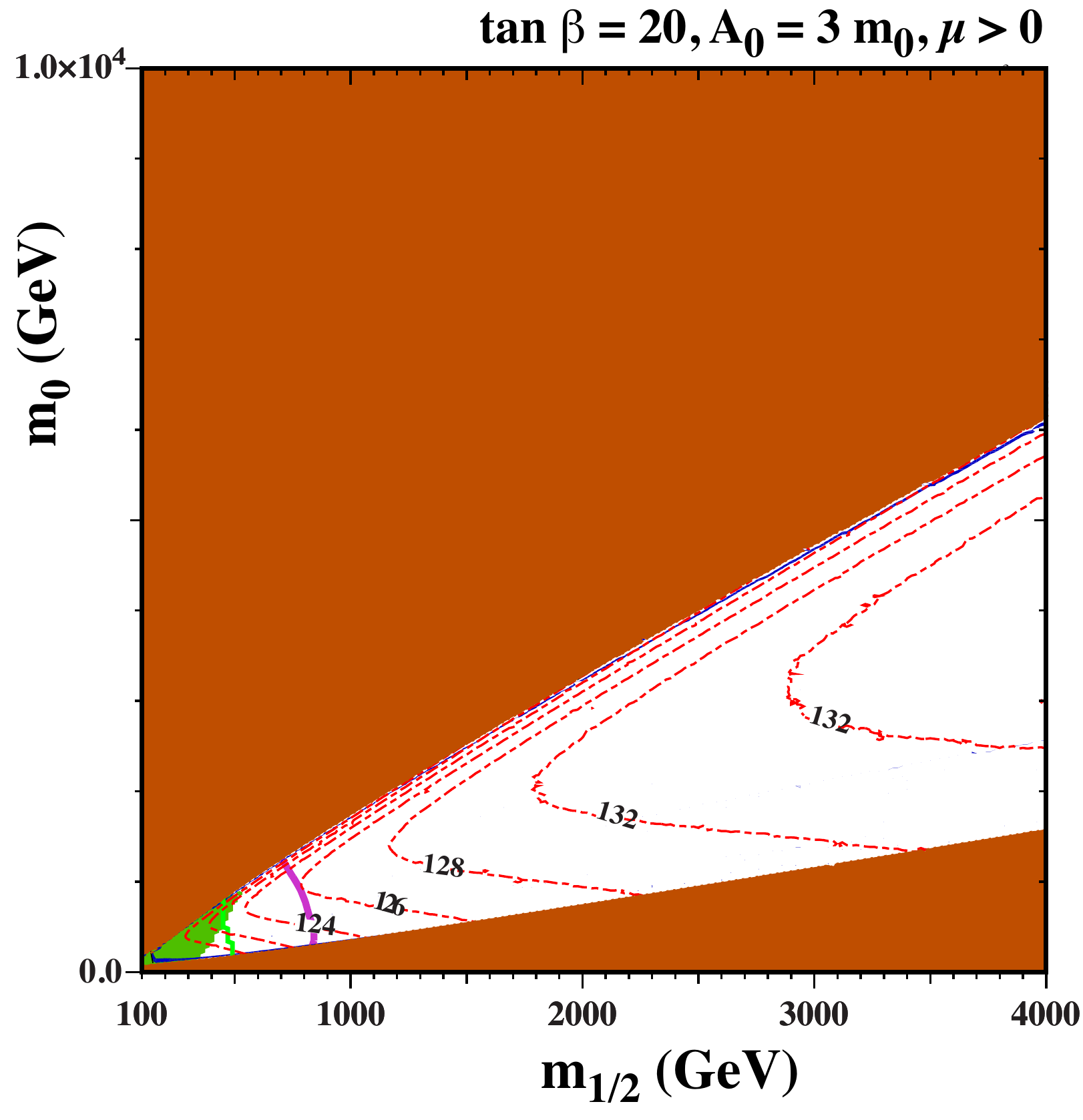} & 
%\hspace{-2cm}
\includegraphics[height=8cm]{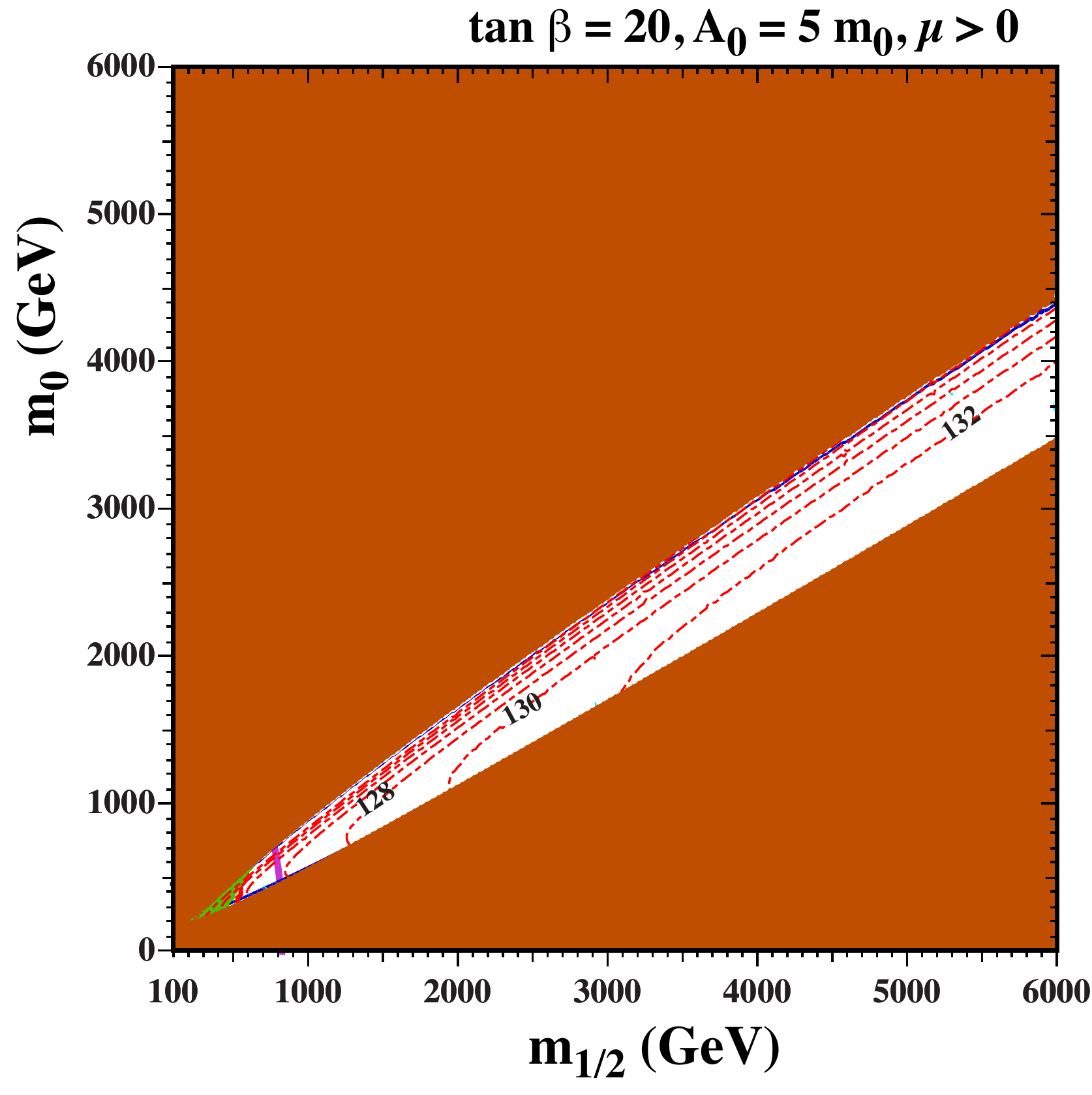} \\
\end{tabular}
\end{center}   
\caption{\label{fig:m12m0planes}\it
The allowed regions in the $(m_{1/2}, m_0)$ planes for $\tb = 20$ and $A_0 = 2.2 \, m_0$
(upper left), $2.5 \, m_0$ (upper right), $3.0 \, m_0$
(lower left) and $5.0 \, m_0$ (lower right). The line styles and shadings are
described in the text. The {\tt FeynHiggs~2.10.0} code is used to calculate contours of $m_h$
that are separated by 2~GeV: the uncertainty in $m_h$ is typically
 $\pm 3$~GeV. Stop coannihilation strips run close to the boundaries
of the brown shaded regions in the upper left corners of all the panels.
In the lower left corners of all the panels there are (green) shaded regions excluded by $b \to s \gamma$,
(green) 95\% exclusion contours from $B_s \to \mu^+ \mu^-$ and (purple) 95\% exclusion
contours from searches for $\ETslash$ events at the LHC.
}
\end{figure}

\begin{figure}
\vspace{-4cm}
\begin{center}
\begin{tabular}{c c}
\includegraphics[height=8cm]{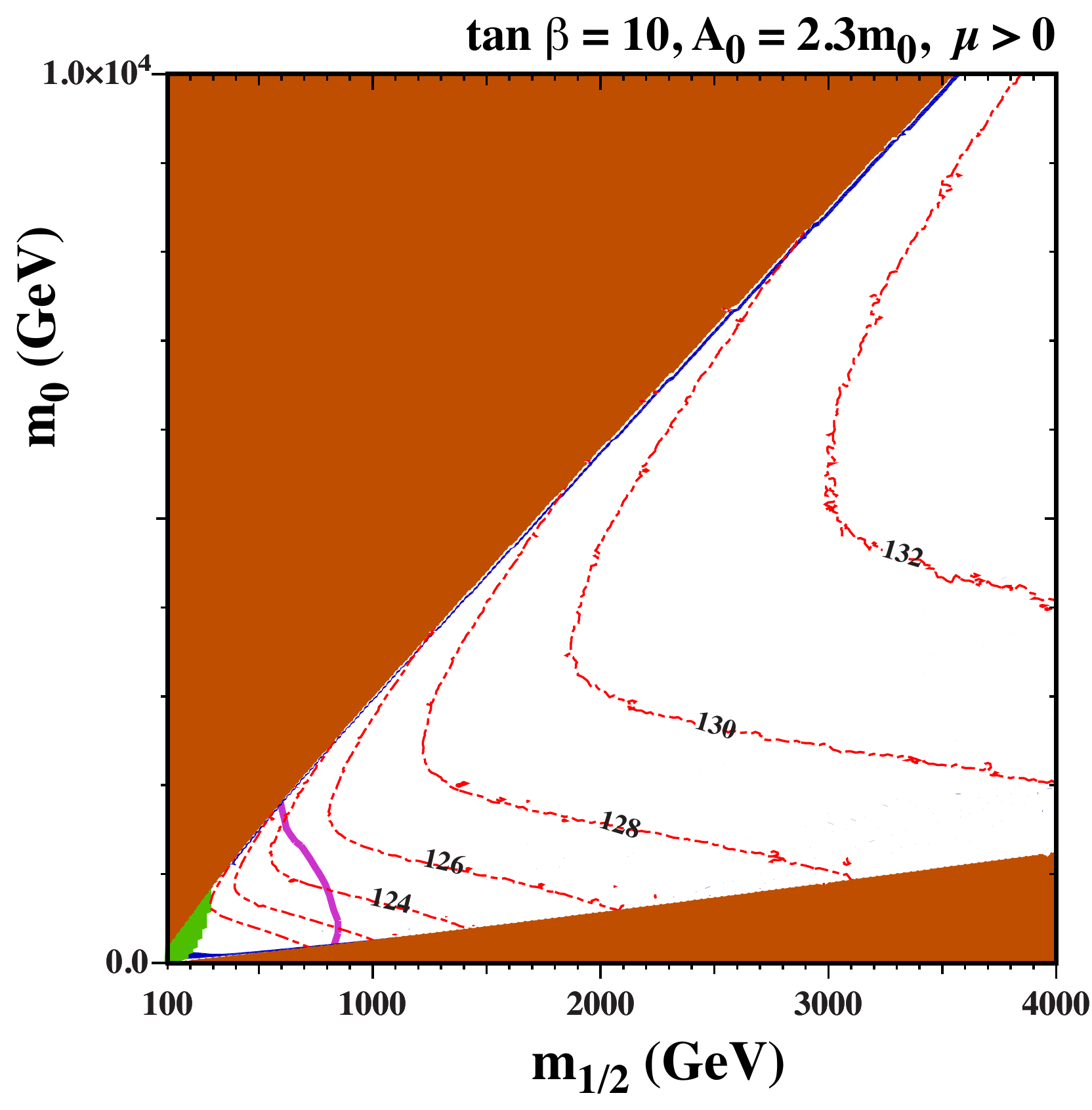} & 
%\hspace{-2cm}
\includegraphics[height=8cm]{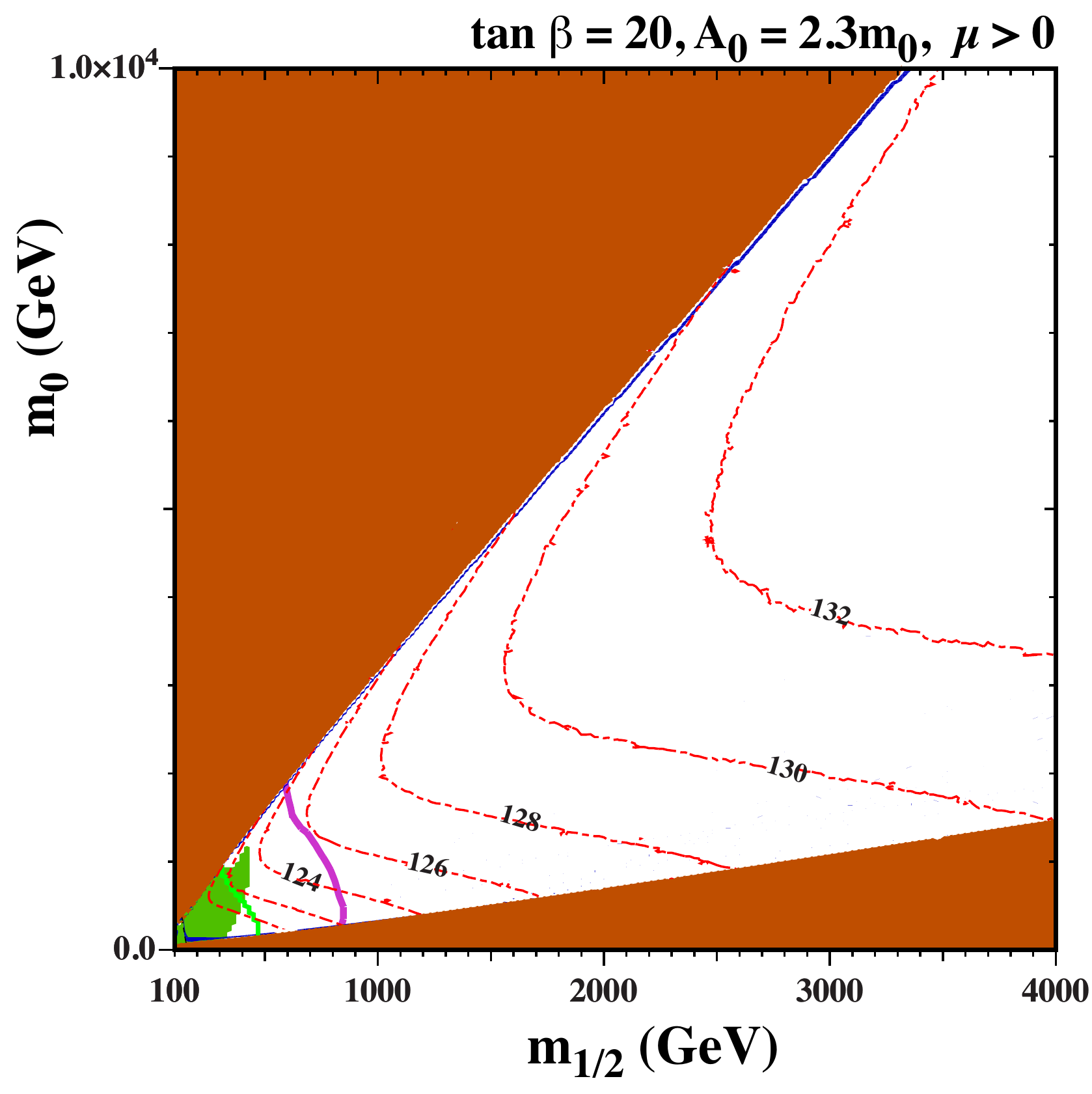} \\
\end{tabular}
\end{center}   
%\vspace{-5cm}
\begin{center}
\begin{tabular}{c c}
\includegraphics[height=8cm]{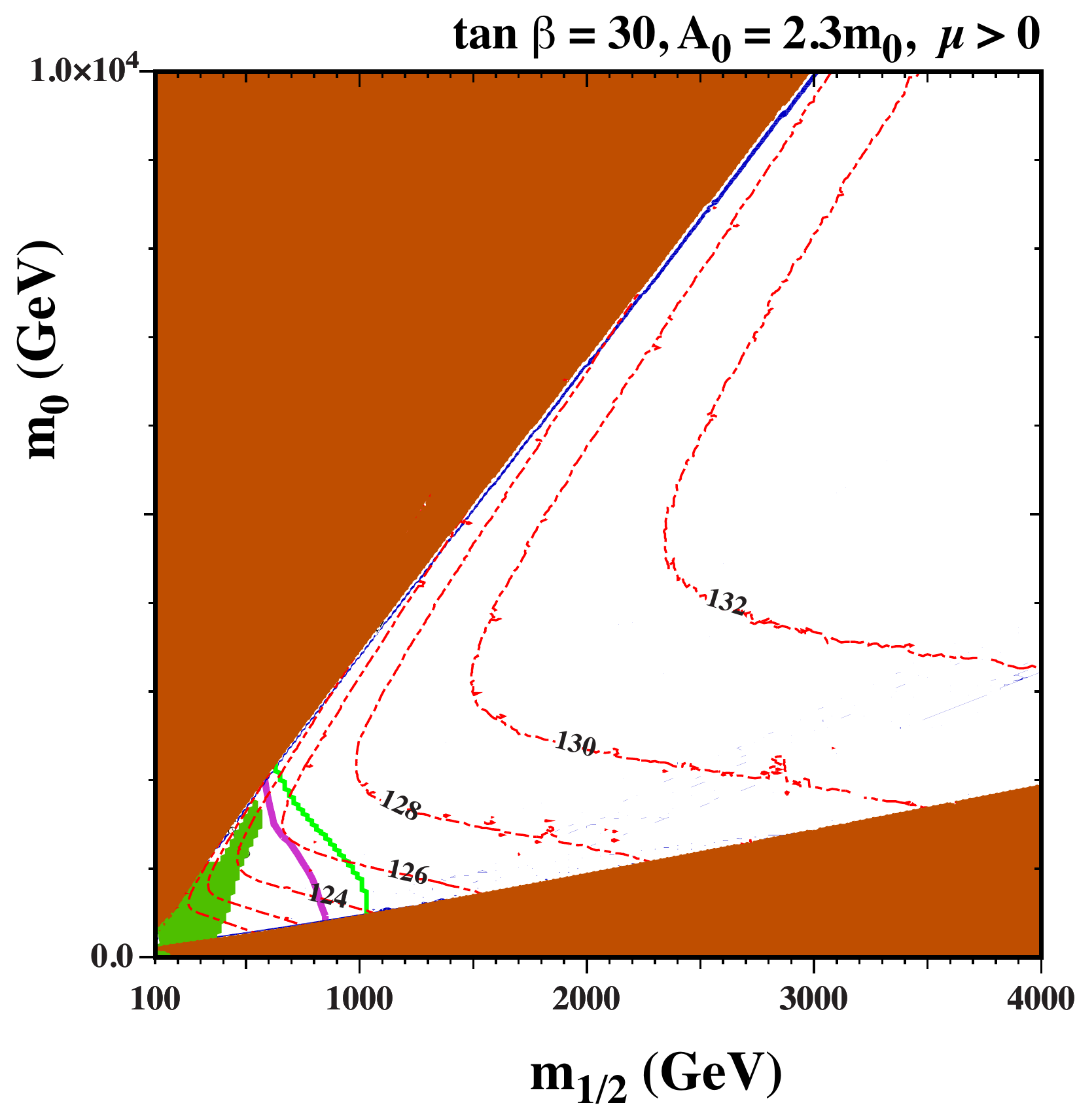} & 
%\hspace{-2cm}
\includegraphics[height=8cm]{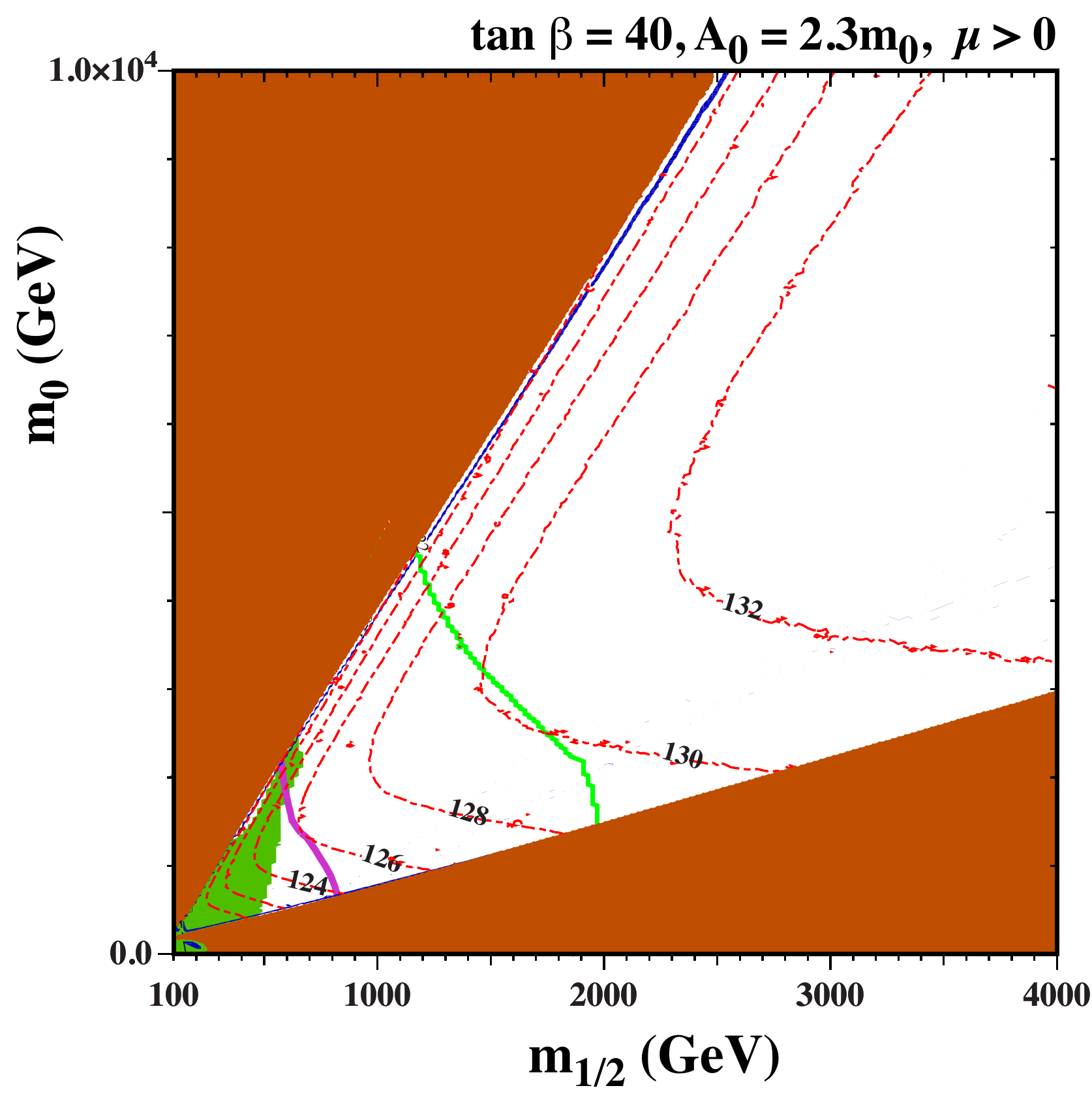} \\
\end{tabular}
\end{center}   
\caption{\label{fig:m12m0planestb}\it
As Fig.~\protect\ref{fig:m12m0planes}, displaying
the allowed regions in the $(m_{1/2}, m_0)$ planes for $A_0 = 2.3 \, m_0$ and $\tb = 10$
(upper left), $\tan \beta = 20$ (upper right), $\tan \beta = 30$
(lower left) and $\tan \beta = 40$ (lower right). The line styles and shadings are
described in the text.  
}
\end{figure}

We display in Fig.~\ref{fig:m12m0planes} some representative CMSSM $(m_{1/2}, m_0)$ planes
for fixed $\tan \beta = 20$, $\mu > 0$ and different values of $A_0/m_0$ that illustrate
the interplay of the various theoretical, phenomenological, experimental and cosmological constraints.
In each panel, any region that does not have a neutral, weakly-interacting LSP is shaded brown. 
Typically there are two such regions which appear as triangular wedges. The wedge in the upper left 
of the $(m_{1/2}, m_0)$ plane contains a stop LSP or tachyonic stop, and the wedge in the lower
right of the plane contains a stau LSP or tachyonic stau.
The dark blue strips running near the boundaries of these regions have a relic LSP density
within the range of the cold dark matter density indicated by astrophysics and 
cosmology~\cite{planck}~\footnote{The widths of these dark matter strips have been enhanced for visibility.
Barely visible in the lower parts of the unshaded wedges between the strips in some panels of Figs.~\ref{fig:m12m0planes}
and \ref{fig:m12m0planestb} are a low density of
points where annihilations of other sparticles coannihilating with the neutralino are enhanced by direct-channel
Higgs poles, reducing $\Omega_\chi h^2$ into the allowed range.}: that near the
boundary of the upper left wedge is due to stop coannihilation, and that near the boundary of the lower
right wedge is due to stau coannhilation.
{\it As we discuss later, the stop coannihilation strips typically extend to much larger values of $m_{1/2}$ than
the stau coannhilation strips, indeed to much larger values of $m_{1/2}$
than those displayed in Fig.~\ref{fig:m12m0planes}, reaching as far as $7000~{\rm GeV} -13000$~GeV
in the models studied.}
The green shaded regions are incompatible with the experimental measurement of $b \to s \gamma$
decay \cite{HFAG}, and the green solid lines are 95 \% CL constraints from the measured rate of $B_s \to \mu^+ \mu^-$ decay \cite{bmm}. 
The solid purple lines show the constraint from the absence of $\ETslash$ events at the LHC at
8~TeV \cite{ATLASsusy}, and the red dot-dashed lines are contours of $m_h$ calculated using {\tt FeynHiggs~2.10.0}, which
have a typical uncertainty $\pm 3$~GeV for fixed input values of $m_{1/2}, m_0, \tan \beta$
and $A_0$ \cite{fh,fh2100}.

%%%%%%%%%%%%%% F I G U R E %%%%%%%%%%%%%%%%%%%%%%%%%%%%%%%%%%%%%%%%%%%%%%%%%%%
%%%%%%%%%%%%%% F I G U R E %%%%%%%%%%%%%%%%%%%%%%%%%%%%%%%%%%%%%%%%%%%%%%%%%%%

In general, we identify stop coannihilation strips in CMSSM $(m_{1/2}, m_0)$ planes for
$2.1 \, m_0 \lappeq A_0 \lappeq 5.5 \, m_0$, and the panels in Fig.~\ref{fig:m12m0planes}
have been chosen to represent the range of possibilities for $\tan \beta = 20$.
The angle subtended by the (brown) stop LSP wedge increases with $A_0/m_0$,
and this wedge meets the (brown) stau LSP wedge and closes the
intermediate (unshaded) neutralino LSP wedge for $A_0 \gappeq 5.5 \, m_0$~\footnote{For
$\tan \beta = 20$ and $A_0 = 5.5 \, m_0$ the neutralino LSP regions is reduced to
a very narrow slit extending from $(m_{1/2}, m_0) = (500, 400)$~GeV to $(4500, 3000)$~GeV.}.
Each of the panels of Fig.~\ref{fig:m12m0planes} also features a stau coannihilation strip
running close to the boundary of the stau LSP wedge,
which extends to $m_{1/2} \sim 1000$~GeV corresponding to $m_\chi \sim 400$~GeV.

Along these strips, the LHC $\ETslash$ constraint excludes $m_{1/2} < 800$~GeV,
but the excluded range of $m_{1/2}$ is reduced for the larger values of $m_0$ along
the stop coannihilation strip.
For the planes shown in Fig.~\ref{fig:m12m0planes}, the stop strip extends far beyond the range
of $m_{1/2}$ shown (see section \ref{pheno} below for more discussion about the endpoint of the stop
strips). However, depending on the ratio, $A_0/m_0$, the strip may conflicted with the measured
value of the Higgs mass.  For example, for $A_0/m_0 = 2.2$, the strip crosses $m_h = 128$ GeV
at $m_{1/2} \simeq 1100$ GeV. As $A_0/m_0$ is increased, the Higgs mass rapidly 
decreases along the strip. When $A_0/m_0 = 2.5$, the strip crosses $m_h = 128$ GeV
at $m_{1/2} \simeq 2600$ GeV and $m_{1/2} \gappeq 1100$ GeV for $m_h > 124$ GeV. For 
$A_0/m_0 = 3.0$,  $m_{1/2} \gappeq 2200$ GeV for $m_h > 124$ GeV and the strip is
allowed to extend to much higher $m_{1/2}$ than shown in the Figure. For $A_0/m_0 = 5.0$, 
only the far end of the strip at large $m_{1/2} \gappeq 4$ TeV is allowed. 
We return later to the impact of the LHC constraint on $m_h$ and other phenomenological
constraints on the stop coannihilation strip.

Fig.~\ref{fig:m12m0planestb} displays the sensitivity of the stop coannihilation strip
to the choice of $\tan \beta$ for the representative choice $A_0 = 2.3 \, m_0$.
Here we see that the opening angle of the stop LSP wedge is rather insensitive
to $\tan \beta$, that of the stau coannihilation strip being more sensitive. Also, we recall
that studies indicate that the LHC $\ETslash$ constraint is essentially independent of $\tan \beta$.
On the other hand, the impacts of the $b \to s \gamma$ and $B_s \to \mu^+ \mu^-$ constraints
increase with $\tan \beta$. They only ever exclude a fraction of the stop coannihilation strip,
but the $B_s \to \mu^+ \mu^-$ constraint does exclude the entire stau coannihilation strip for
$\tan \beta = 40$. The $m_h$ contours calculated  using {\tt FeynHiggs~2.10.0}
are quite similar for $\tan \beta = 10, 20$ and 30. However, we find smaller values of $m_h$
for $\tan \beta = 40$, a feature whose implications we discuss in more detail later.

%%%%%%%%%%%%%% F I G U R E %%%%%%%%%%%%%%%%%%%%%%%%%%%%%%%%%%%%%%%%%%%%%%%%%%%
%%%%%%%%%%%%%% F I G U R E %%%%%%%%%%%%%%%%%%%%%%%%%%%%%%%%%%%%%%%%%%%%%%%%%%%

\subsection{$(\tan \beta, A_0)$ Planes}

\begin{figure}[!htbp]
\centering
\includegraphics[width=0.6\textwidth]{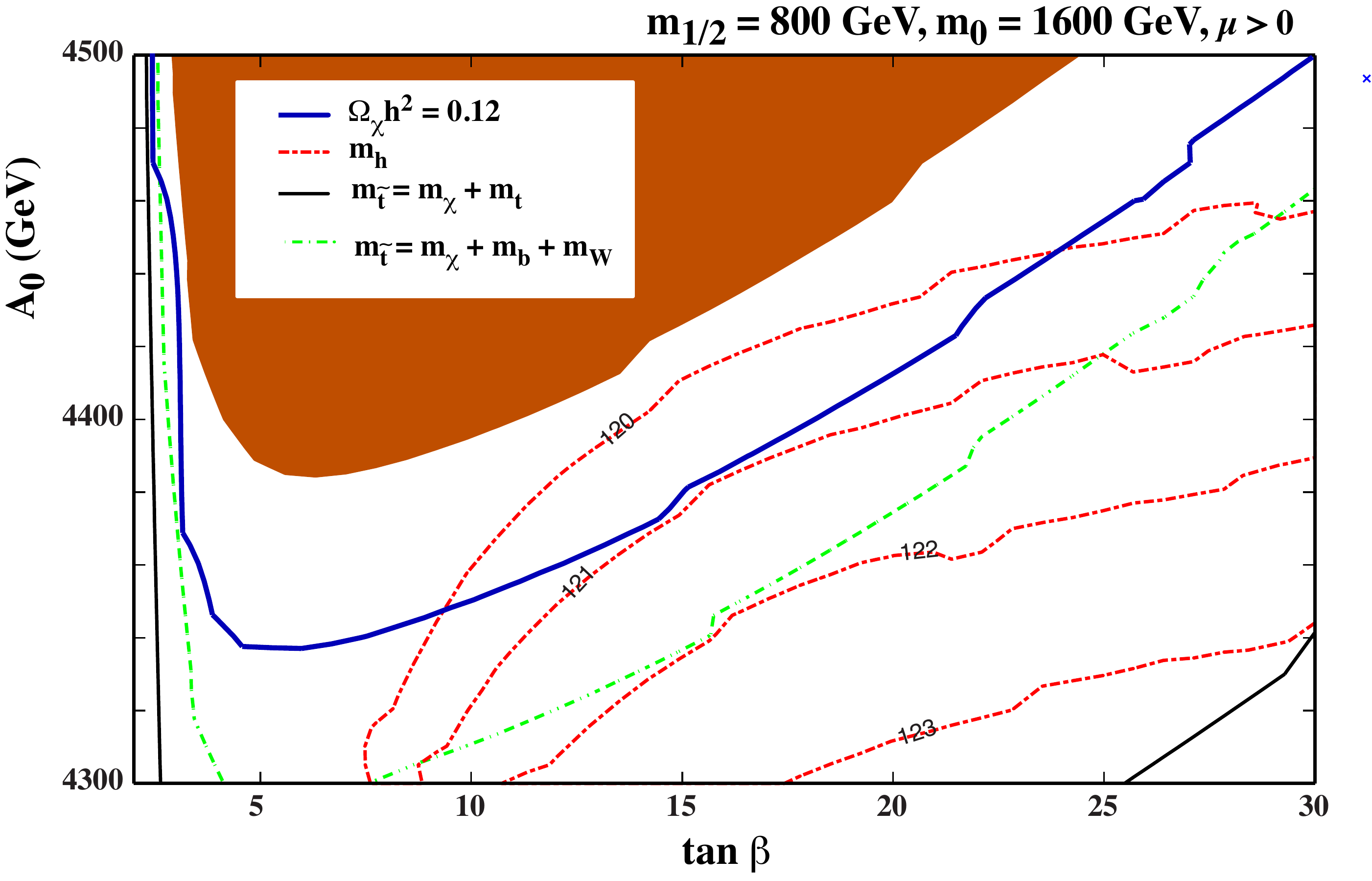}%
\\
\includegraphics[width=0.6\textwidth]{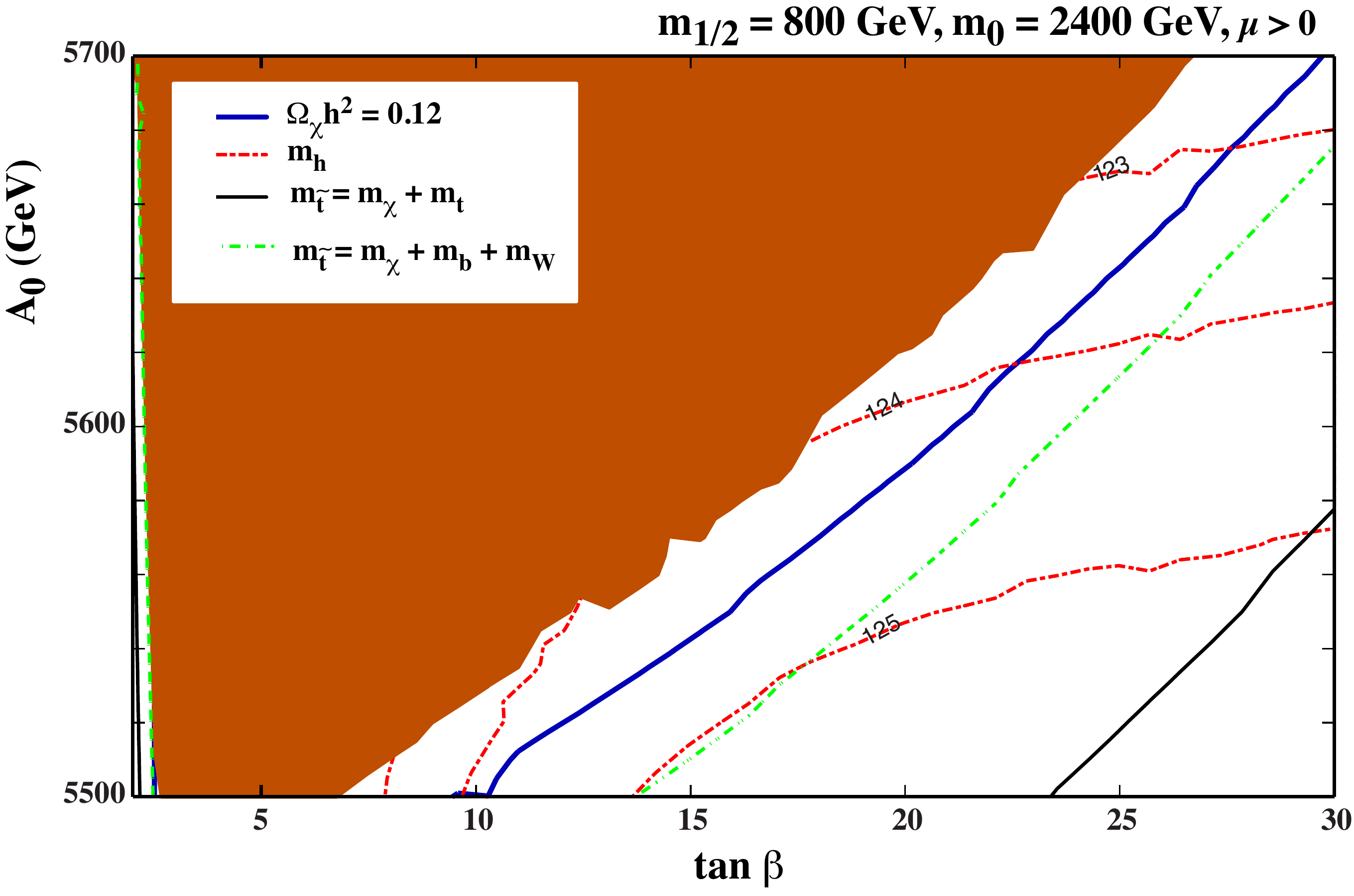}%
\\
\includegraphics[width=0.6\textwidth]{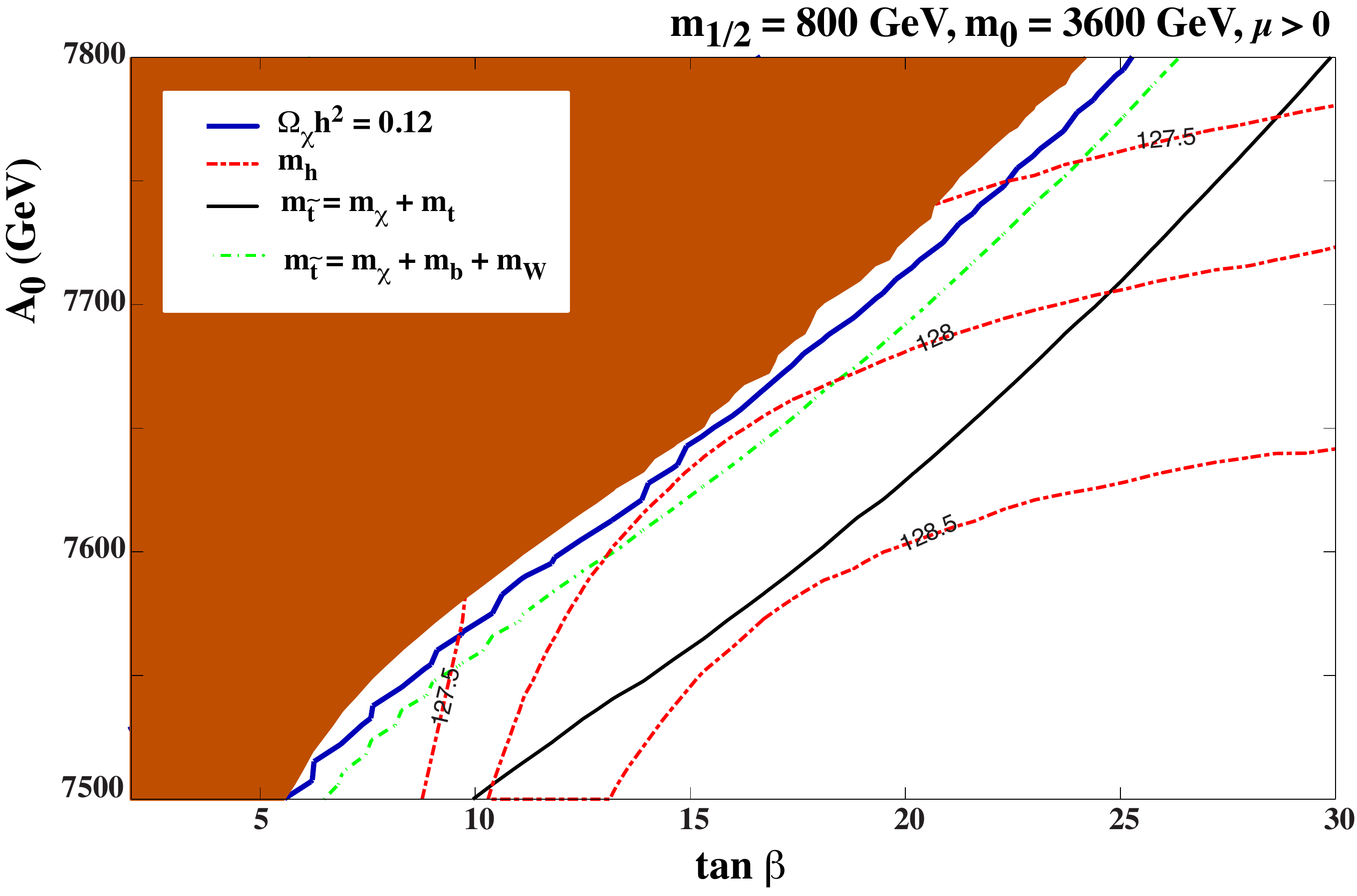}
\caption{\it The CMSSM $(\tan\beta, A_0)$ planes for $(m_{1/2}, m_0) = (800, 1600/2400/3600)$~GeV
in the top/middle/bottom panels, respectively. The (brown) shaded is excluded because $m_{\tilde t_1} < m_\chi$. Also
shown are the contours $m_{\tilde t_1} = m_\chi + m_b + m_W$ (green, dash-dotted) 
and $m_{\tilde t_1} = m_\chi + m_t$ (black, solid). The solid blue line is the strip where $\Omega_\chi h^2 = 0.12$
and the red dash-dotted lines are contours of $m_h$ calculated with {\tt FeynHiggs~2.10.0}.}
\label{fig:tb_A_scan1}
\end{figure}

In view of the dependences of the stop coannihilation strips on the values of
$\tan \beta$ and $A_0$, we display in Fig.~\ref{fig:tb_A_scan1} 
examples of $(\tan \beta, A_0)$ planes in the CMSSM for fixed $m_{1/2}$ and $m_0$.
In the (brown) shaded region at the top of each panel, the ${\tilde t_1}$ is lighter than the $\chi$, so there
is no weakly-interacting neutral dark matter. Running below this boundary, the solid (blue) line
is the contour where $\Omega_\chi h^2 = 0.12$. The other roughly parallel contours are
$m_{\tilde t_1} = m_\chi + m_b + m_W$ (green, dash-dotted) and
$m_{\tilde t_1} = m_\chi + m_t$ (black, solid). Finally, the red dash-dotted lines are contours of $m_h$
calculated using {\tt FeynHiggs~2.10.0}. In each panel, we see that the calculated value of $m_h$ increases with
increasing $\tan \beta$ and decreases with increasing $A_0$, and comparing the panels for
$m_0 = 1600$~GeV (top), 2400~GeV (middle) and 3600~GeV (bottom) we see that $m_h$
also increases with $m_0$.

\begin{figure}[!htbp]
\centering
\includegraphics[width=0.6\textwidth]{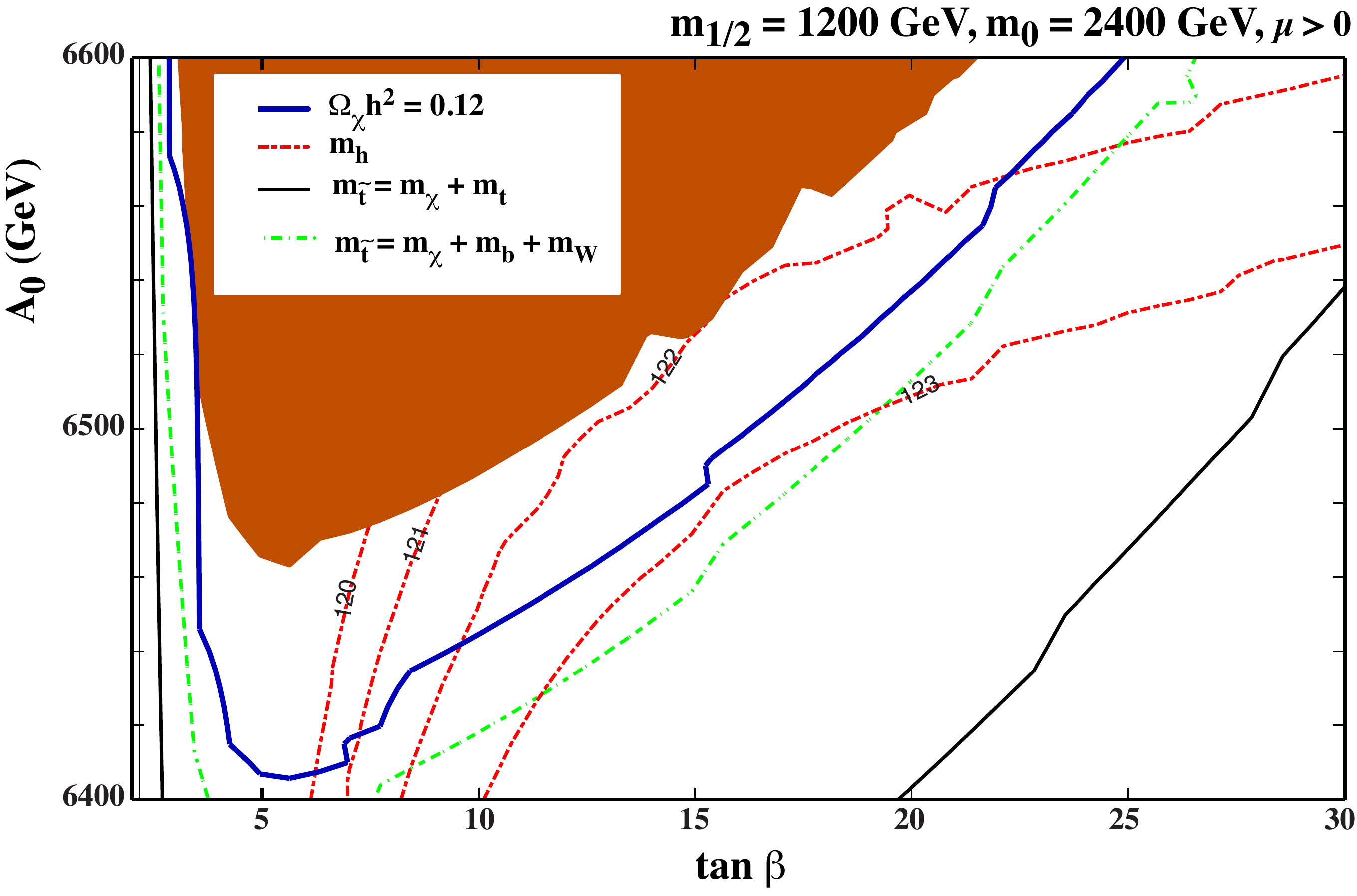}%
\\
\includegraphics[width=0.6\textwidth]{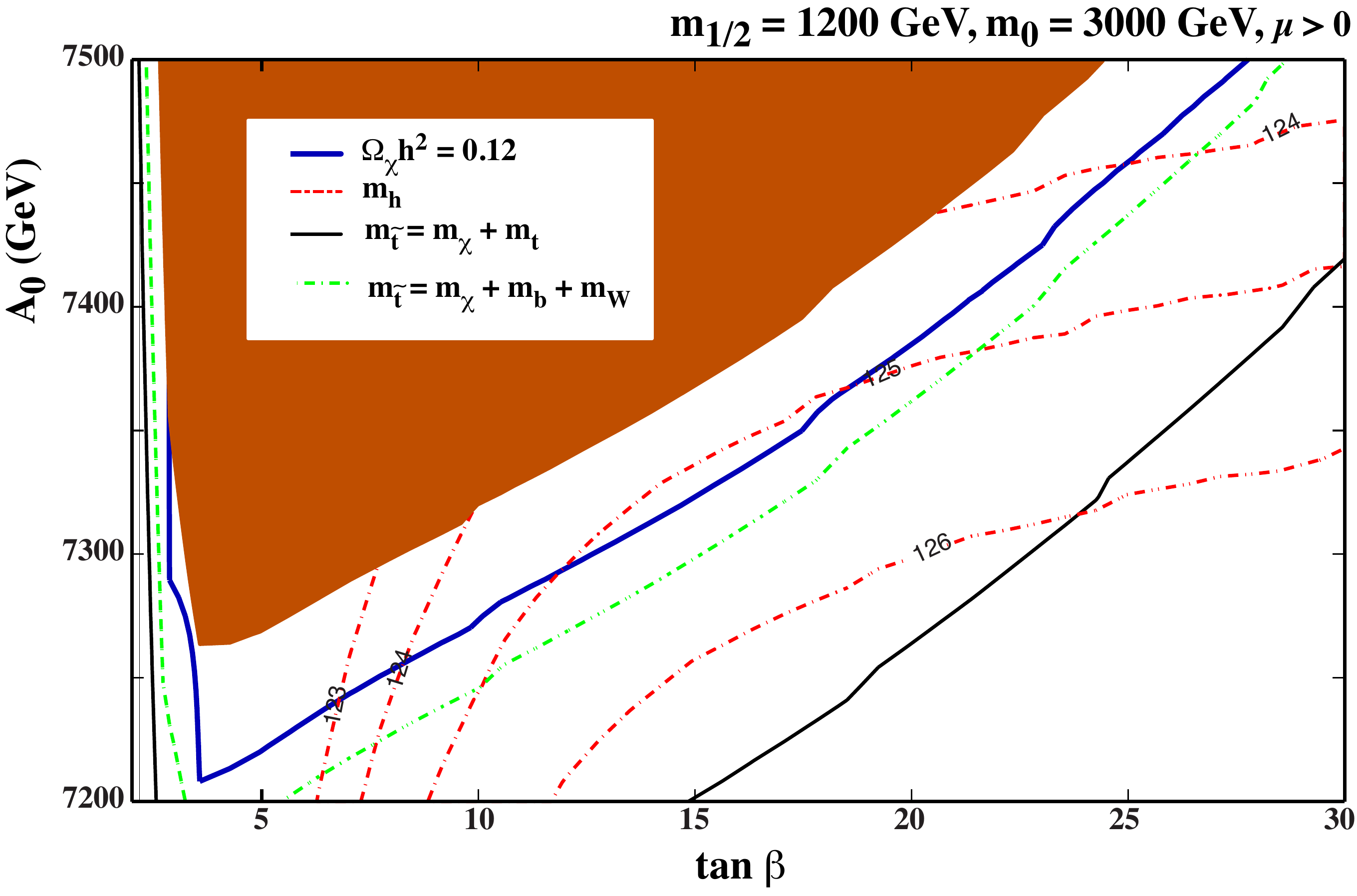}%
\\
\includegraphics[width=0.6\textwidth]{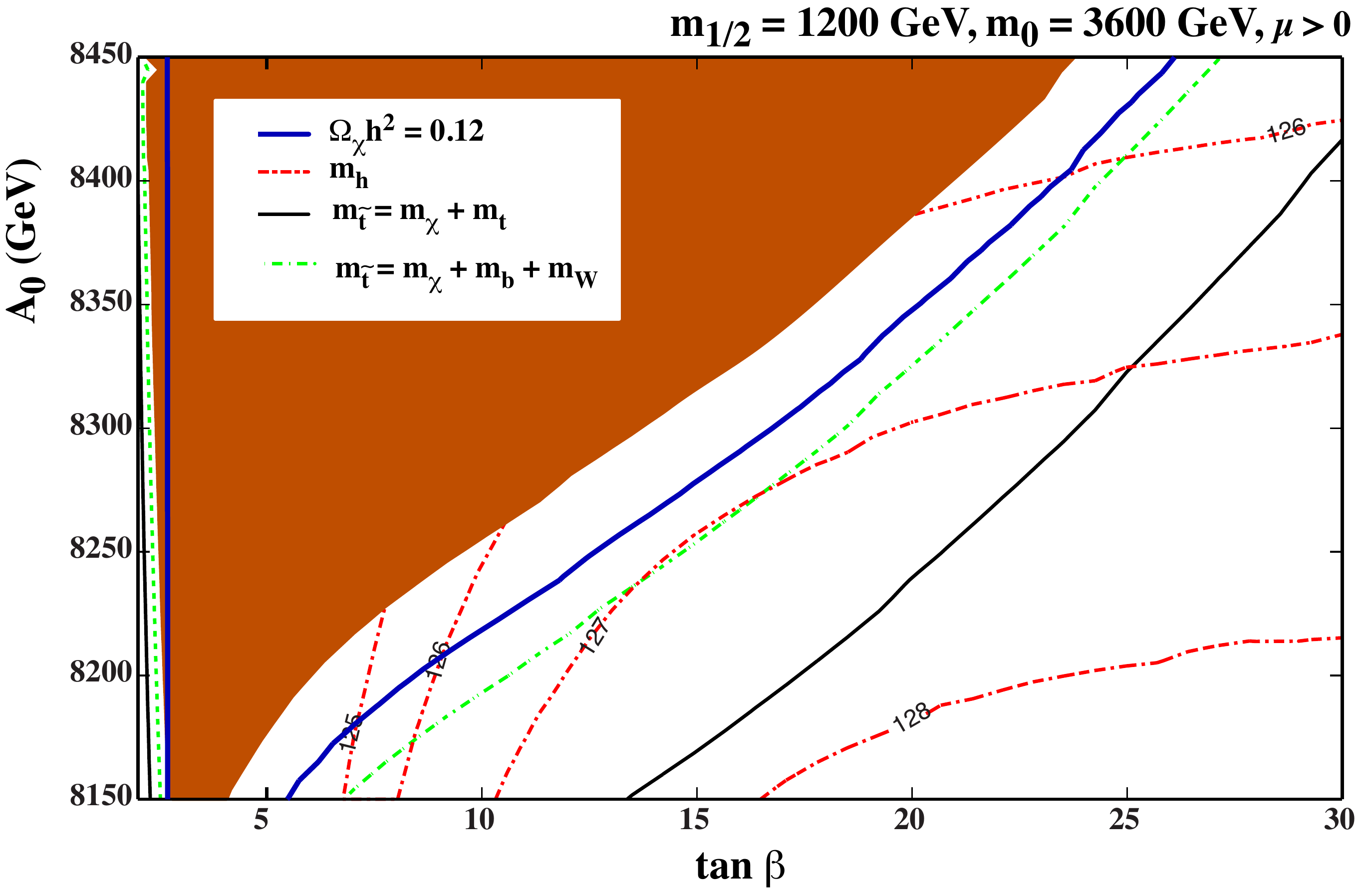}
\caption{\it As in Fig.~\protect\ref{fig:tb_A_scan1}, but for fixed
$(m_{1/2}, m_0) = (1200, 2400/3000/3600)$ in the top/middle/bottom panels.}
\label{fig:tb_A_scan2}
\end{figure}

We see in the top panel of Fig.~\ref{fig:tb_A_scan1} for the
combination $(m_{1/2}, m_0) = (800, 1600)$~GeV that $m_h < 121$~GeV along all the
$\Omega_\chi h^2 = 0.12$ contour, so the LHC Higgs mass measurement rules out this
combination of $m_{1/2}$ and $m_0$ for any value of $\tan \beta$ and $A_0$. %OKJE
On the other hand, we see in the middle panel for $(m_{1/2}, m_0) = (800, 2400)$~GeV
that $m_h > 122.5$~GeV (and hence is compatible with the measured value of $m_h$ after allowing for the theoretical
uncertainty $\sim 3$~GeV in the {\tt FeynHiggs~2.10.0} calculation) along all the displayed portion of the dark matter contour
extending from $(\tan \beta, A_0) = (10, 5500~{\rm GeV})$ to $(28, 5700~{\rm GeV})$,
corresponding to $A_0/m_0 \sim 2.4$. Finally, in the bottom panel of Fig.~\ref{fig:tb_A_scan1}
we see that along all the displayed portion of the dark matter contour extending from
$(\tan \beta, A_0) = (6, 7500~{\rm GeV})$ to $(25, 7800~{\rm GeV})$
corresponding to $A_0/m_0 \sim 2.2$ we have
$127~{\rm GeV} < m_h < 128$~GeV, which is also compatible with the experimental 
measurement within the estimated theoretical uncertainties~\footnote{We note that the ATLAS search
for jets + $\ETslash$ events, the measurement by CMS and LHCb of $B_s \to \mu^+ \mu^-$ decay
and the experimental constraint on $b \to s \gamma$ do not constrain any of the strip regions
shown in Figs.~\ref{fig:tb_A_scan1}, \ref{fig:tb_A_scan2}, \ref{fig:mhalf_A_scan} and \ref{fig:m0_A_scan}.}.

Fig.~\ref{fig:tb_A_scan2} displays analogous $(\tan\beta, A_0)$ planes for $(m_{1/2}, m_0) = (1200, 2400/3000/3600)$~GeV
in the top/middle/bottom panels, respectively. We see in the top panel that $m_h$ is compatible
with the experimental value within the estimated theoretical uncertainty of $\sim 3$~GeV only for $\tan \beta \sim 15$ where
{\tt FeynHiggs~2.10.0} yields a nominal value $m_h \simeq 122.5$~GeV. On the other hand,
we see in the middle panel, where $m_0$ is increased to 3000~GeV, that LHC-compatible
values of $m_h$ are found for all values of $\tan \beta \in (5, 27)$, and the same holds true in
the bottom panel where $m_0 = 3600$~GeV. Value of $A_0/m_0$ in the displayed regions of the stop
coannihilation strips range from $\sim 2.3$ to $\sim 2.7$.

\subsection{$(m_{1/2}, A_0)$ Planes}

\begin{figure}[!htbp]
\centering
\includegraphics[width=0.6\textwidth]{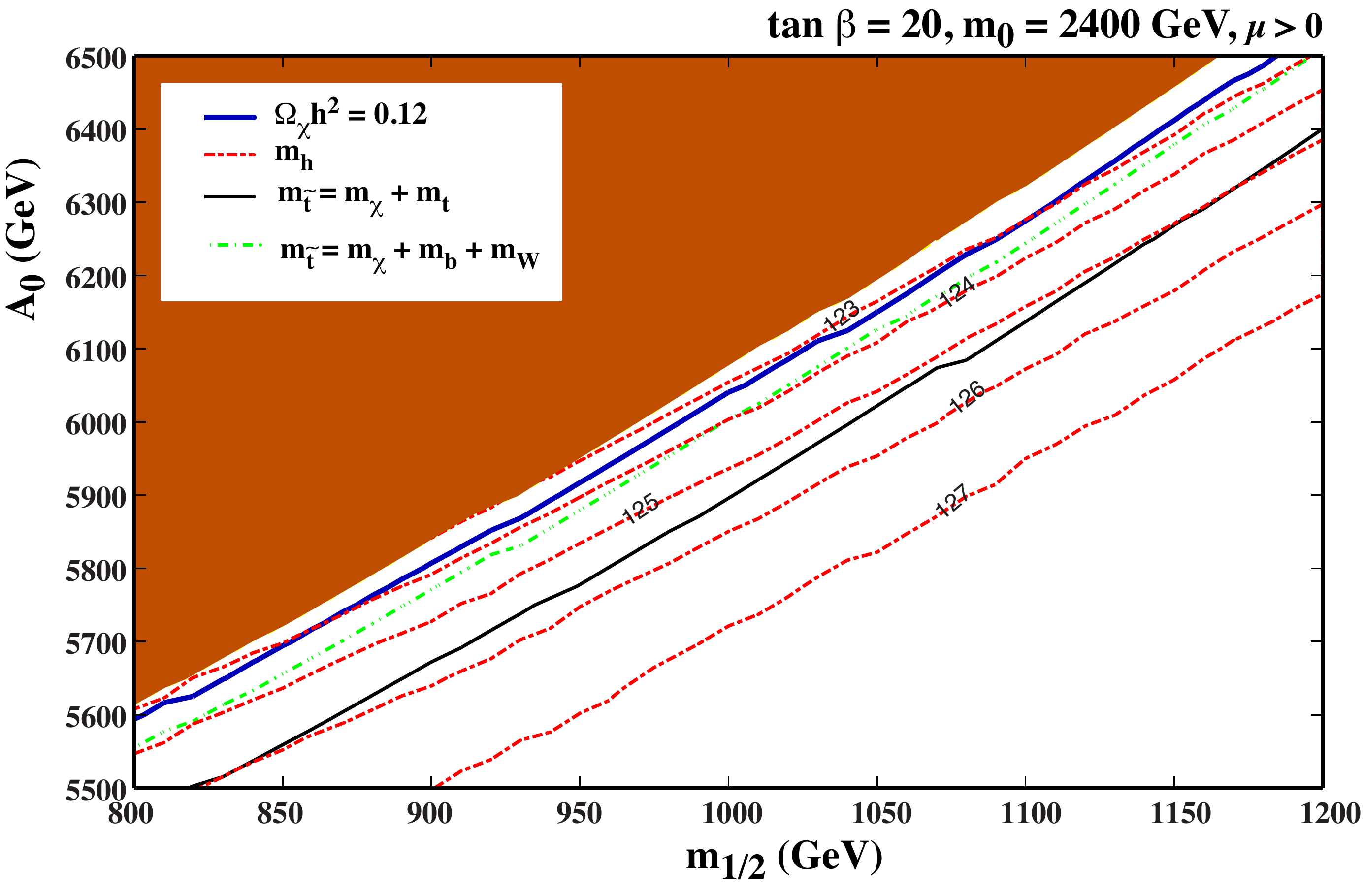}%
\\
\includegraphics[width=0.6\textwidth]{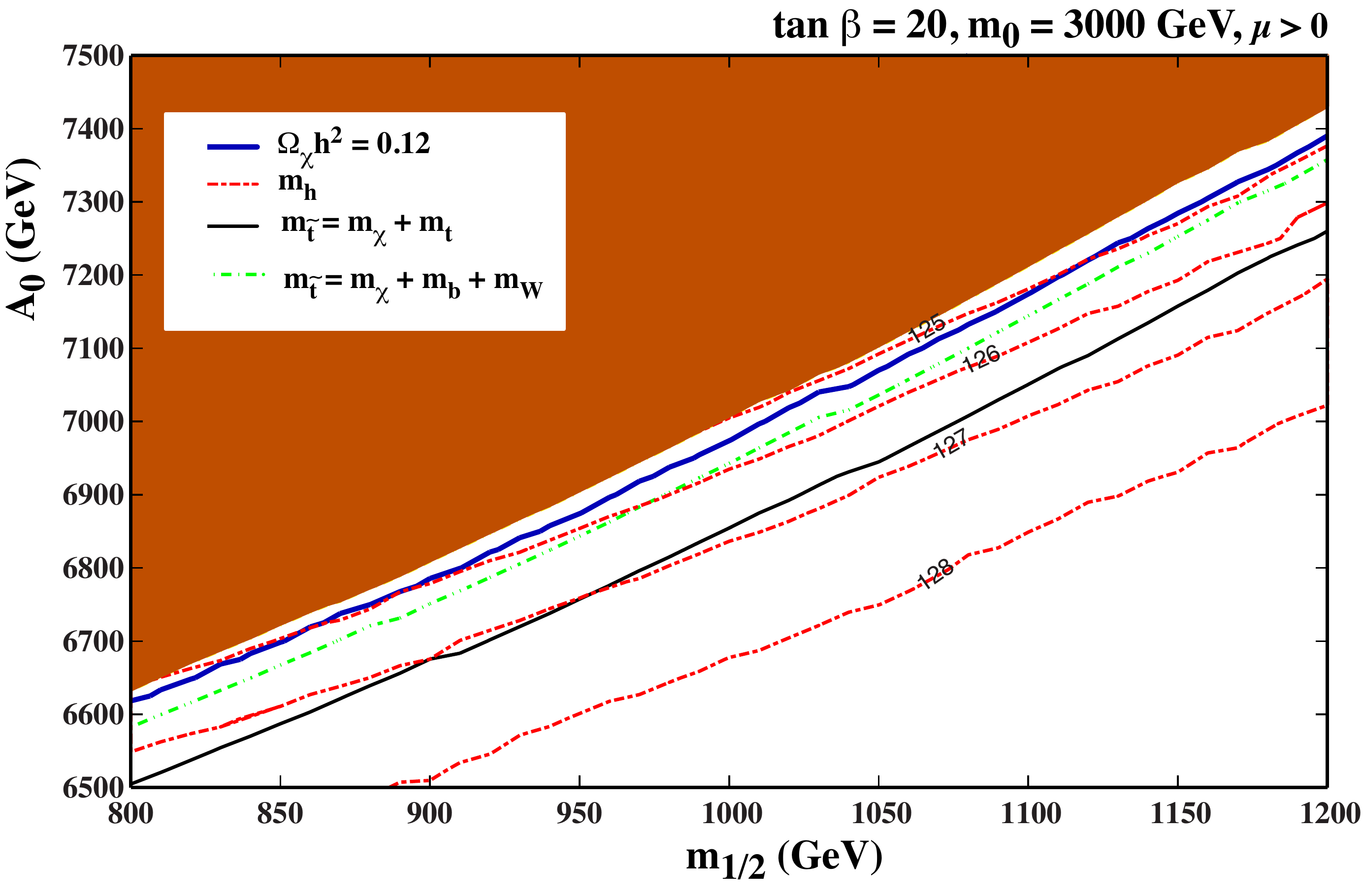}%
\\
\includegraphics[width=0.6\textwidth]{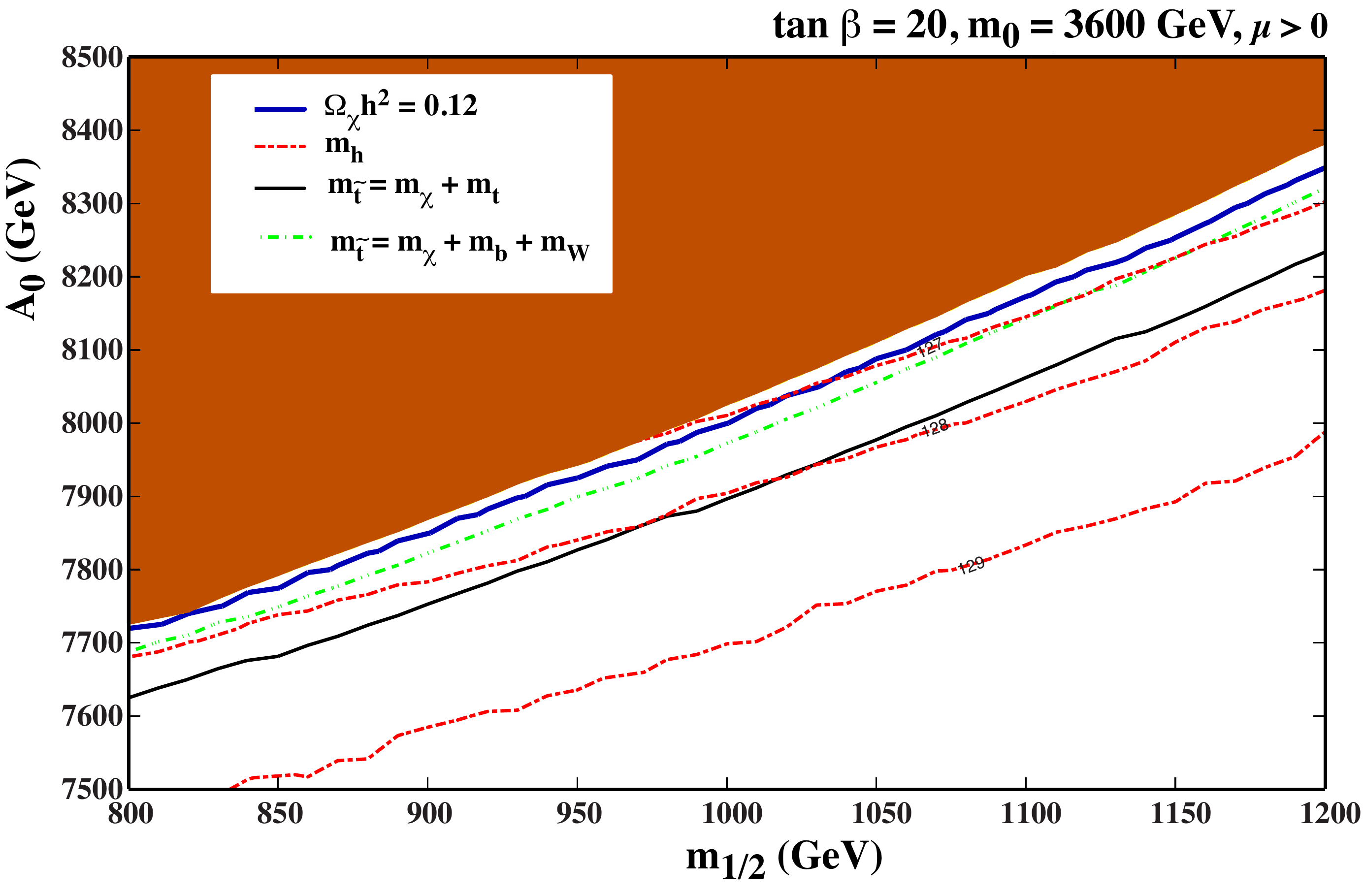}
\caption{\it Using the same line styles as in Fig.~\protect\ref{fig:tb_A_scan1}, $(m_{1/2}, A_0)$ planes for fixed
$(\tan \beta, m_0) = (15, 2400/3000/3600)$~GeV in the top/middle/bottom
panels.}
\label{fig:mhalf_A_scan}
\end{figure}

Fig.~\ref{fig:mhalf_A_scan} displays some $(m_{1/2}, A_0)$ planes for fixed
$(\tan \beta, m_0) = (15, 2400/3000/3600)$~GeV in the top/middle/bottom
panels, showing the same mass and relic density contours as in the previous figures.
In each of the three panels, we see that $m_h$ decreases as we move along
the strip to higher $m_{1/2}$.  In the top panel, $m_h$ falls below 123 GeV at $m_{1/2} \sim 1100$ GeV
and lower values of $m_{1/2}$ are preferred. Since the relic density and Higgs mass contours
are nearly parallel, 
in each panel of the lower two panels, we find LHC-compatible values of $m_h$ along all the displayed portion of the relic density contour from $m_{1/2} \in (800, 1200)$~GeV.

\subsection{$(m_{0}, A_0)$ Planes}

\begin{figure}[!htbp]
\centering
\includegraphics[width=0.6\textwidth]{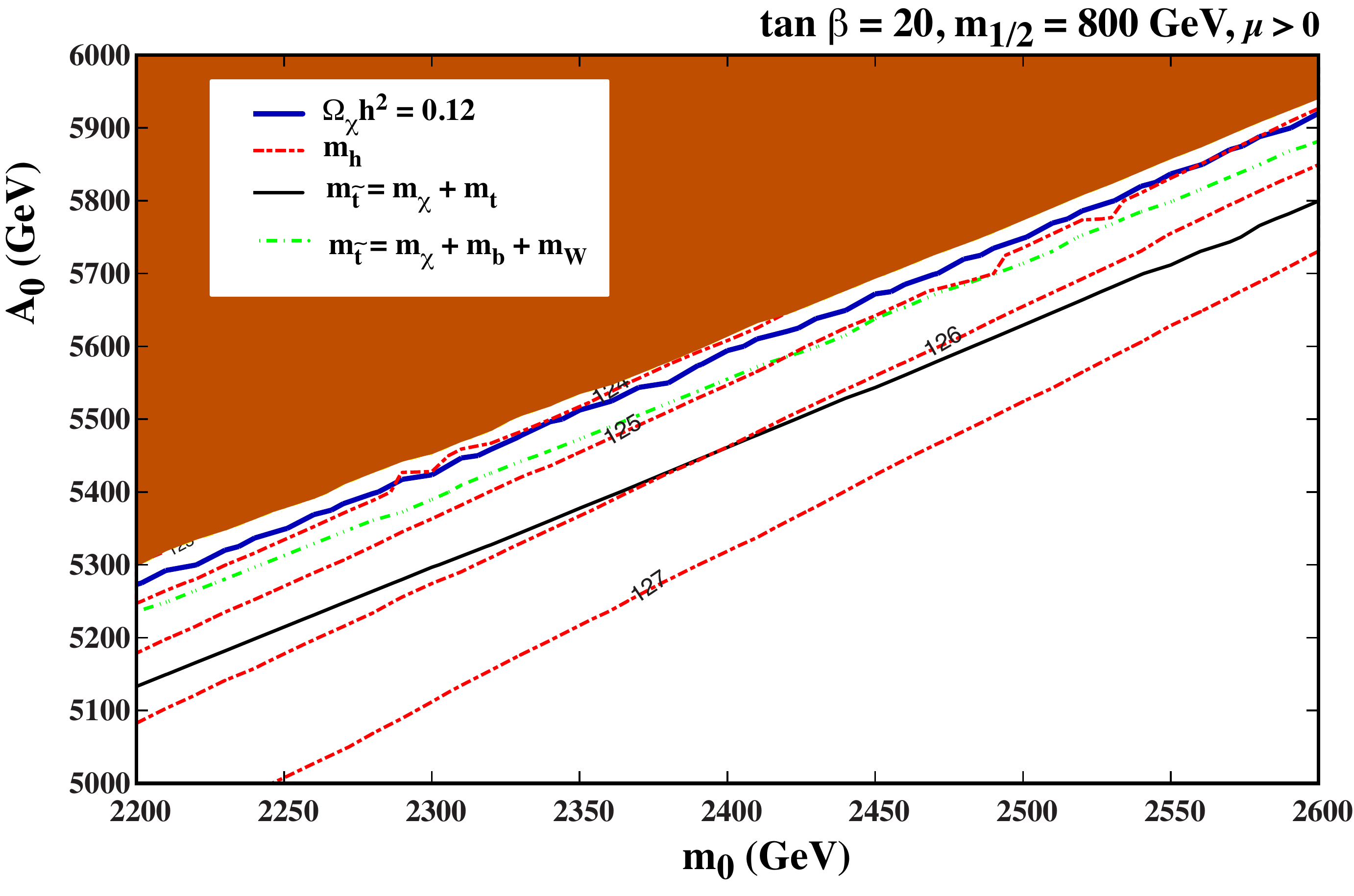}%
\\
\includegraphics[width=0.6\textwidth]{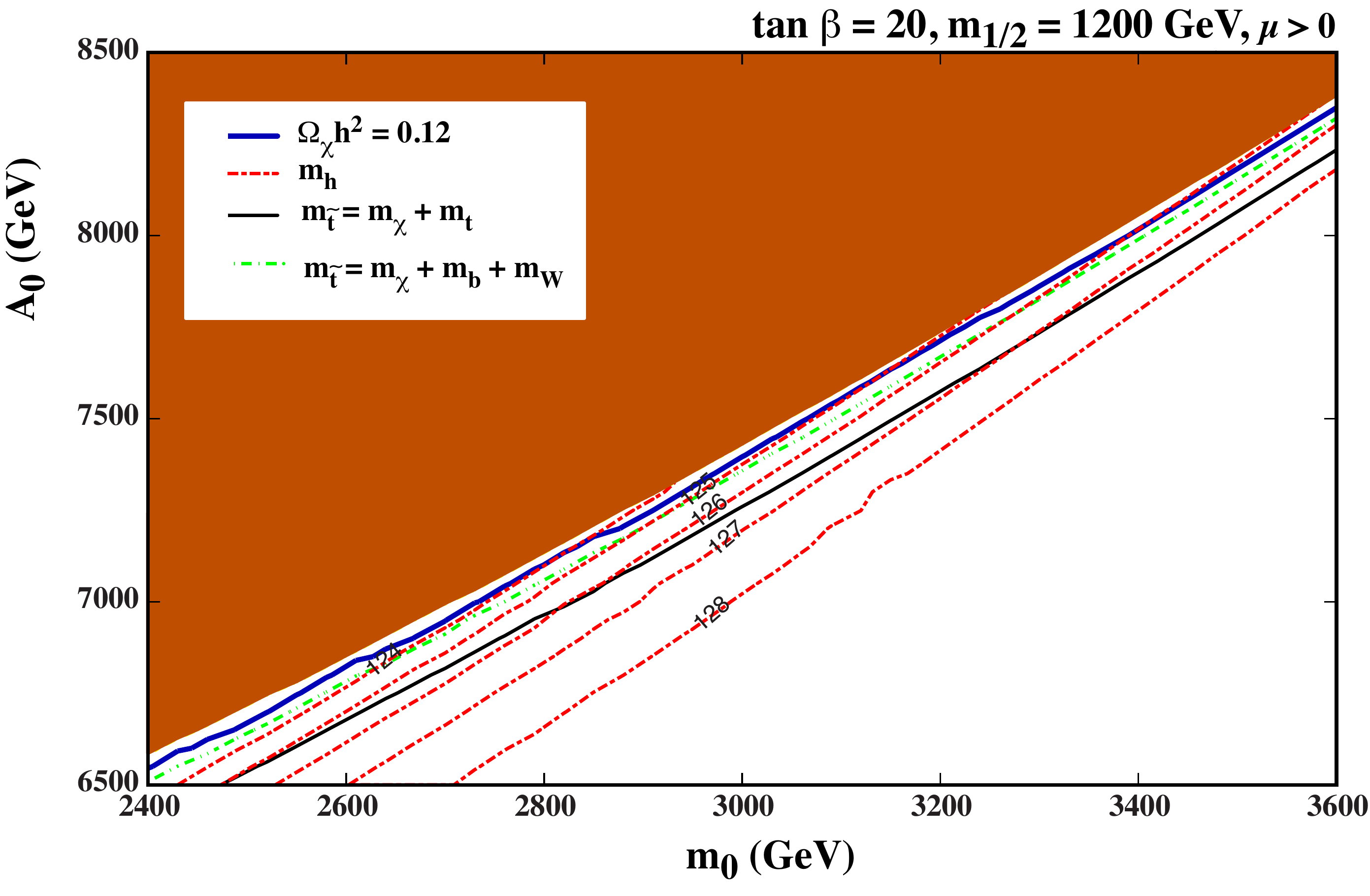}
\caption{\it Using the same line styles as in Fig.~\protect\ref{fig:tb_A_scan1}, $(m_{0}, A_0)$ planes for fixed
$(\tan \beta, m_0) = (15, 800/1200)$~GeV in the upper/lower panels.}
\label{fig:m0_A_scan}
\end{figure}

Fig.~\ref{fig:m0_A_scan} displays some $(m_{0}, A_0)$ planes for fixed
$(\tan \beta, m_{1/2}) = (15, 800/1200)$~GeV in the upper/lower
panels, showing the same mass and relic density contours as in the previous figures.
The relic density strip now tends to larger $m_h$ as $m_0$ is increased.
In the upper panel, we find LHC-compatible values of $m_h$ along all the displayed portion of
the relic density contour from $m_{0} \in (2200, 2600)$~GeV, and similarly in the lower
panel for $m_{0} \in (2400, 3600)$~GeV.

\section{Phenomenology along Stop Coannihilation Strips}
\label{pheno}

Having established the context for our study of stop coannihilation strips, 
we now consider in more detail phenomenological constraints and
possible experimental signatures along these strips. In general,
the value of $\delta m \equiv m_{\tilde t_1} - m_\chi$ plays an important r\^ole in this
phenomenology, falling to zero at the tip of the strip. Typical
values of $\delta m$ can be inferred from Figs.~\ref{fig:tb_A_scan1}, \ref{fig:tb_A_scan2}, \ref{fig:mhalf_A_scan}
and \ref{fig:m0_A_scan}, where we see that the $m_h$-compatible regions of the $\Omega_\chi h^2 = 0.12$
strip generally have $m_\chi + m_c < m_{\tilde t_1} < m_\chi + m_b + m_W$. However, we emphasize that
smaller values of $\delta m$ would be allowed if the neutralino LSP provided only a fraction of the
astrophysical cold dark matter.

\subsection{Strips for fixed $A_0/m_0$}

\begin{figure}[h!]
\centering
\includegraphics[width=0.49\textwidth]{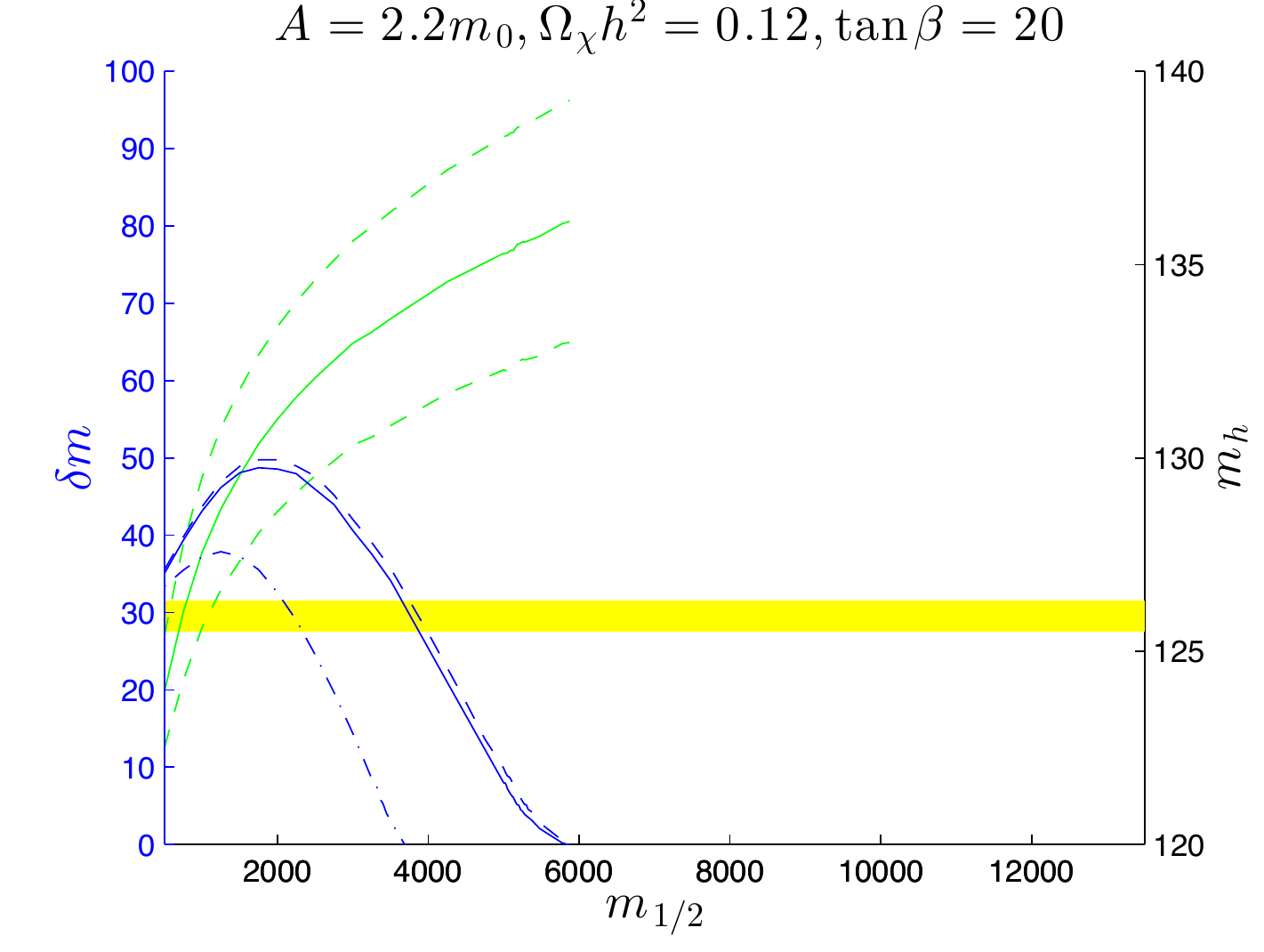}%
\hspace{0.004\textwidth}%
\includegraphics[width=0.49\textwidth]{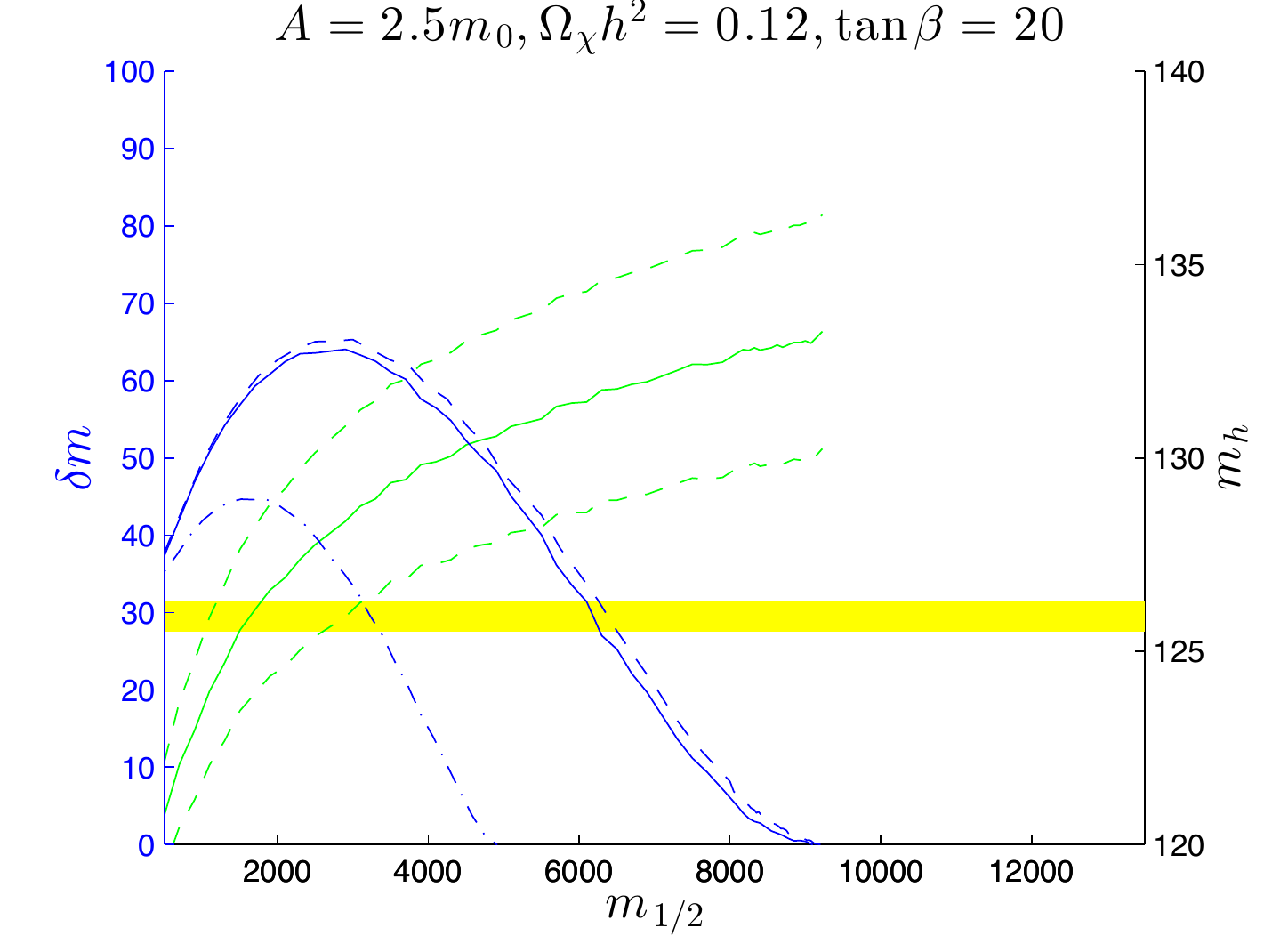}
\\
\includegraphics[width=0.49\textwidth]{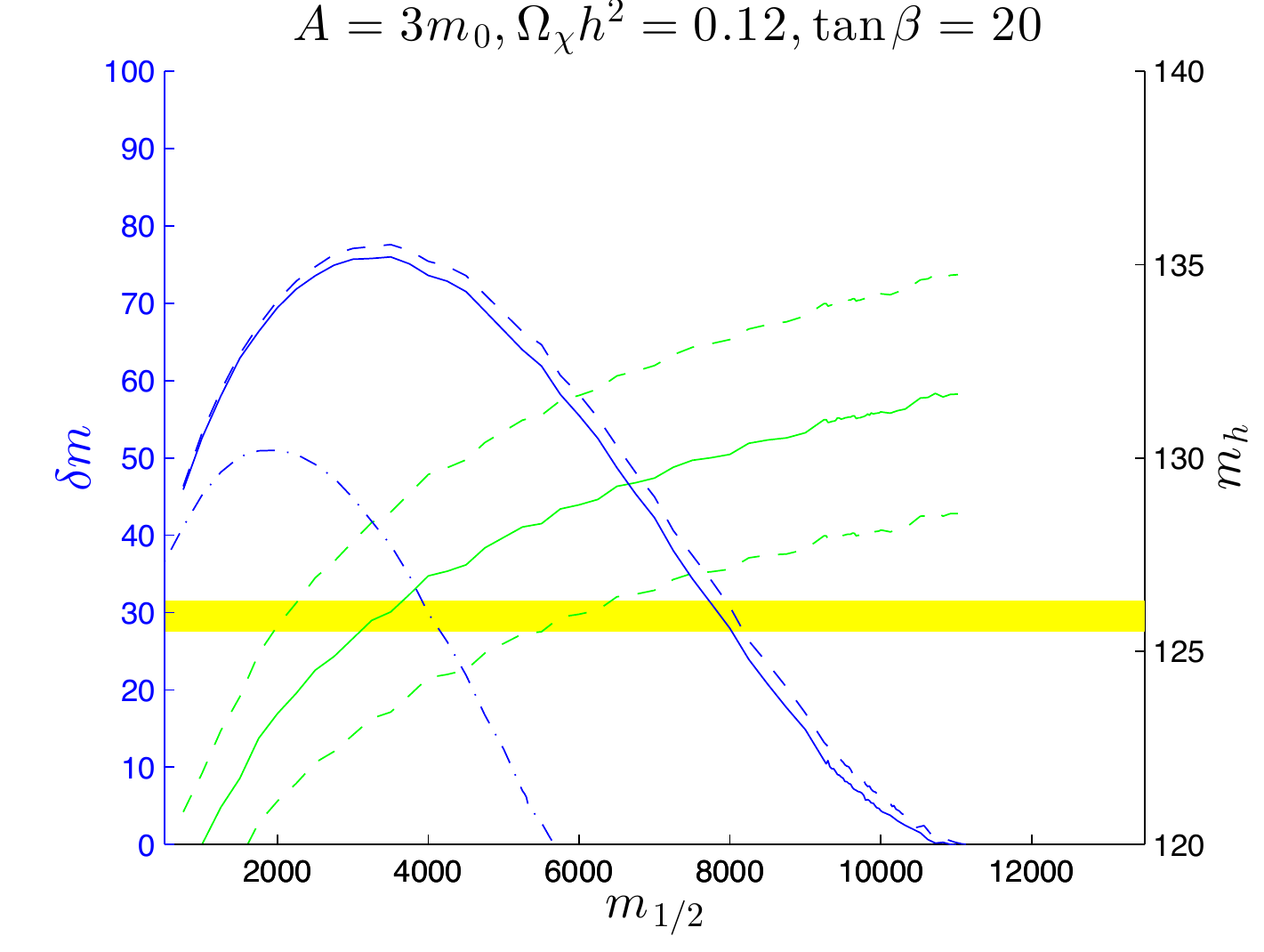}%
\hspace{0.004\textwidth}%
\includegraphics[width=0.49\textwidth]{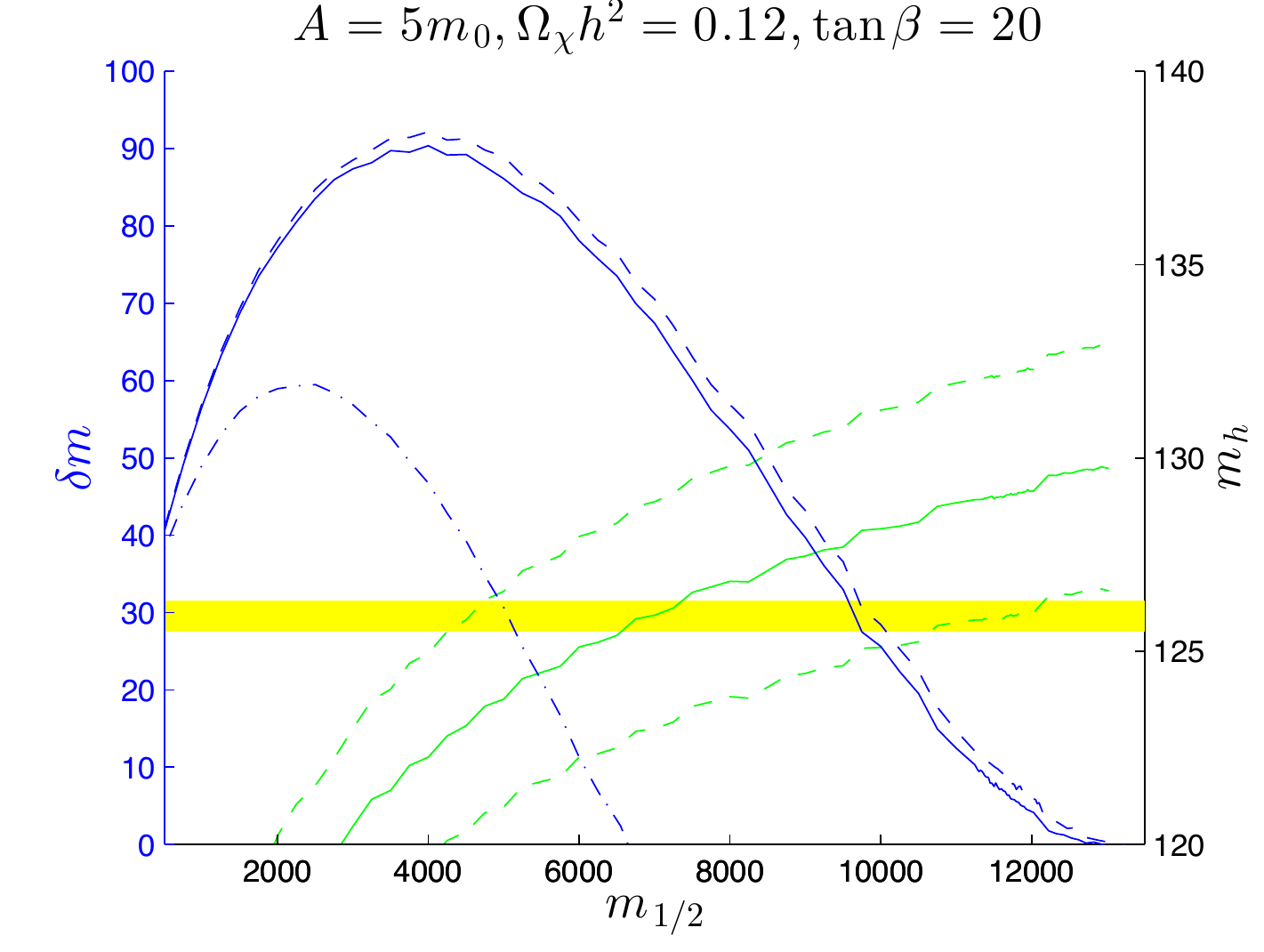}
\caption{\it The mass difference $\delta m=m_{\tilde t_1}-m_\chi$ and the Higgs mass $m_h$
(all masses in GeV units) as
functions of $m_{1/2}$ along the coannihilation strip where $\Omega_{\chi} h^2=0.12$, 
for $\tan\beta=20$ and $A=2.2 \, m_{0}, 2.5 \, m_0, 3.0 \, m_0$ and $5.0 \, m_0$. 
The solid blue lines show the values of $\delta m$
incorporating the Sommerfeld corrections. The dashed blue lines show $\delta m$ with 
$\Omega_{\chi} h^2=0.125$ and the dot dashed blues line show $\delta m$ without the Sommerfeld
correction. The green lines show the values of $m_h$, with
the dashed lines representing the uncertainty range given by {\tt FeynHiggs~2.10.0}.}
\label{fig:diff_scan_A}
\end{figure}

Fig.~\ref{fig:diff_scan_A} shows $\delta m = m_{\tilde t_1}-m_\chi$ and $m_h$ as
functions of $m_{1/2}$ along the coannihilation strip where $\Omega_{\chi}h^2=0.12$, 
for $\tan \beta = 20$ and $A_0 = 2.2 \, m_{0}, 2.5 \, m_0, 3.0 \, m_0$ and $5.0 \, m_0$.
The solid blue lines show the values of $\delta m$
incorporating the Sommerfeld corrections, and the lower dashed blue lines show the
values of $\delta m$ that would be required in the absence of the Sommerfeld corrections.
The inclusion of the Sommerfeld effects increases significantly $\delta m$ for generic
values of $m_{1/2}$, and also extends significantly the length of the stop coannihilation strip.
For $A_0 = 2.2 \, m_0$, we see that $\delta m$ rises to a maximum $\sim 50$~GeV at $m_{1/2} \sim 2000$~GeV,
before falling to zero at $m_{1/2} \sim 6000$~GeV, corresponding to $m_{\tilde t_1} = m_\chi
\sim 3000$~GeV. However, these values are not universal, with a maximal value of $\delta m > 60$~GeV
being attained at $m_{1/2} \sim 3000$~GeV for $A_0 = 2.5 \, m_0$ and the tip of the coannihilation
strip increasing to $\sim 9000$~GeV, corresponding to $m_{\tilde t_1} = m_\chi \sim 4600$~GeV.
These values increase further to $\delta m > 75 (90)$~GeV at $m_{1/2} = 3500 (4000)$~GeV with
the tip at $m_{1/2} = 11000 (13000)$~GeV for $A_0 = 3 (5) \, m_0$, corresponding to
$m_{\tilde t_1} = m_\chi \sim 5500 (6500)$~GeV
This non-universality reflects the model-dependence of the ${\tilde t_1} - {\tilde t_2} - h$ coupling
noted in (\ref{t1t2h}). The upper dashed blue lines in Fig.~\ref{fig:diff_scan_A}
show the values of $\delta m$ that would be required for
$\Omega_{\chi}h^2=0.125$, 2$\sigma$ above the central value for $\Omega_{\chi} h^2$.
We see that the astrophysical uncertainty in $\Omega_{\chi} h^2$ does not impact significantly the
length of the stop coannihilation strip.

The yellow bands in Fig.~\ref{fig:diff_scan_A} represent the current measurement of $m_h$,
with its experimental error, and the green lines show the values of $m_h$
calculated with {\tt FeynHiggs~2.10.0}, where
the dashed lines represent the estimated uncertainty range also determined using {\tt FeynHiggs~2.10.0}.
We note that only parts of the stop coannihilation strips are compatible with the LHC measurement of $m_h$, 
even after including the {\tt FeynHiggs~2.10.0}
uncertainty.  For $A_0/m_0 = 2.2$, we are restricted to $m_{1/2} \lappeq 1000$ GeV. The allowed range
jumps to $1000~{\rm GeV} \lappeq m_{1/2} \lappeq 3000$~GeV
for $A_0 = 2.5 \, m_0$, to the range $(2000, 6000$~GeV
for $A_0 = 3 \, m_0$ and the range $(4000, 12000)$~GeV for $A_0 = 5 \, m_0$.

\begin{figure}[h!]
\centering
\includegraphics[width=0.49\textwidth]{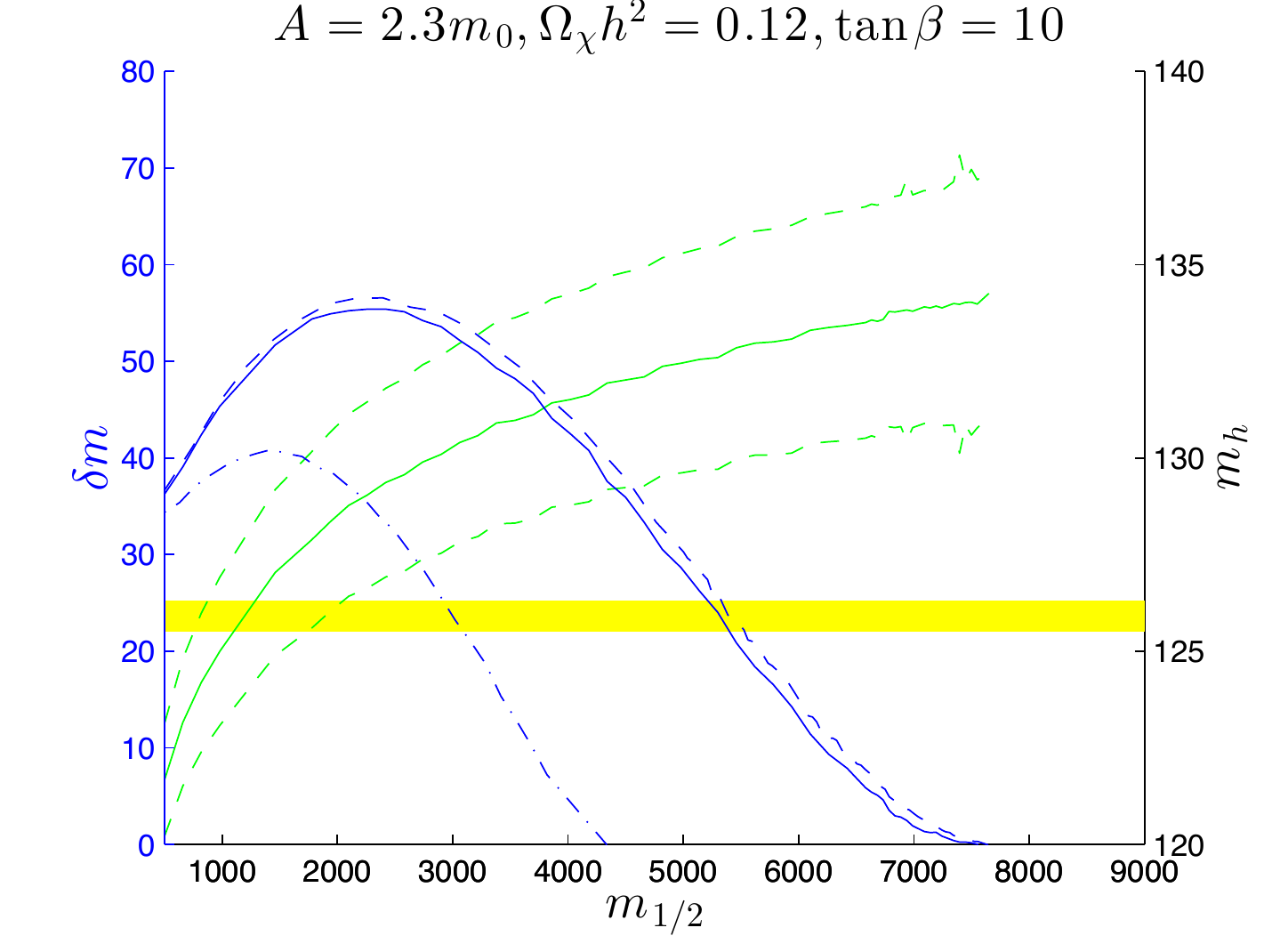}%
\hspace{0.004\textwidth}%
\includegraphics[width=0.49\textwidth]{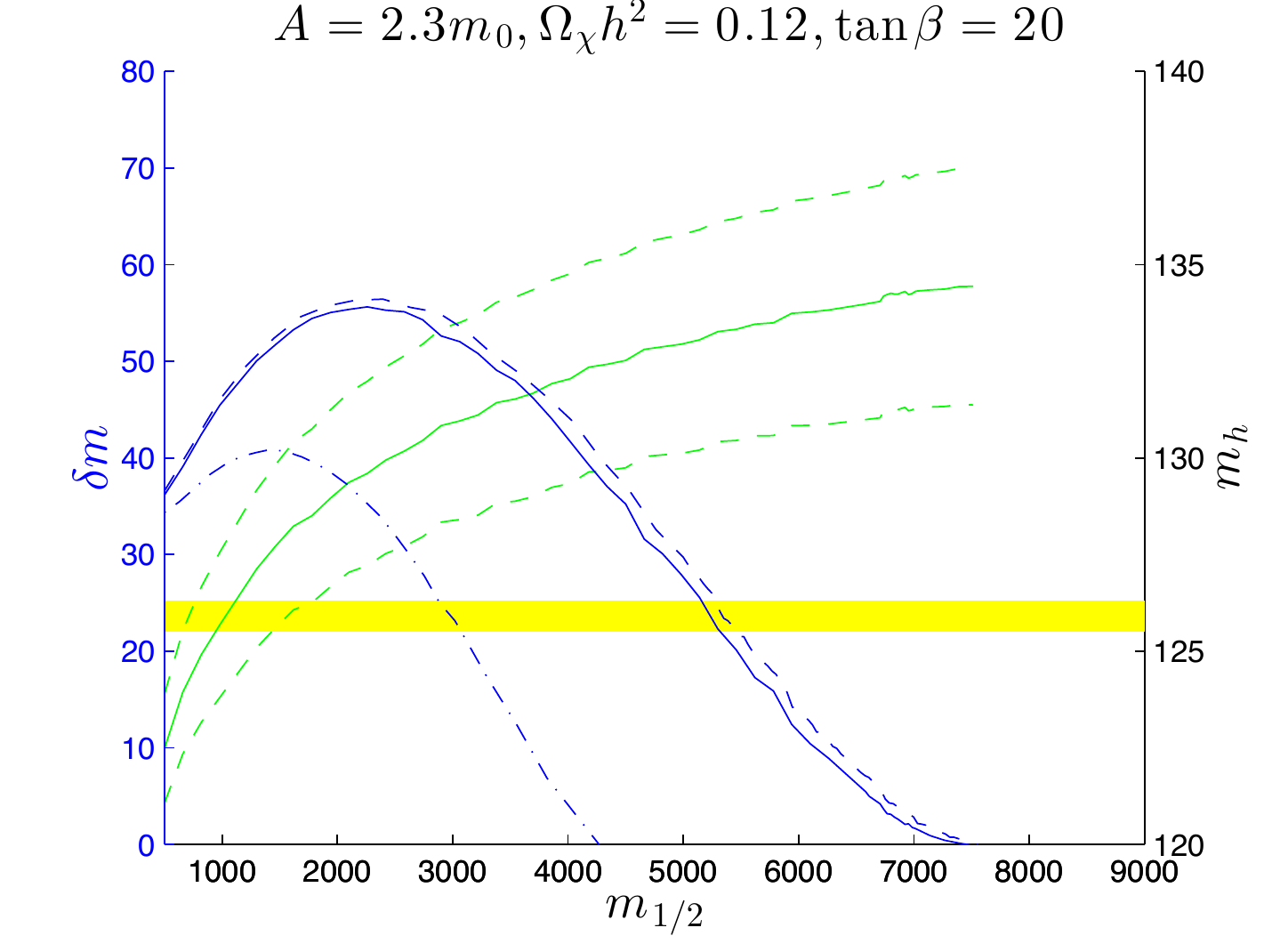}
\\
\includegraphics[width=0.49\textwidth]{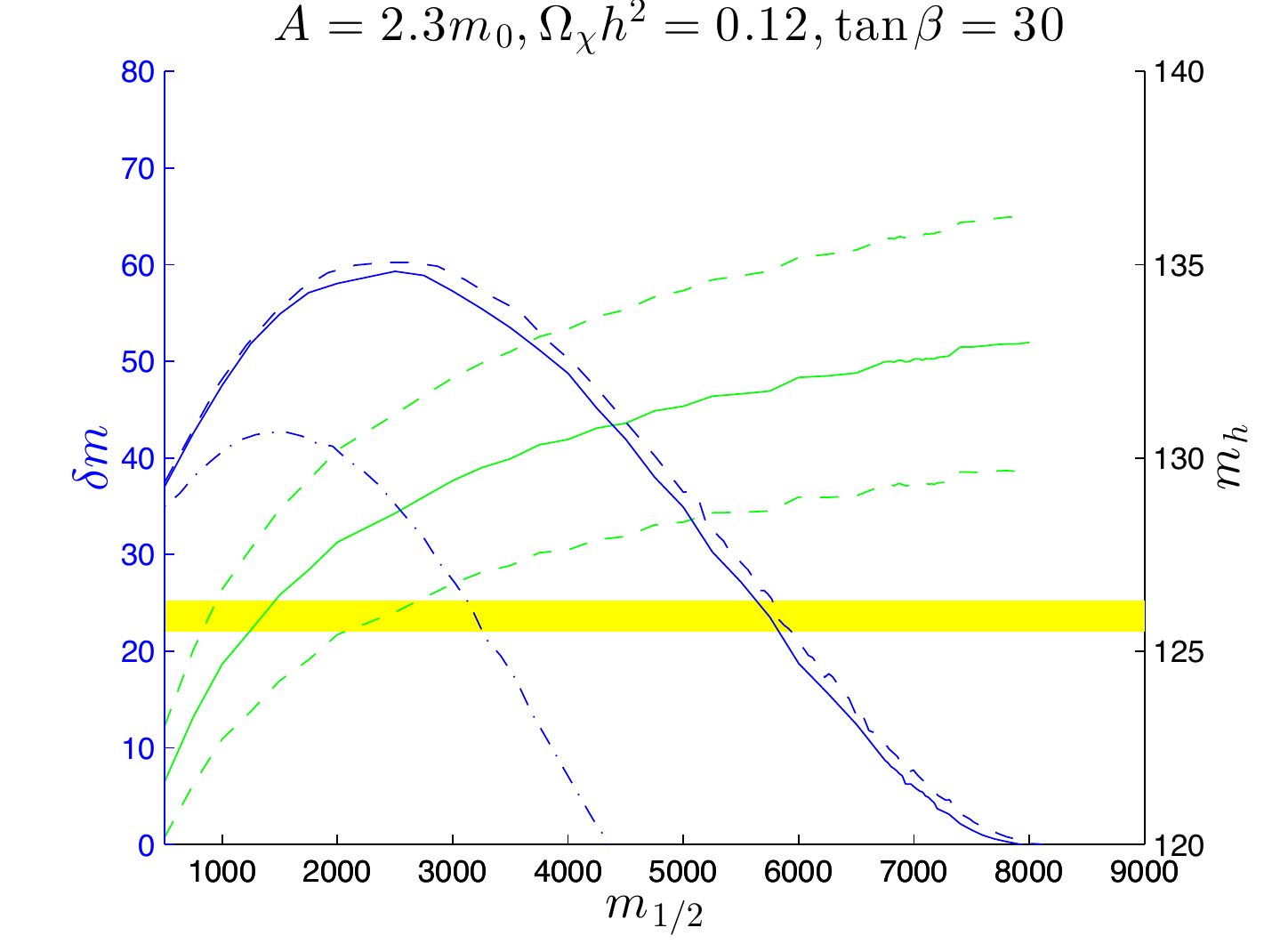}%
\hspace{0.004\textwidth}%
\includegraphics[width=0.49\textwidth]{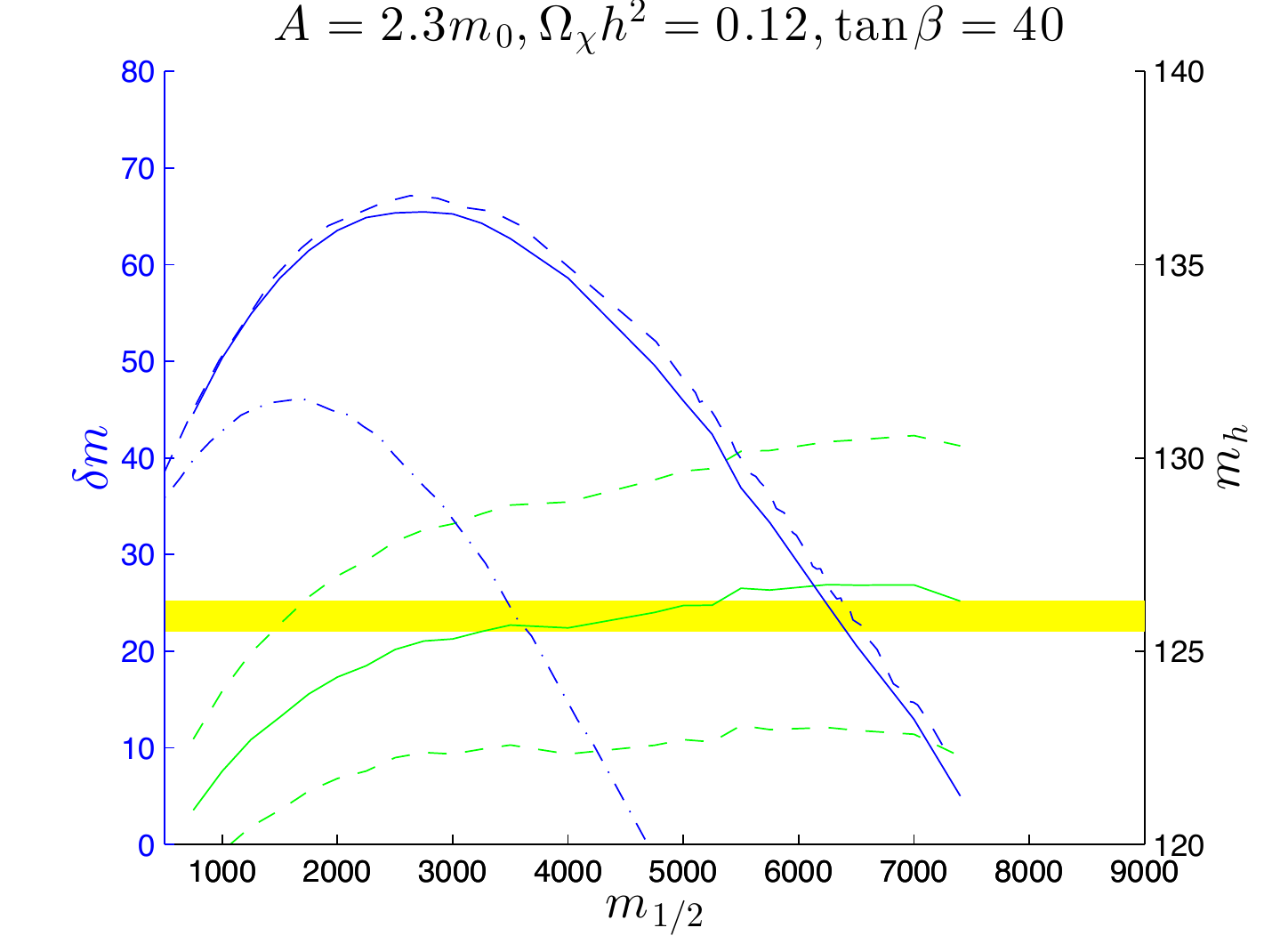}
\caption{\it As in Fig.~\protect\ref{fig:diff_scan_A}, but for
$A=2.3 \, m_{0}$ and $\tan\beta=10,20,30$ and $40$.}
\label{fig:diff_scan_A=2.3m0}
\end{figure}

Fig.~\ref{fig:diff_scan_A=2.3m0} shows the mass difference $\delta m = m_{\tilde t_1}-m_\chi$
and $m_h$ as functions of $m_{1/2}$ along the stop coannihilation strips for $A_0 = 2.3 \, m_0$ and
$\tan \beta = 10, 20, 30$ and 40. For this value of $A_0$ the maximum values of $\delta m$ exceed
50~GeV for $\tan \beta = 10, 20$ and 30, and are attained for values of $m_{1/2} \gappeq 2000$~GeV.
For $\tan \beta = 40$, the maximum value of $\delta m$ is above 60~GeV, and is achieved for
$m_{1/2} \sim 3000$~GeV. Correspondingly, the tips of the stop coannihilation strips are not universal,
extending from $\sim 7500$~GeV for $\tan \beta = 10$ and 20 to $\sim 8000$~GeV for $\tan \beta = 30$
and $\sim 8500$~GeV for $\tan \beta = 40$.
The strips for $\tan \beta = 10$ and $20$ are compatible with $m_h$
only for $m_{1/2} \lappeq 2000$~GeV,  and that for $\tan \beta = 30$ is compatible for
$m_{1/2} \lappeq 2500$~GeV, whereas the full coannihilation strip for $\tan \beta = 40$ above 1500~GeV is compatible
with $m_h$ within the theoretical uncertainties.

We display in Table~\ref{table:endpoints} the principal parameters characterizing
the endpoints of the stop coannihilation strips in the CMSSM for $A_0 = 2.2 \, m_0, 2.5 \, m_0, 3 \, m_0$
and $5 \, m_0$ and $\tan \beta = 20$, and for $A_0 = 2.3 \, m_0$ and
$\tan \beta = 10, 20, 30$ and 40, noting their values of $m_0$ and $m_{1/2}$ and the
corresponding values of $m_\chi = m_{\tilde t_1}$ as well as other parameters that are important for
determining the endpoints.

\begin{table}
\begin{center}
\begin{tabular}{ |c||c|c|c|c||c||c|c|c|c| }
  \hline
  Parameter & \multicolumn{4}{|c||}{$\tan \beta = 20$} & Parameter & \multicolumn{4}{|c|}{$A=2.3\, m_0$} \\
  \hline
  \hline
  $A_0/m_0$ & $2.2$ & $2.5$ & $3.0$ & $5.0$ & $\tan \beta$ & $10$ & $20$ & $30$ & $40$ \\
      \hline
   $m_{1/2}$ & $5900$ & $9200$  &$11000$  &$13000$ & $m_{1/2}$ & $7600$ & $7500$ & $8000$ & 7600\\
\hline
  $m_0$ &$24800$ &$19400$ & $15300$ &$8800$ & $m_0$ & $20900$ &$22200$ & $26900$ &  38600\\
\hline
$A_0$ & $54600$ &$48500$ & $45900$ &$44200$ & $A_0$ & $48000$ &$51100$ & $61900$ & 88800\\
  \hline \hline
  $\mu$ &$18600$ &$18800$ & $19400$ &$20300$ & $\mu$ & $18200$ &$18500$ & $21100$ & 27000\\
  \hline
  $A_t$ &$25700$ &$30100$ &$32600$  &$35600$ & $A_t$ & $27300$ &$27900$ & $31100$ & 36200  \\
  \hline
    $\sin\alpha$ &$-0.060$  &$-0.059$ &$-0.059$  &$-0.059$ & $\sin\alpha$ & $-0.11$ &$-0.059$ & $-0.042$ & -0.034\\
  \hline
  $m_{\tilde{t}_2}$ &$17500$ &$16600$ &$16200$  &$16100$ & $m_{\tilde{t}_2}$ & $17100$ &$16900$ & $18100$ & 20300\\ 
  \hline  \hline
   $m_\chi = m_{\tilde t_1}$ &$3000$  &$4600$ &$5500$ &$6500$ & $m_\chi = m_{\tilde t_1}$ & $3800$ &$3800$ & $4000$ & 3900 \\
   \hline
   $m_h$ &$136.1$ &$133.3$ &$131.7$ &$129.8$ & $m_h$ &$134.2$ &$134.5$ &$133.1$ & 126.2 \\
  \hline
\end{tabular}
\end{center}
\caption{\it Parameters characterizing the endpoints of the stop coannihilation strips in different
CMSSM scenarios with fixed $\tan \beta$ and varying $A_0/m_0$ (left columns)
and with fixed $A_0/m_0$ and varying $\tan \beta$ (right columns).
The values of $m_{1/2}$, $m_0$ and $A_0$ are specified at the GUT scale, 
whereas the other parameters are specified at the weak scale. Mass parameters are given
in GeV and, with the exception of $m_h$, quoted to 100~GeV accuracy.}
\label{table:endpoints}
\end{table}

\subsection{Strips for fixed $m_0/m_{1/2}$}

We have also considered coannihilation strips for fixed values of $m_0/m_{1/2}$
and $\tan \beta$, i.e., rays in the $(m_{1/2}, m_0)$ plane. The values of $A_0/m_0$  are adjusted
point-by-point along such lines to obtain the desired value of $\Omega_\chi h^2$.

\begin{figure}[h!]
\centering
\includegraphics[width=0.49\textwidth]{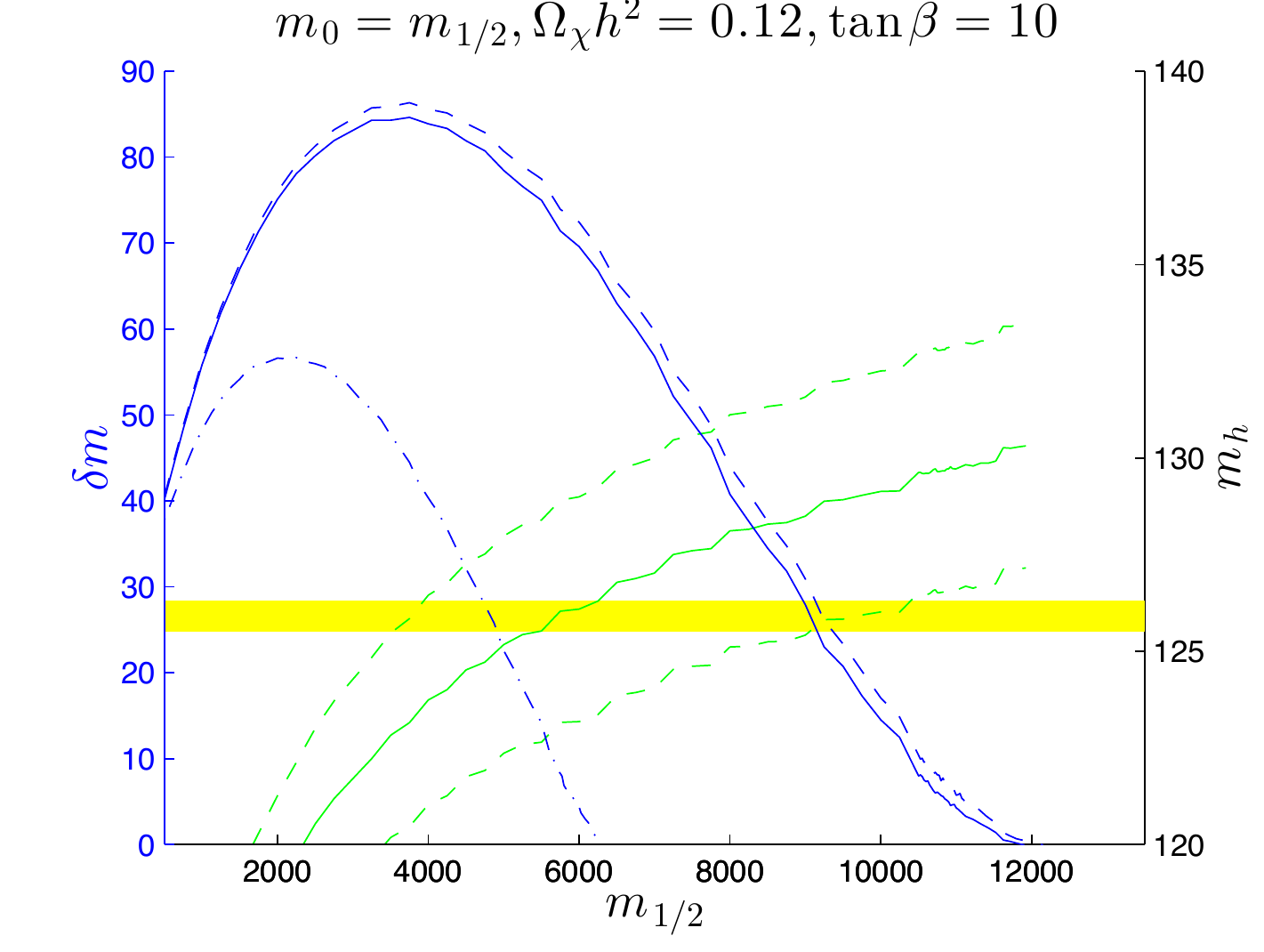}%
\hspace{0.004\textwidth}%
\includegraphics[width=0.49\textwidth]{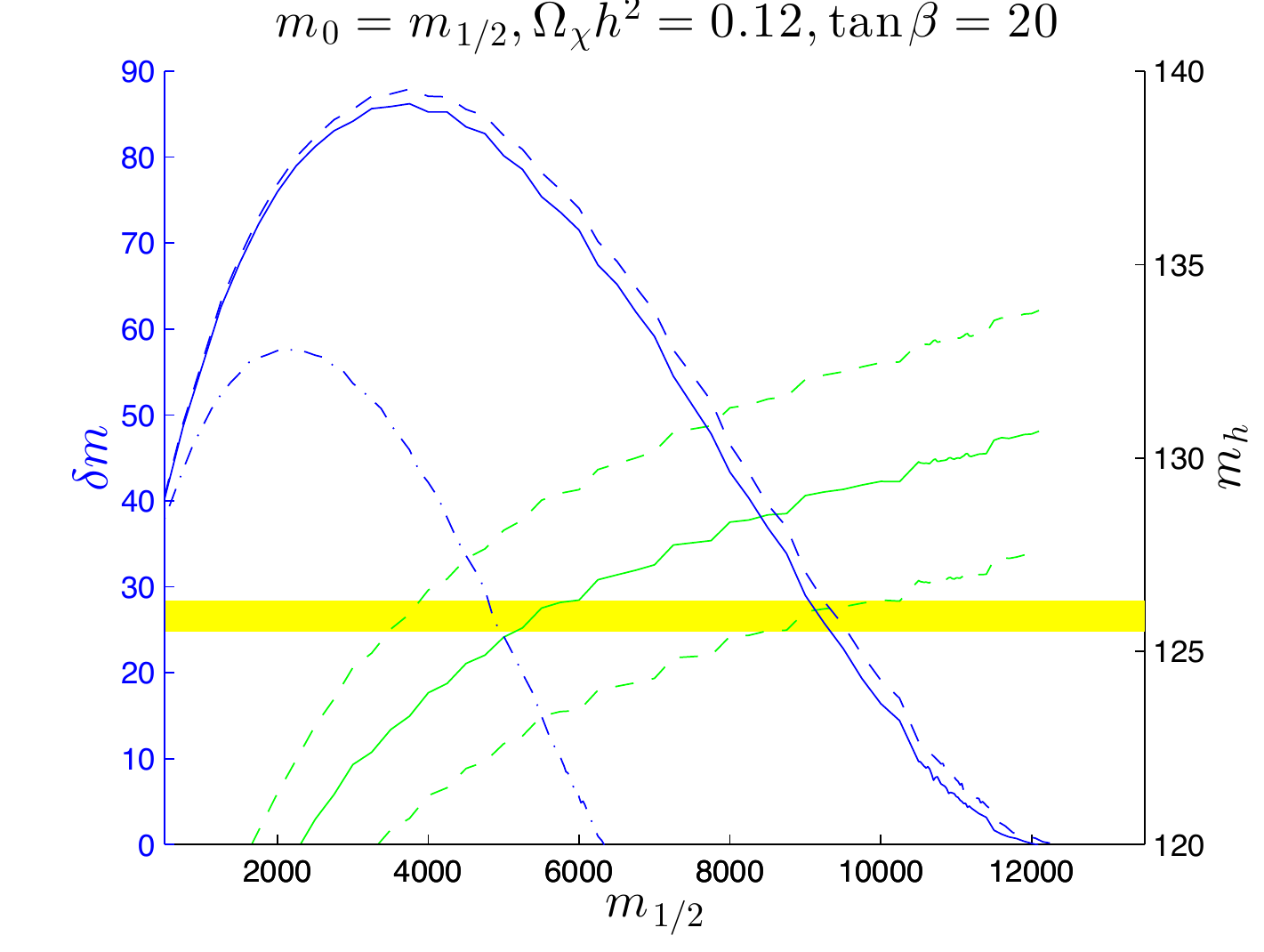}
\\
\includegraphics[width=0.49\textwidth]{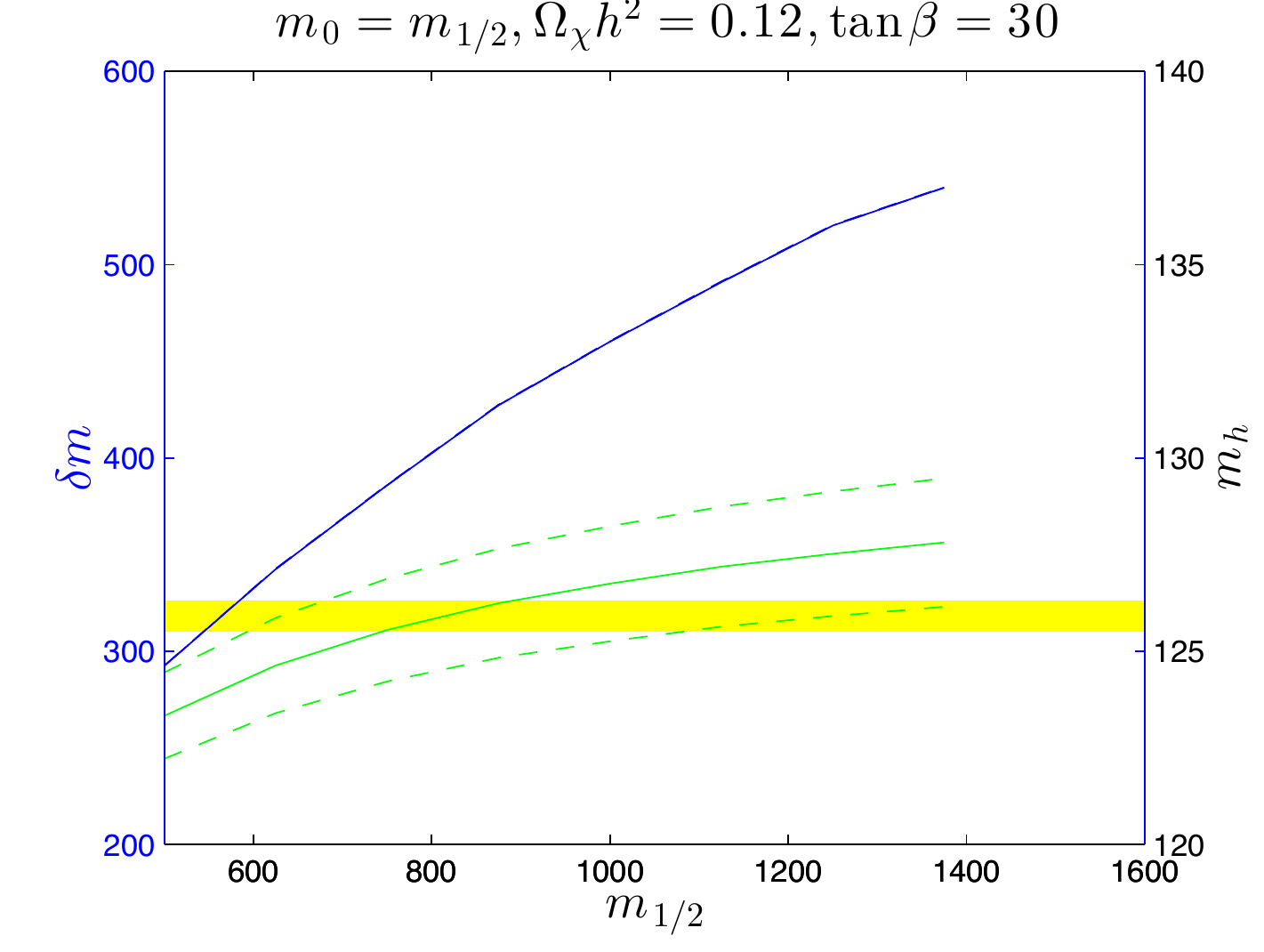}%
\hspace{0.004\textwidth}%
\includegraphics[width=0.49\textwidth]{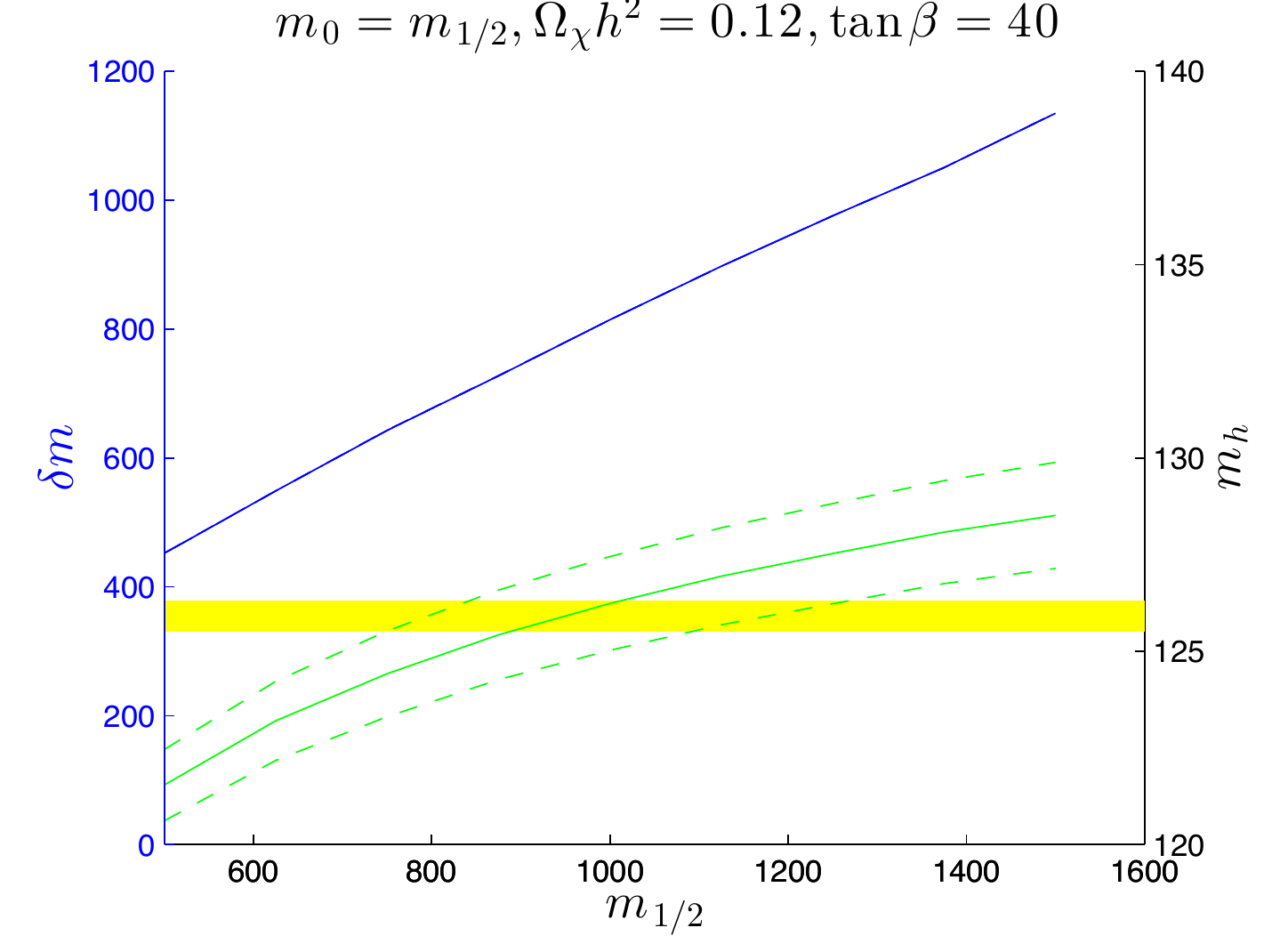}
\\
\caption{\it As in Fig.~\protect\ref{fig:diff_scan_A}, but for
$m_{0} = m_{1/2}$ and $\tan\beta=10, 20, 30$ and $40$.}
\label{fig:diff_scan_m0=mhalf}
\end{figure}

Fig.~\ref{fig:diff_scan_m0=mhalf} shows the behaviours of $\delta m$ and $m_h$ along
coannihilation strips for fixed $m_0 = m_{1/2}$ for the choices $\tan \beta = 10, 20, 30$ and 40. 
In the upper left panel for $\tan \beta = 10$ we see that $\delta m$ is maximized at $\sim 83$~GeV
for the nominal value $\Omega_\chi h^2 = 0.120$, when $m_{1/2} \sim 4000$~GeV.
This value of $\delta m$ is just below the threshold for ${\tilde t_1} \to \chi + b + W$ decay.
The end-point of this strip is at $m_{1/2} \sim 12000$~GeV corresponding to $m_\chi = m_{\tilde t_1} \sim 5900$~GeV, 
and the portion of the strip with $m_{1/2} \in (4000, 10000)$~GeV
has a value of $m_h$ compatible with the LHC measurement within the {\tt FeynHiggs~2.10.0}
uncertainties. The upper right panel for $\tan \beta = 20$ is quite similar, with $\delta m$ rising
slightly higher, but still below $m_\chi + M_W + m_b$ for $\Omega_\chi h^2 = 0.120$. The lower
panels for $\tan \beta = 30$ and 40 are very different. Indeed, in these cases the appropriate relic density
is found along the stau coannihilation strip, and the ends of the blue lines in these panels mark
the tips of the corresponding stau coannihilation strips. In $\tan \beta = 30$ case, all the strip
with $m_{1/2} \gappeq 600$~GeV is compatible with the measured value of $m_h$, and in the
$\tan \beta = 40$ case the portion with $750~{\rm GeV} \lappeq m_{1/2} \lappeq 1250$~GeV is compatible.
However, in both cases the portions with $m_{1/2} \lappeq 800$~GeV are excluded by the ATLAS jets + $\ETslash$
constraint, and the $B_s \to \mu^+ \mu^-$ constraint excludes the portion of the $\tan \beta = 30$ strip
with $m_{1/2} \lappeq 100$~GeV and all of the $\tan \beta = 40$ strip.

\begin{figure}[!htbp]
\centering
\includegraphics[width=0.49\textwidth]{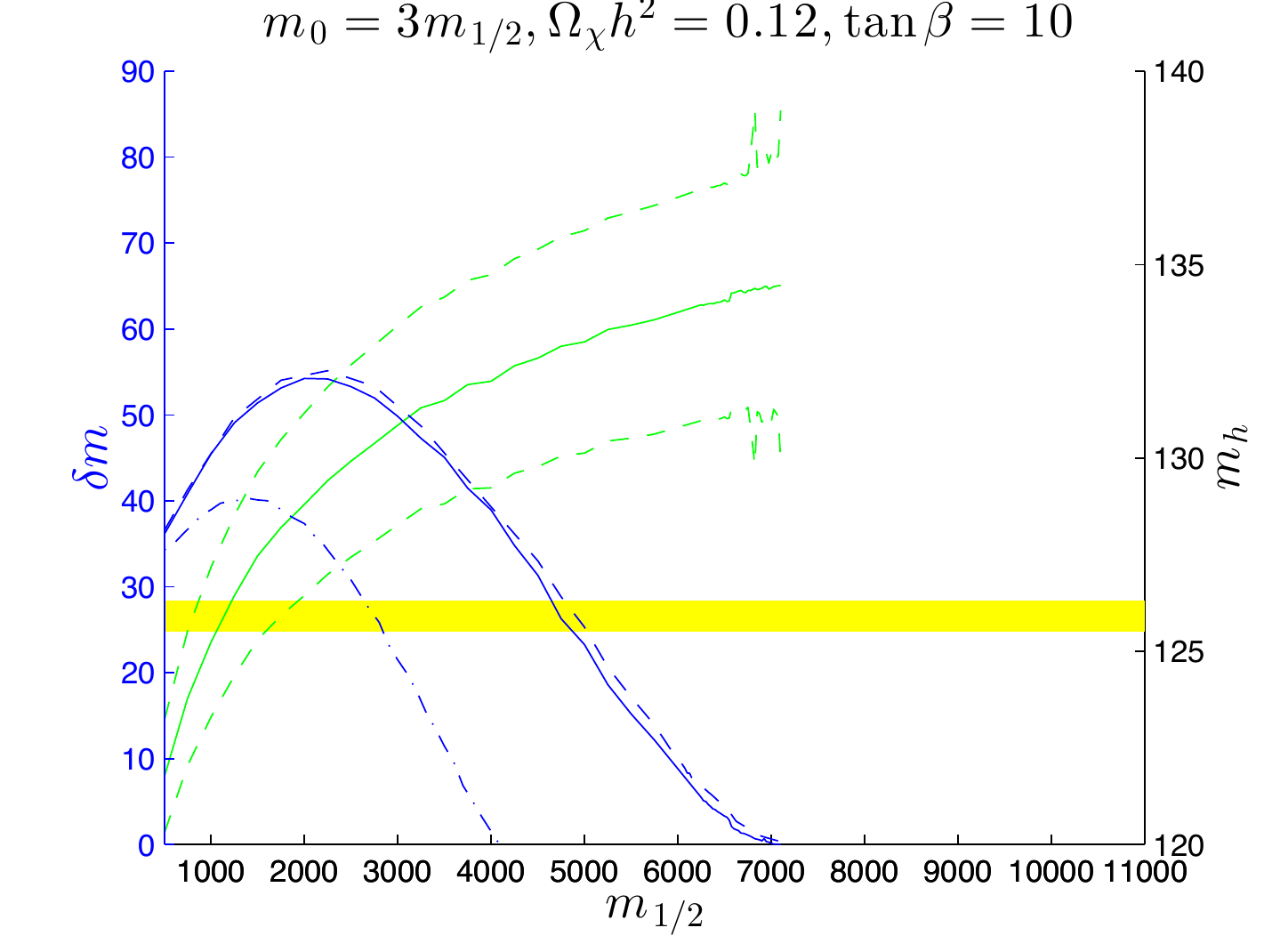}%
\hspace{0.004\textwidth}%
\includegraphics[width=0.49\textwidth]{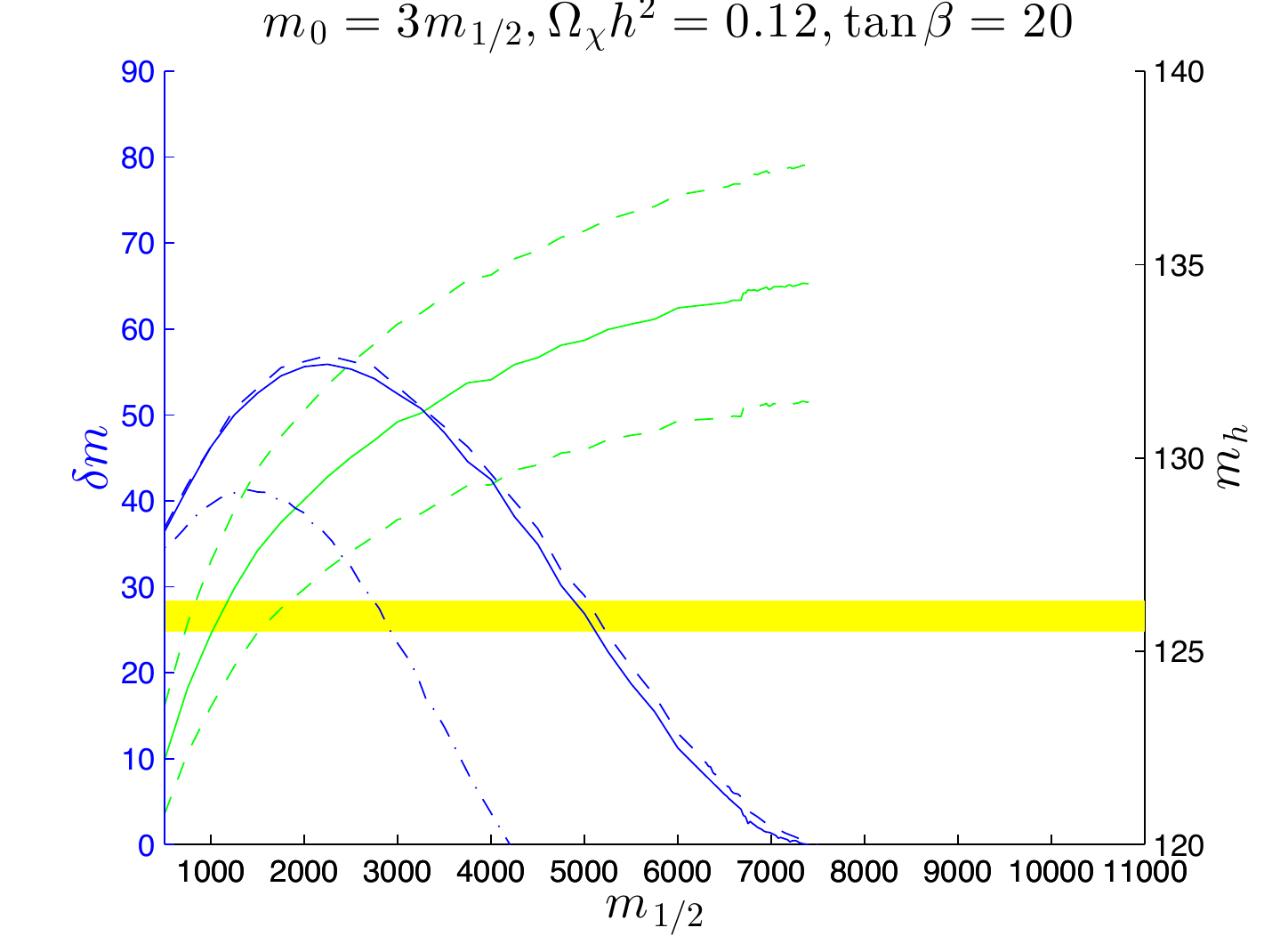}
\\
\includegraphics[width=0.49\textwidth]{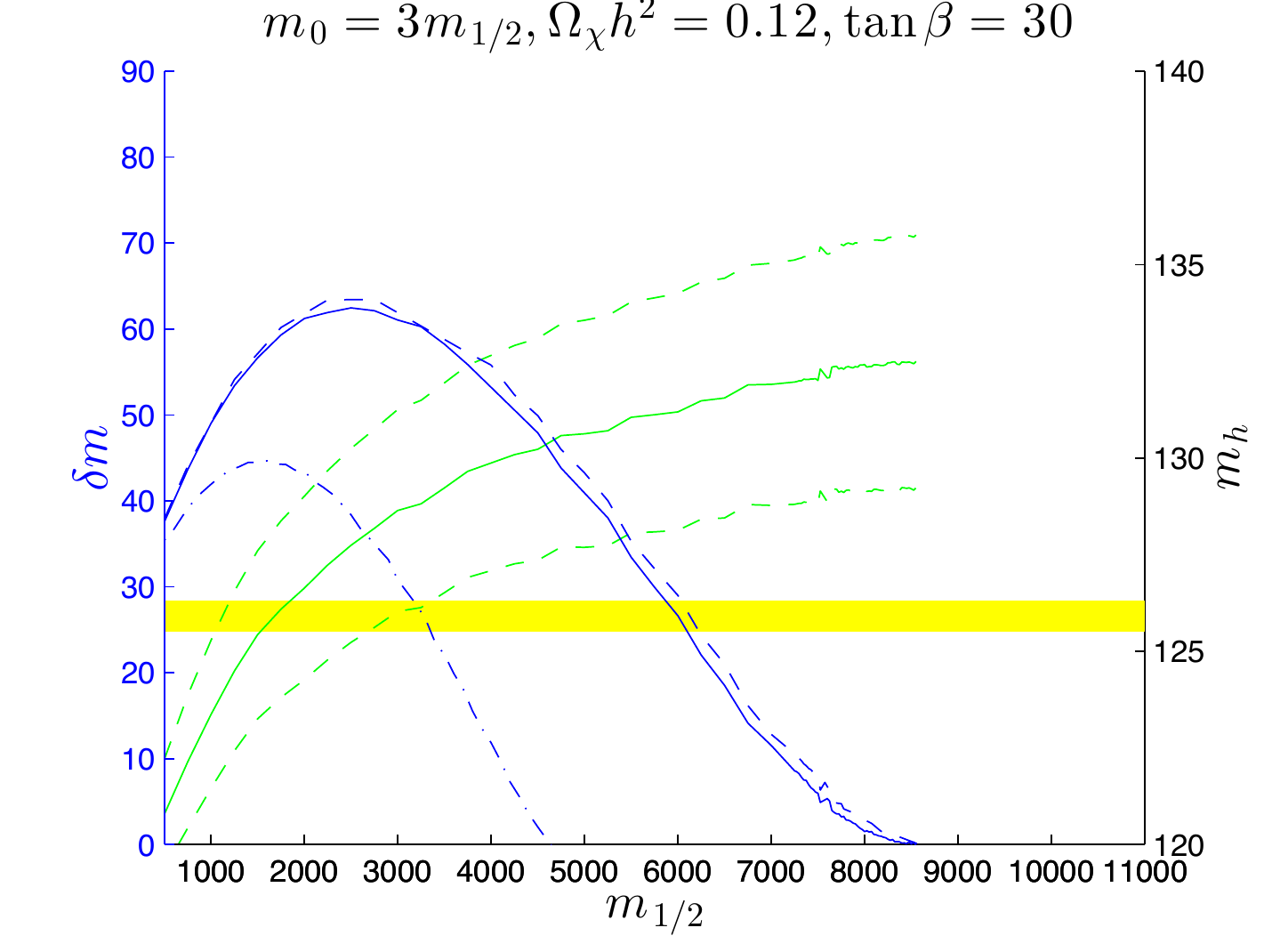}%
\hspace{0.004\textwidth}%
\includegraphics[width=0.49\textwidth]{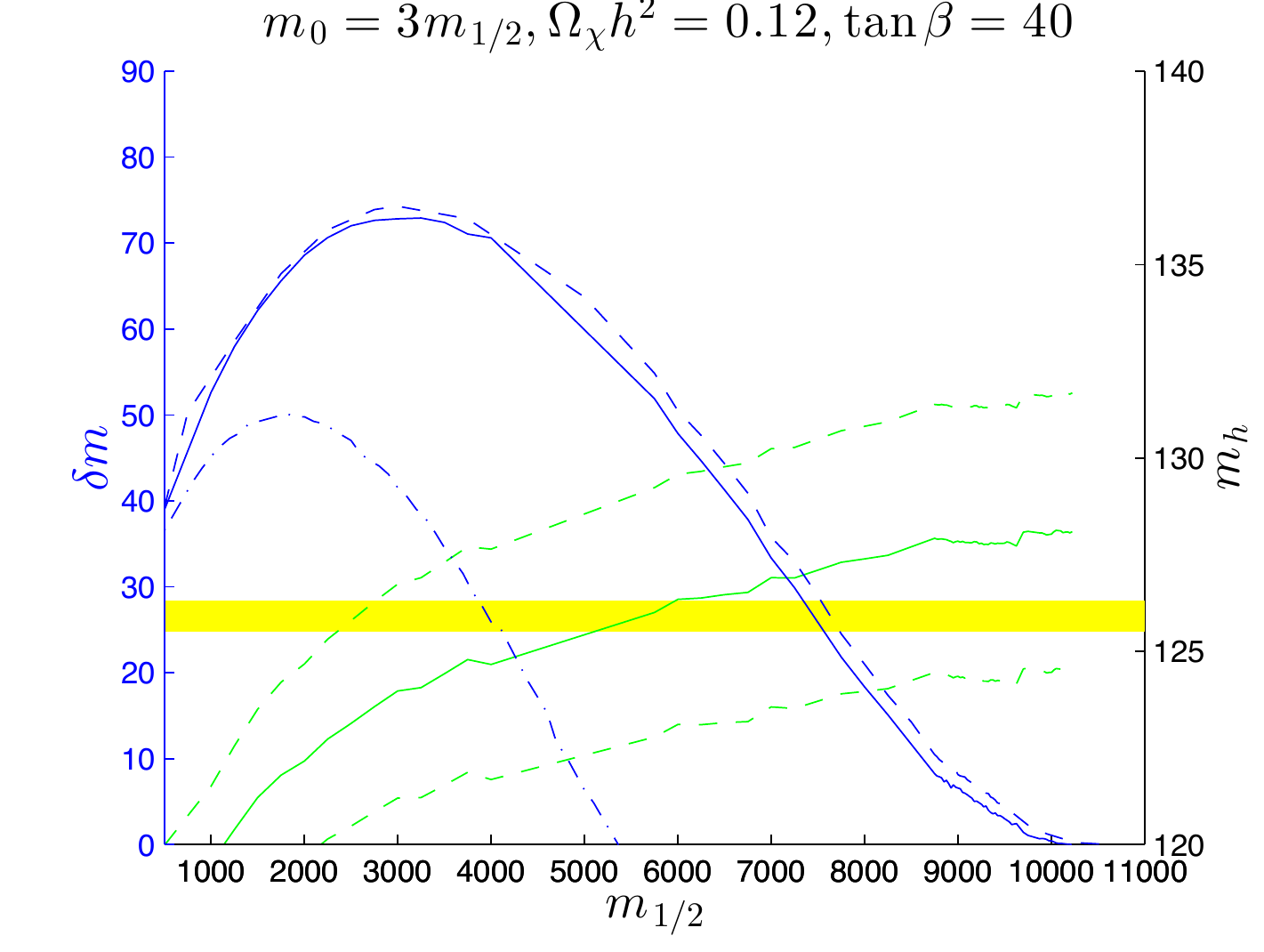}
\\
\caption{\it As in Fig.~\protect\ref{fig:diff_scan_A}, but for
$m_{0} = 3 \, m_{1/2}$ and $\tan\beta=10, 20, 30$ and $40$.}
\label{fig:diff_scan_m0=3mhalf}
\end{figure}

Fig.~\ref{fig:diff_scan_m0=3mhalf} shows the behaviours of $\delta m$ and $m_h$ along the corresponding
stop coannihilation strips for fixed $m_0 = 3 \, m_{1/2}$ for the choices $\tan \beta = 10, 20, 30$ and 40. 
In these cases, we see again that the maximum value of $\delta m$ increases with $\tan \beta$ from
$\sim 53$~GeV at $m_{1/2} \sim 2000$~GeV when $\tan \beta = 10$ to $\sim 70$~GeV at
$m_{1/2} \sim 3000$~GeV when $\tan \beta = 40$. Likewise, the tip of the coannihilation strip
extends from $\sim 7000$~GeV when $\tan \beta = 10$ to $\sim 10000$~GeV when $\tan \beta = 40$.
In the cases $\tan \beta = 10$ and 20, the calculated value of $m_h$ is compatible with
the value measured at the LHC for $1000~{\rm GeV} \lappeq m_{1/2} \lappeq 2000$~GeV, rising to $\lappeq 3000$~GeV
when $\tan \beta = 30$ and the range $\gappeq 2000$~GeV when $\tan \beta = 40$.

Table~\ref{table:endpointsm12m0} lists relevant parameters of the endpoints of
the stop coannihilation strips for $m_0/m_{1/2} = 1$ and $\tan \beta = 10$ and 20, and for
$m_0/m_{1/2} = 3$ and $\tan \beta = 10, 20, 30$ and 40.
\begin{table}
\begin{center}
\begin{tabular}{ |c||c|c||c|c|c|c| }
  \hline
  Parameter & \multicolumn{2}{|c||}{$m_0/m_{1/2} = 1$} & \multicolumn{4}{|c|}{$m_0/m_{1/2} = 3$} \\
  \hline
  \hline
  $\tan\beta$ & $10$ & $20$ & $10$ & $20$ & $30$ & $40$ \\
      \hline
   $m_{1/2}$ & $11900$ &$12100$ & $7100$ &$7400$ & $8600$ &$10200$ \\
\hline
  $m_0$ & $11900$ &$12100$ & $21300$ &$22200$ & $25700$ &$30700$ \\
\hline
$A_0$ & $43500$ &$44700$ & $48100$ &$50900$ & $60000$ &$73200$ \\
  \hline \hline
  $\mu$ & $19700$ &$19800$ & $18000$ &$18400$ & $20900$ &$24500$ \\
  \hline
  $A_t$ & $33600$ &$34100$ & $26400$ &$27600$ & $31600$ &$36900$  \\
  \hline
    $\sin\alpha$ & $-0.11$ &$-0.059$ & $-0.11$ &$-0.059$ & $-0.042$ &$-0.033$ \\
  \hline
  $m_{\tilde{t}_2}$ & $16500$ &$16100$ & $17000$ &$16800$ & $17800$ &$18900$ \\ 
  \hline  \hline
   $m_\chi = m_{\tilde t_1}$ & $5900$ &$6000$ & $3500$ &$3700$ & $4300$ &$5200$ \\
   \hline
   $m_h$ &$130.3$ &$130.7$ &$134.5$ &$134.6$ &$132.7$ &$128.6$ \\
  \hline
\end{tabular}
\end{center}
\caption{\it As in Table~\protect\ref{table:endpoints}, but for CMSSM scenarios with
fixed $m_0/m_{1/2} = 1$ and $3$.}
\label{table:endpointsm12m0}
\end{table}

\subsection{Stop Decay Signatures along the Coannihilation Strip}

We now consider the stop decay signatures along the coannihilation strips
discussed in the previous Section. Generally speaking, one expects the
two-body decays ${\tilde t_1} \to \chi + c$ to dominate as long as $\delta m > m_D \sim 1.87$~GeV~\cite{mp}.
Below this threshold, the dominant two-body decay processes are ${\tilde t_1} \to \chi + u$, which
would lead to decays of a mesino ${\tilde t_1}{\bar q} \to \chi +$ non-strange mesons
and of a sbaryon ${\tilde t_1} qq \to \chi +$ baryon, etc..
Four-body decays ${\tilde t_1} \to \chi + b + \ell + \nu$ and ${\tilde t_1} \to \chi + b + u + {\bar d}$
are also important as long as $\delta m > m_B \sim 5.3$~GeV, together with
${\tilde t_1} \to \chi + b + c + {\bar s}$ when $\delta m > m_{B_s} + m_D \sim m_{B} + m_{D_s} \sim 
m_{B_c} + m_K \sim 7$~GeV. Above this threshold, the total four-body decay rate
$\sim 9 \Gamma({\tilde t_1} \to \chi + b + \ell + \nu)$.

\begin{figure}[h!]
\centering
%\vspace{-5cm}
\includegraphics[width=0.49\textwidth]{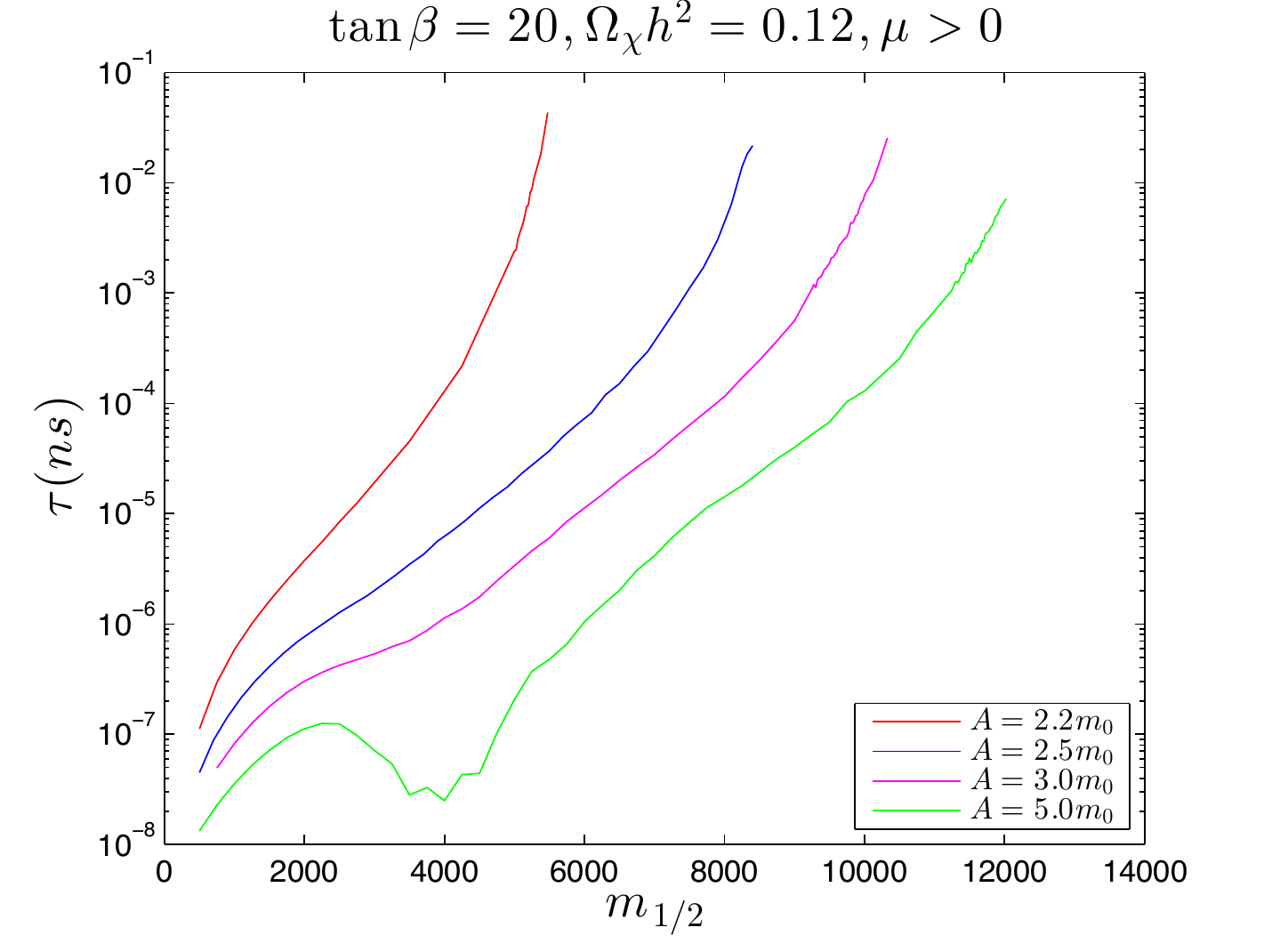}%
\hspace{0.004\textwidth}%
%\vspace{4cm}
\includegraphics[width=0.49\textwidth]{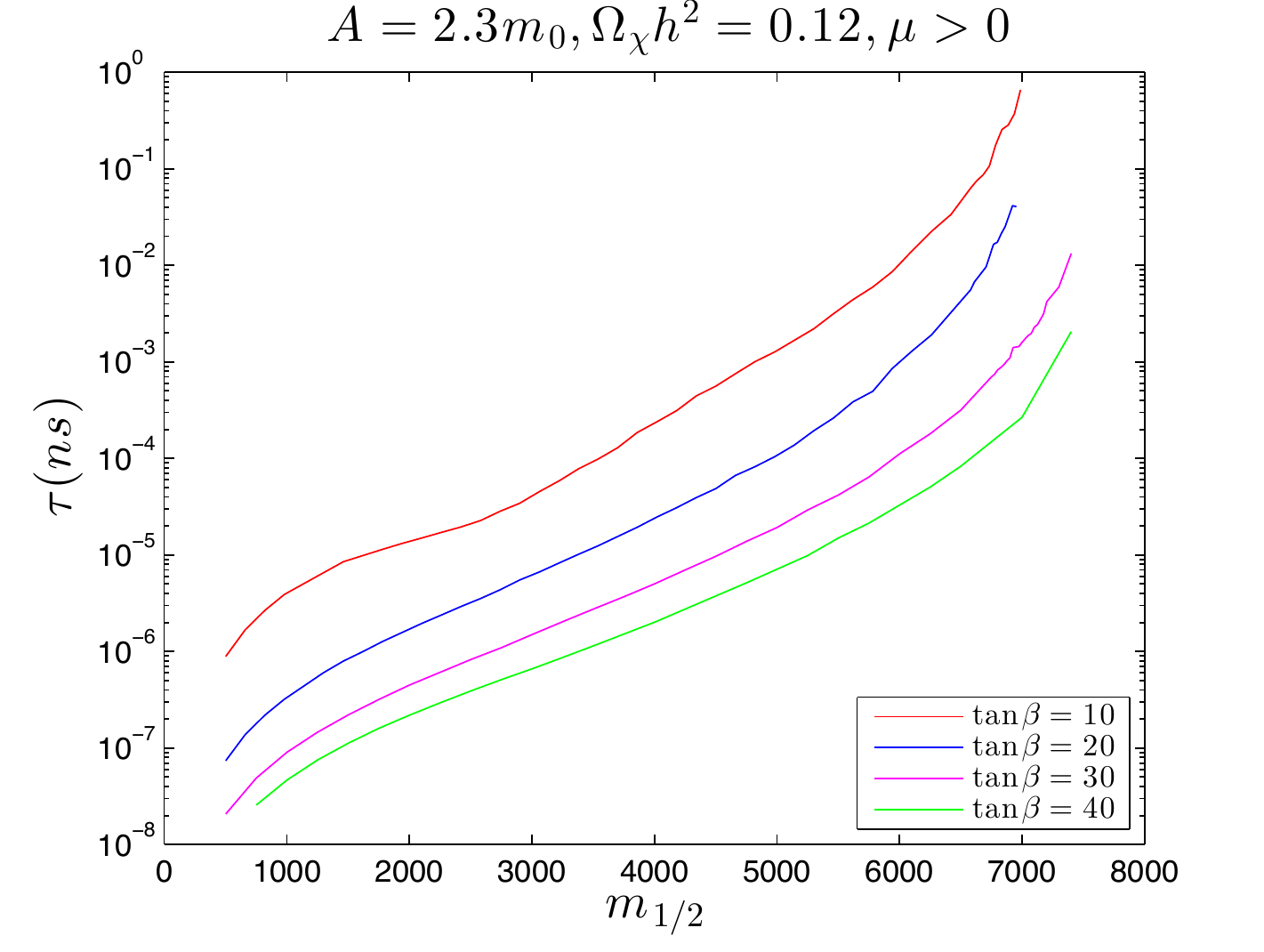} \\ %
%\vspace{-8cm}
\includegraphics[width=0.49\textwidth]{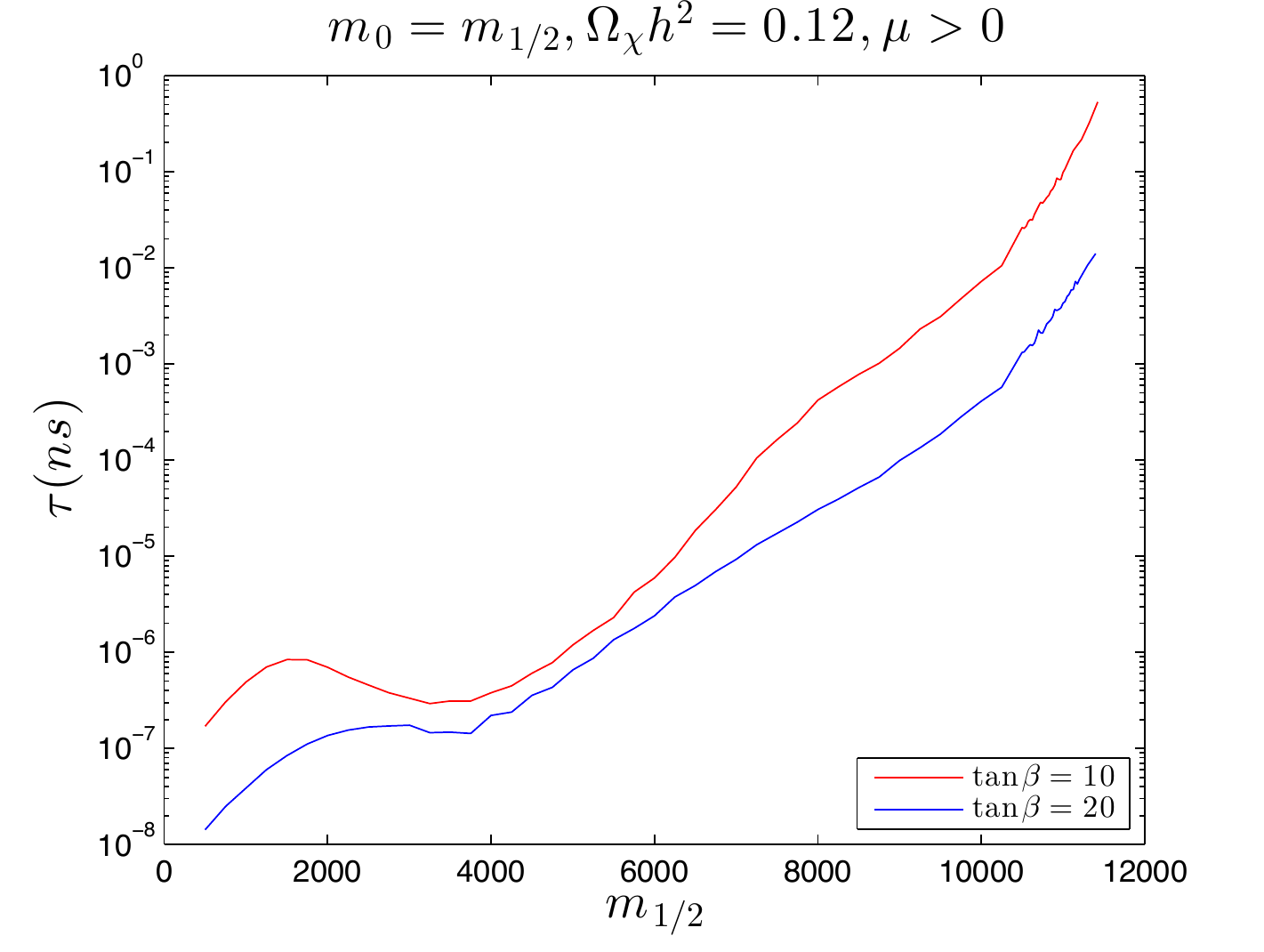}%
\hspace{0.004\textwidth}%
\includegraphics[width=0.49\textwidth]{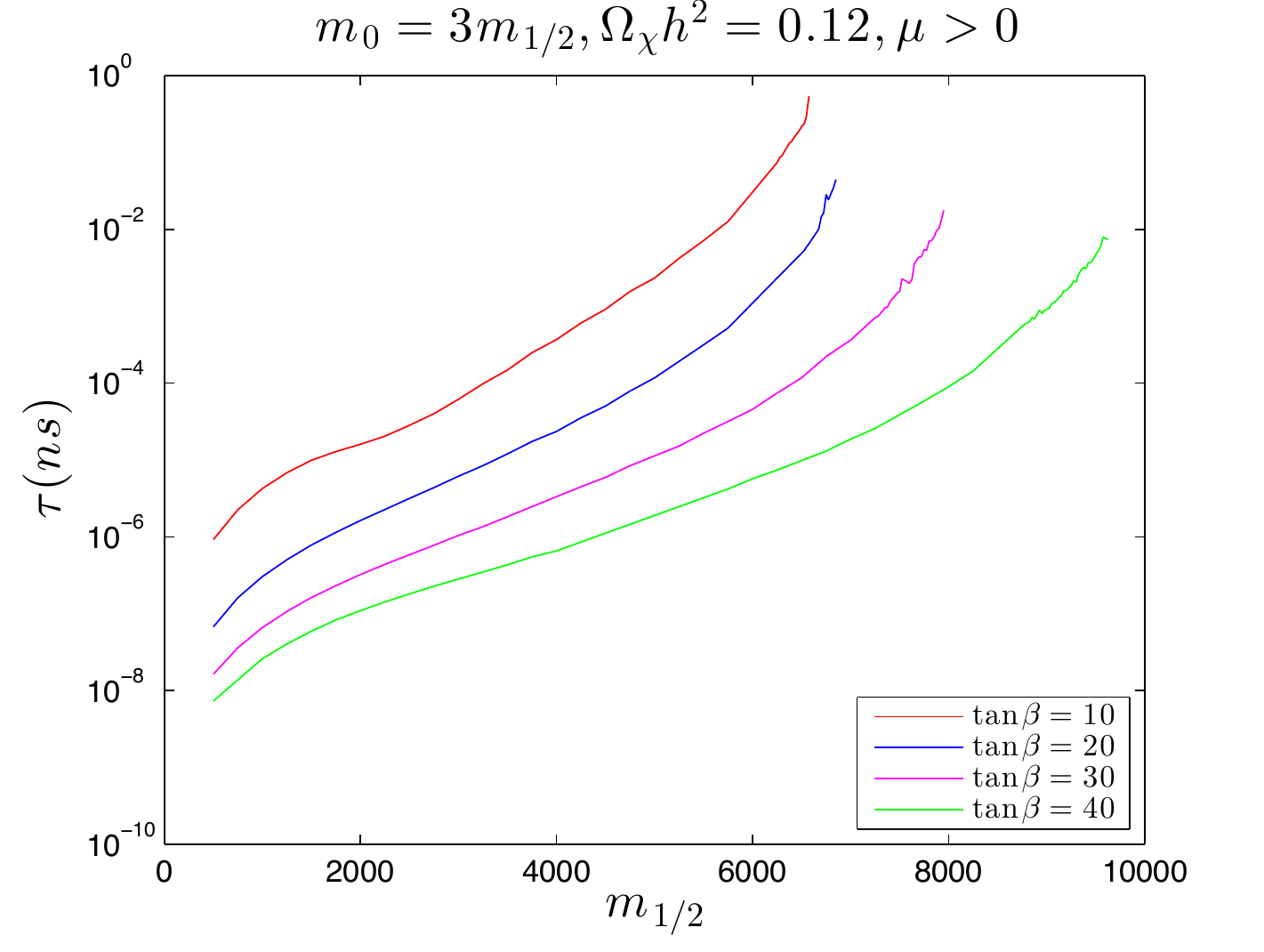}%
\caption{\it The total ${\tilde t_1}$ lifetime along the stop coannihilation strips (upper left) for $\tan \beta = 20$ and
$A_0 = 2.2 \,m_0$ (red), $2.5 \, m_0$ (blue), $3.0 \, m_0$ (purple) and $5.0 \, m_0$ (green),
(upper right) for $A_0 = 2.3 \, m_0$ when $\tan \beta = 10$ (red), $20$ (blue), $30$ (purple)
and $40$ (green), (lower left) for $m_0 = m_{1/2}$ and $\tan \beta = 10$ (red) and $\tan \beta = 20$ (blue),
and (lower right) for $m_0 = 3 \, m_{1/2}$ and $\tan \beta = 10$ (red), $20$ (blue), $30$ (purple)
and $40$ (green). The lines are restricted to the ranges of $m_{1/2}$ where $\delta m > m_D \sim 1.87$~GeV.}
\label{fig:lifetime}
\end{figure}

Fig.~\ref{fig:lifetime} displays calculations of the total ${\tilde t_1}$ lifetime along the stop
coannihilation strips for $\tan \beta = 20$ and $A_0 = 2.2 \, m_0, 2.5 \, m_0, 3 \, m_0$
and $5 \, m_0$ (upper left panel), and for $A_0 = 2.3 m_0$ with $\tan \beta = 10, 20, 30$ and 40 (upper right panel),
truncated to the ranges where $\delta m > m_D \sim 1.87$~GeV. 
In general, we see that the lifetime $\tau_{\tilde t_1}$
increases as $m_{1/2}$ increases monotonically towards the end of the coannihilation strip, reaching
$\tau_{\tilde t_1} \sim 1$~ns near the end of the strip for $A_0 = 2.3 \, m_0$ and $\tan \beta = 10$~\footnote{Exceptions
are seen in the left panels of Fig.~\ref{fig:lifetime}. The dips in the lifetime
arise because $\delta m \sim m_B + m_W$, as seen in the lower right panel of
Fig.~\ref{fig:diff_scan_A} and the upper panels of Fig.~\ref{fig:diff_scan_m0=mhalf}.}. The lifetime would be further enhanced
when $\delta m < m_D$, by a CKM matrix element factor ${\cal O}(20)$ as well as by phase-space
suppression, but we do not discuss this possibility in detail.
In the lower left panel of Fig.~\ref{fig:lifetime} we display the corresponding calculations of the total ${\tilde t_1}$ lifetime
for the stop coannihilation strips with $m_0 = m_{1/2}$ and $\tan \beta = 10$ and $20$, and in the lower right
panel the lifetime along the $m_0 = 3 \, m_0$ strips for $\tan \beta = 10, 20, 30$ and 40. We see that again
$\tau_{\tilde t_1} \sim 1$~ns near the end of the strip for $m_0 = 3 \, m_{1/2}$ and $\tan \beta = 10$.

\begin{figure}[h!]
\centering
%\vspace{-3cm}
\includegraphics[width=0.49\textwidth]{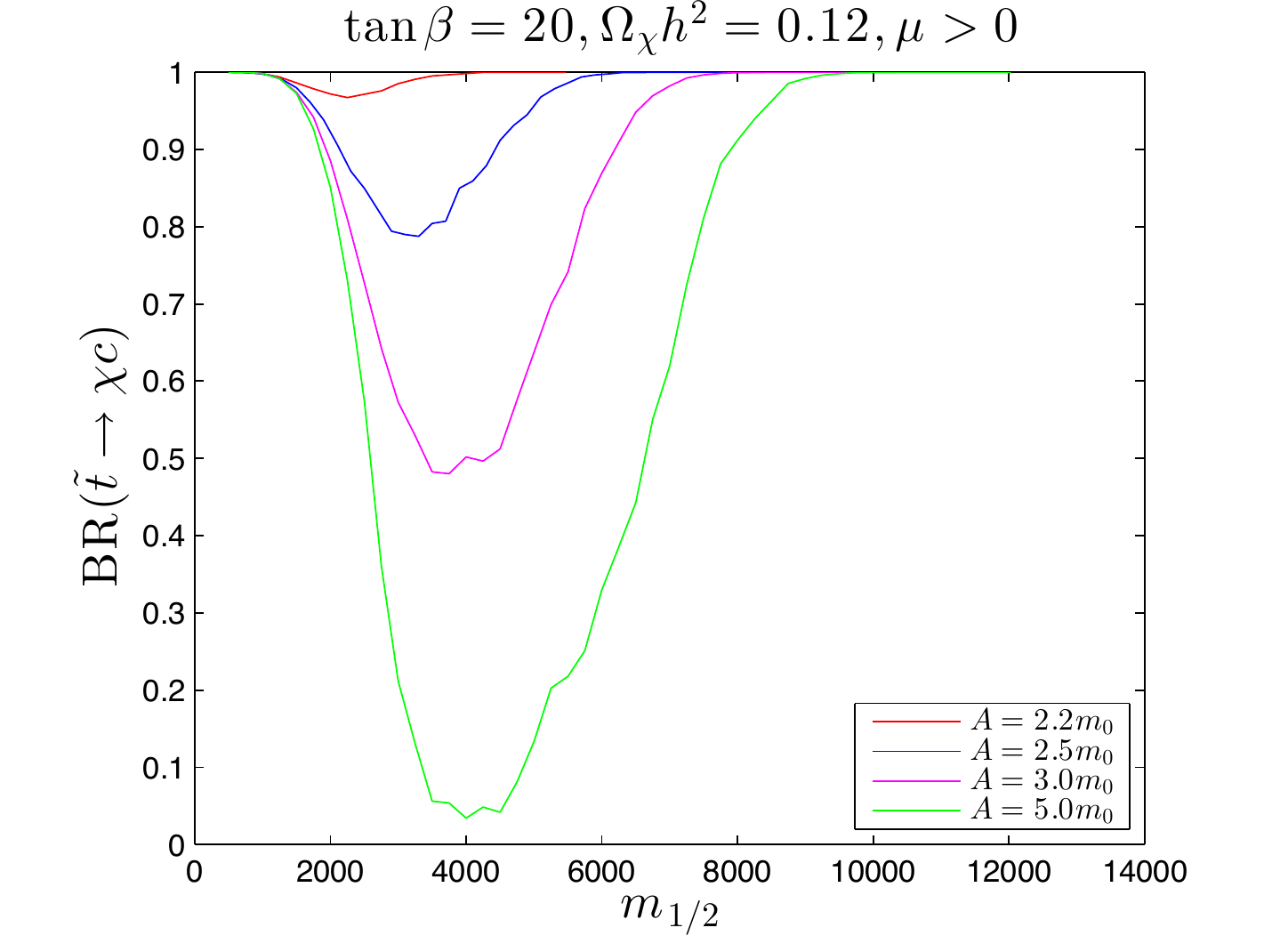}%
\hspace{0.004\textwidth}%
%\vspace{4cm}
\includegraphics[width=0.49\textwidth]{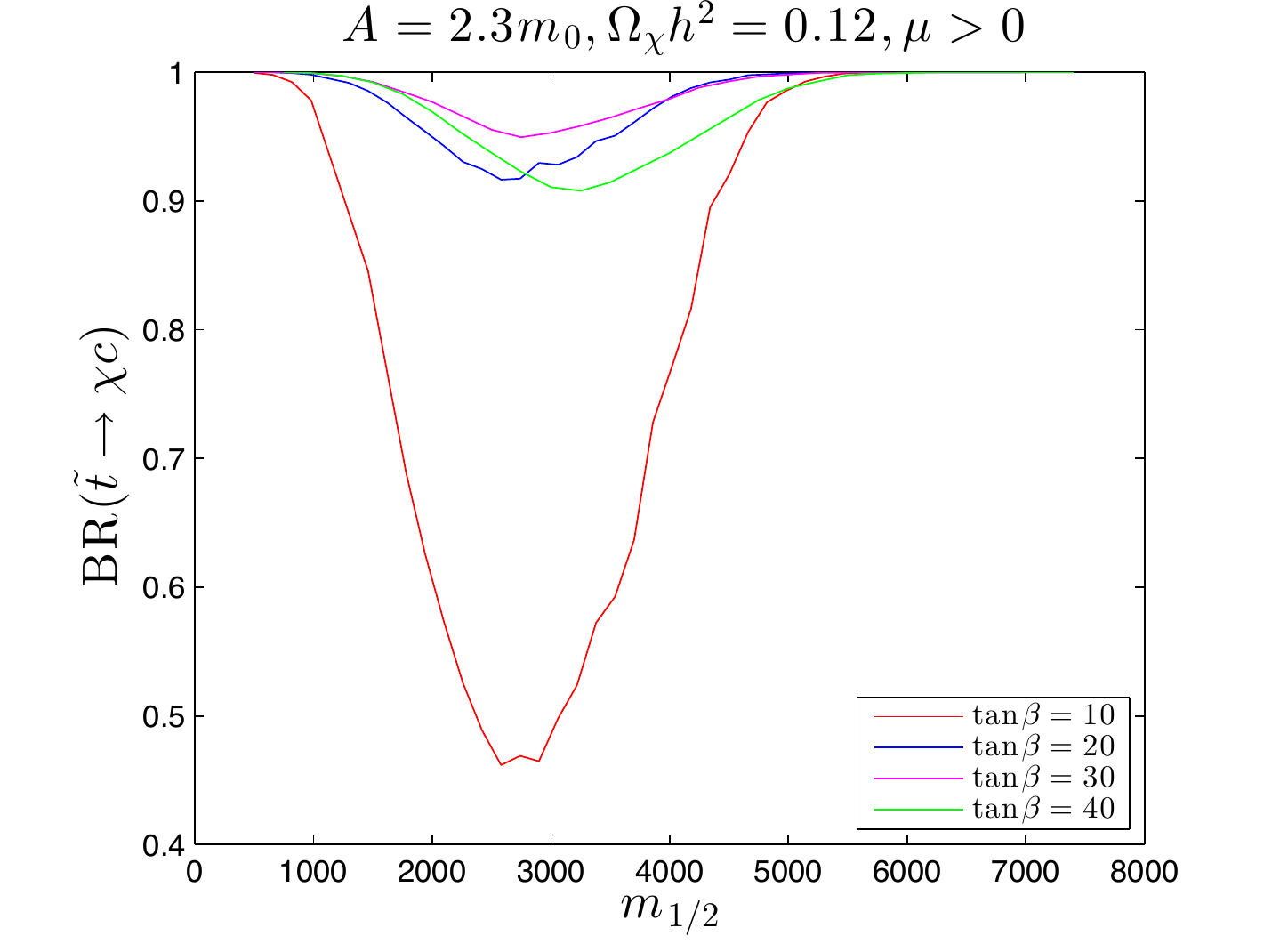} \\
%\vspace{-8cm}
\includegraphics[width=0.49\textwidth]{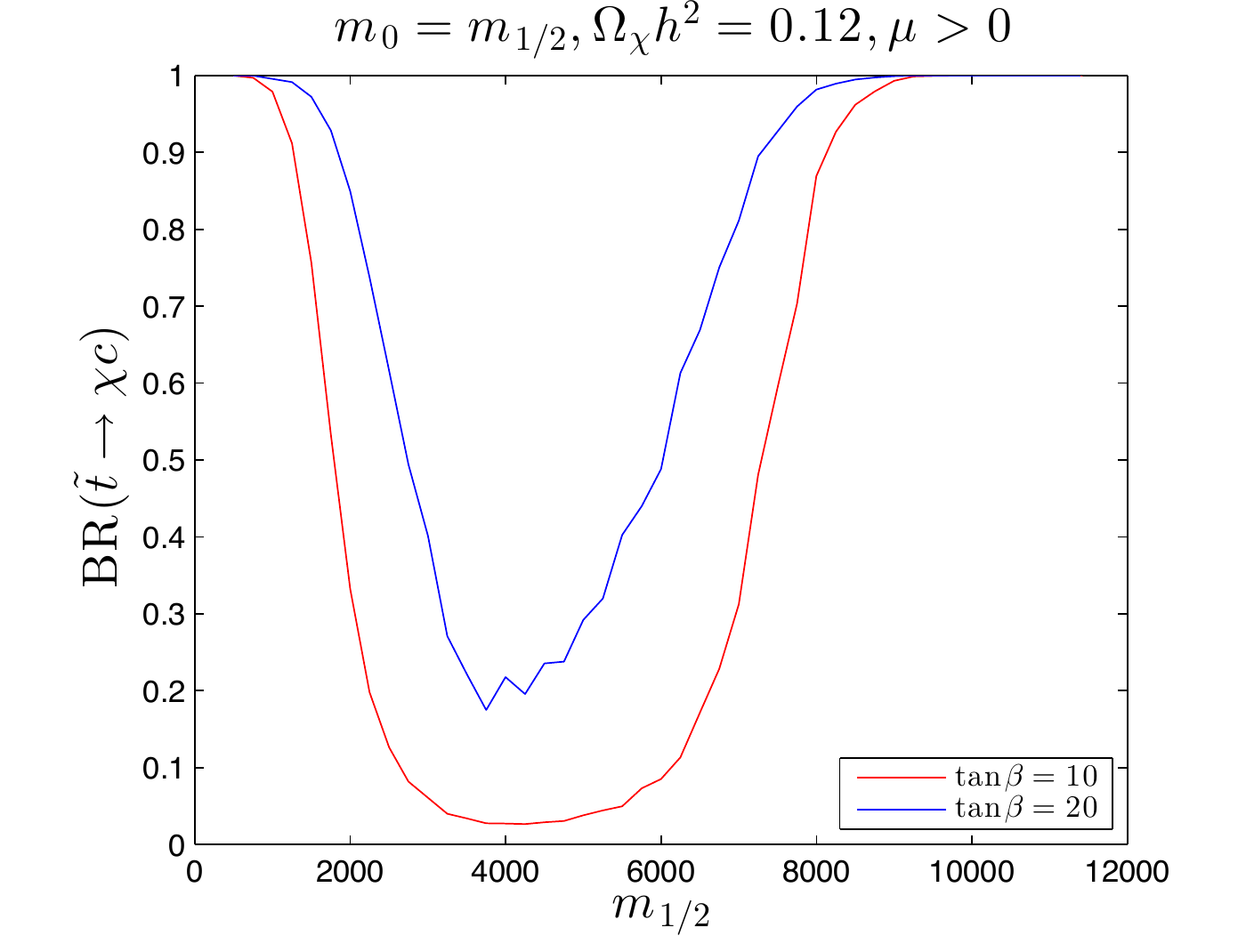}%
\hspace{0.004\textwidth}%
\includegraphics[width=0.49\textwidth]{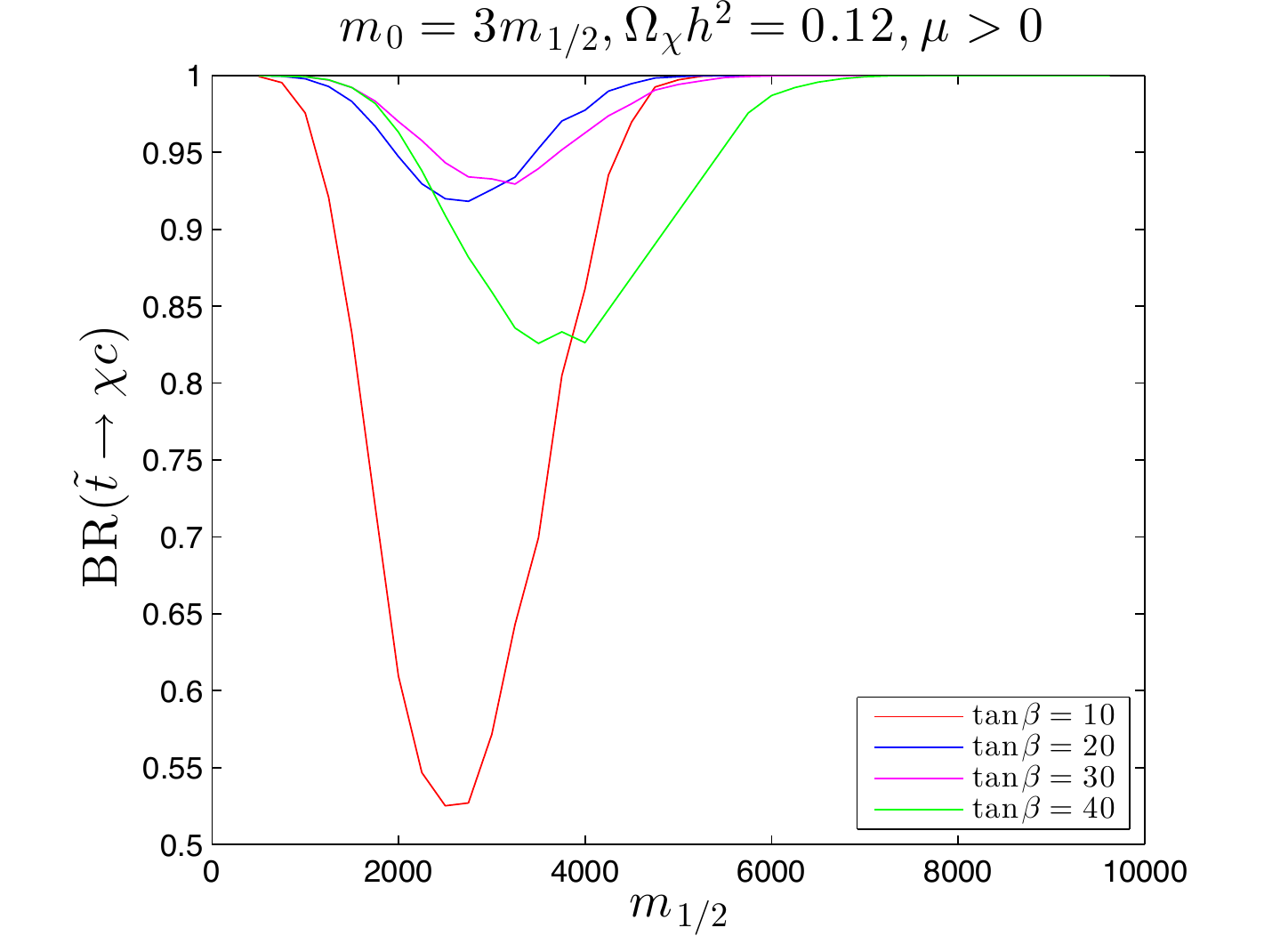}%
\caption{\it The branching ratios for ${\tilde t_1} \to \chi + c$ decay in the same models
as in Fig.~\protect\ref{fig:lifetime} and using the same colours for the lines.}
\label{fig:BRs}
\end{figure}

Fig.~\ref{fig:BRs} displays calculations of the ${\tilde t_1} \to \chi + c$ branching ratio
along the stop coannihilation strips for $\tan \beta = 20$ and $A_0 = 2.2 \, m_0, 2.5 \, m_0, 3 \, m_0$
and $5 \, m_0$ (upper left panel), for $A_0 = 2.3 m_0$ with $\tan \beta = 10, 20, 30$ and 40 (upper right panel)
for $m_0 = m_{1/2}$ and $\tan \beta = 10$ and $20$ (lower left panel), and
for $m_0 = 3 \, m_0$ and $\tan \beta = 10, 20, 30$ and 40 (lower right panel),
again truncated to the ranges where $\delta m > m_D \sim 1.87$~GeV. 
We see that the two-body decay ${\tilde t_1} \to \chi + c$ is usually more important than the four-body decays 
${\tilde t_1} \to \chi + b + f + {\bar f'}$, but with important exceptions
such as when $\tan \beta = 20, A_0 = 5.0 \, m_0$ for $3000~{\rm GeV} \lappeq m_{1/2} \lappeq 7000$~GeV
and when $m_0 = m_{1/2}$ and $\tan \beta = 20$ for $2000~{\rm GeV} \lappeq m_{1/2} \lappeq 7500$~GeV.
As a general rule, two-body dominance is reduced for intermediate values of $m_{1/2}$
where $\delta m$ is largest and the four-body phase space opens up, in which case
four-body decay signatures may become interesting as well as two-body decays.
Indeed, for $3000~{\rm GeV} \lappeq m_{1/2} \lappeq 5000$~GeV when $\tan \beta = 20$ and $A_0 = 5.0 \, m_0$
and when $m_0 = m_{1/2}$ and $\tan \beta = 20$, $\delta m > m_B + m_W$ so that the three-body decay
${\tilde t_1} \to \chi + b + W$ is formally accessible. In our treatment of this case we calculate
${\tilde t_1} \to \chi + b + (W^* \to f + {\bar f'})$, where $W^*$ denotes an (in general) off-shell
$W$ boson represented by a Breit-Wigner line shape. This yields a larger (and more accurate) decay
rate than calculating naively the three-body decay to $b$ and an on-shell $W$ boson, and we find
that BR(${\tilde t_1} \to \chi + b + f + {\bar f'}$) may exceed BR(${\tilde t_1} \to \chi + c$) by over an order
of magntitude.

\section{Summary and Conclusions}

We have shown in this paper that the existence of a long stop coannihilation strip
where the relic neutralino density $\Omega_\chi h^2$ falls within the cosmological range is
generic in the CMSSM for $2.2 \, m_0 \lappeq A_0 \lappeq 5.5 \, m_0$. It is essential
for calculating the length of this strip and the mass difference $\delta m = m_{\tilde t_1} - m_\chi$
along the strip to include Sommerfeld effects. The two
annihilation processes that are most important for determining the length of this
strip are ${\tilde t_1} {\tilde t_1}^* \to$ 2 gluons via t-channel ${\tilde t_1}$ exchange and s-channel
gluon exchange, which are completely model-independent, and ${\tilde t_1} {\tilde t_1}^* \to$ 2
Higgs bosons, which is more model-dependent. Specifically, the cross-section for the latter
process is mediated by ${\tilde t_2}$ in the cross channel, and hence depends on $m_{\tilde t_2}$
and on the ${\tilde t_1} - {\tilde t_2} - h$ coupling $C_{\tilde{t}_1-\tilde{t}_2-h}$ (\ref{t1t2h}) in the combination
$C_{\tilde{t}_1-\tilde{t}_2-h}/m_{\tilde{t}_2}$. We therefore expect that the location of the
end-point of the stop coannihilation strip should depend primarily on this ratio.

\begin{figure}[!htbp]
\centering
\includegraphics[width=0.69\textwidth]{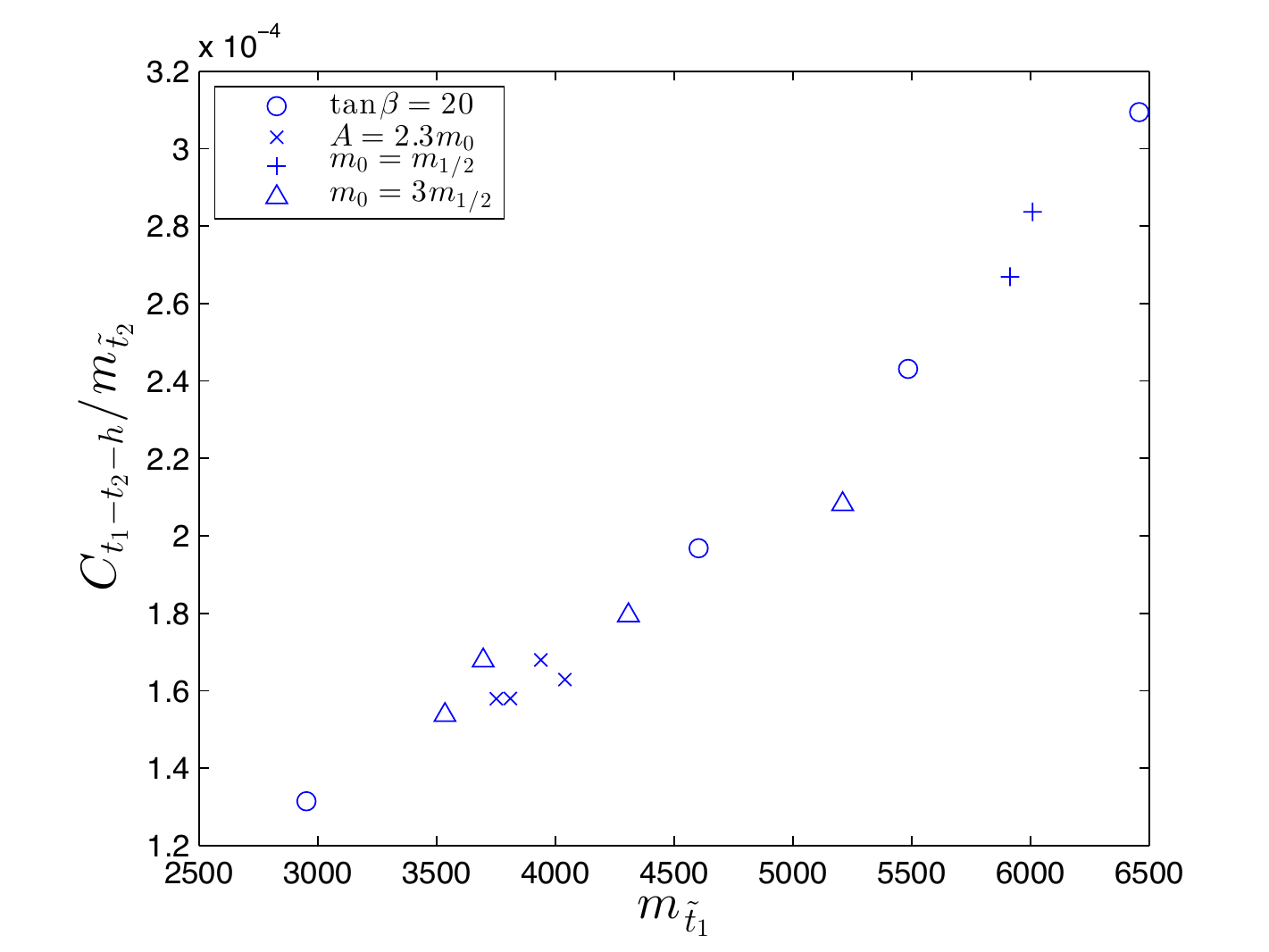}%
\caption{\it A scatter plot of the end-point values of $m_\chi = m_{\tilde t_1}$ vs the quantity
$C_{\tilde{t}_1-\tilde{t}_2-h}/m_{\tilde{t}_2}$ for the models with parameters listed in
Tables~\protect\ref{table:endpoints} and \protect\ref{table:endpointsm12m0}.}
\label{fig:scatter}
\end{figure}

In Tables~\ref{table:endpoints} and \ref{table:endpointsm12m0} we have listed the parameters
of the end-points in the various cases we have studied, including those appearing in the
expression for $C_{\tilde{t}_1-\tilde{t}_2-h}$ (\ref{t1t2h}). In Fig.~\ref{fig:scatter} we display
a scatter plot of the end-point values of $m_\chi = m_{\tilde t_1}$ vs the quantity
$C_{\tilde{t}_1-\tilde{t}_2-h}/m_{\tilde{t}_2}$. We see that, to a good approximation,
the end-point of the stop coannihilation strip is indeed a simple, monotonically-increasing
function of $C_{\tilde{t}_1-\tilde{t}_2-h}/m_{\tilde{t}_2}$. As seen in Fig.~\ref{fig:scatter},
in the models we have studied the maximum value of $m_\chi = m_{\tilde t_1}$
compatible with the cosmological dark matter constraint is $\sim 6500$~GeV. As
seen in the Tables, these scenarios
yield large values of $m_h$ as calculated using {\tt FeynHiggs~2.10.0}, 
but when $\tan \beta = 40$ the end-points are compatible with the
measured value of $m_h$ within the calculational uncertainty of $\sim 3$~GeV. It
seems possible that larger values of $m_\chi = m_{\tilde t_1}$ would be possible in models with larger values of
$C_{\tilde{t}_1-\tilde{t}_2-h}/m_{\tilde{t}_2}$.

We infer that a high-mass end-point for a stop coannihilation strip is likely to be a
general feature of a broad class of models. Its appearance is not restricted to the CMSSM and
closely-related models such as the NUHM~\cite{nuhm}, and its location depends primarily on the
combination $C_{\tilde{t}_1-\tilde{t}_2-h}/m_{\tilde{t}_2}$. However, the extent of the stop
coannihilation strip might be increased further in models in which other sparticles are (almost)
degenerate with the ${\tilde t_1}$ and $\chi$. This might occur, for instance, in circumstances
under which the lighter sbottom ${\tilde b_1}$ or one or more squarks of the first two
generations happened to be nearly degenerate with  the ${\tilde t_1}$ and $\chi$, but this
is unlikely to be a generic model feature.

We note also that the dominant ${\tilde t_1}$ decay mode along the stop coannihilation
strip is likely to be ${\tilde t_1} \to \chi + c$, since the mass difference $\delta m = m_{\tilde t_1} - m_\chi
< m_B + m_W$ in general and four-body decays ${\tilde t_1} \to \chi + b + f + \bar{f^\prime}$ are strongly
suppressed by phase space. This is likely to be a generic feature of stop coannihilation strips.
We also note that the ${\tilde t_1}$ lifetime may approach a nanosecond near the tip of the stop coannihilation
strip, which is also likely to be a generic feature.

We conclude that the stop coannihilation strip may be distinctive as well as generic.

\section*{Acknowledgements}

The work of J.E. was
supported in part by the London Centre for Terauniverse
Studies (LCTS), using funding from the European
Research Council via the Advanced Investigator Grant
267352, and by the UK STFC
via the research grant ST/J002798/1. The work of K.A.O. and J.Z.
was supported in part by DOE grant DE-FG02-94-ER-40823 at the University of Minnesota.

\end{document}